\documentclass[12pt,a4paper]{article}

\def\jgpapersize{}

\usepackage{ulem}
\usepackage{amsmath}
\usepackage{amsthm}
\usepackage{amssymb}
\usepackage[utf8]{inputenc}
\usepackage{alphabeta}
\usepackage{graphicx}
\usepackage{tikz}

\usepackage{hyperref}
\hypersetup{pdftex,colorlinks=true,allcolors=blue}
\usepackage{hypcap}
\usepackage{pgfplots}

\usetikzlibrary{decorations.markings}
\tikzset{->-/.style={decoration={
  markings,
  mark=at position #1 with {\arrow{>}}},postaction={decorate}}}
\tikzset{-<-/.style={decoration={
  markings,
  mark=at position #1 with {\arrow{<}}},postaction={decorate}}}
\usetikzlibrary{external}

\def\defn#1{{\bf #1}}

\def\Real{{\mathbb R}}

\def\norm#1{\left|#1\right|}
\def\innerprod(#1,#2){{\left<#1\,,\,#2\right>}}

\def\qquadtext#1{\qquad\textup{#1}\qquad}
\def\qquadand{\qquadtext{and}}
\def\quadtext#1{\quad\textup{#1}\quad}
\def\quadand{\quadtext{and}}
\def\pfrac#1#2{\frac{\partial #1}{\partial #2}}

\def\dfrac#1#2{\frac{d #1}{d #2}}

\def\Dfrac#1#2{\frac{D #1}{d #2}}

\def\sumdoubleind#1#2{
\begin{array}[t]{c}
\mbox{\LARGE$\sum$}\\
\mbox{\scriptsize$#1$}\\[-0.3em]
\mbox{\scriptsize$#2$}
\end{array}}

\usepackage{fancyhdr}
\fancypagestyle{firststyle}
{
   \fancyhf{}
   
\fancyfoot[C]{1}
\fancyfoot[R]{\footnotesize\bf File: \jobname.tex}
}
\newenvironment{bulletlist}{\begin{list}{$\bullet$}
{
\setlength{\itemindent}{0 em}
\setlength{\itemsep}{1pt}
\setlength{\labelsep}{0.5em}
\setlength{\labelwidth}{15em} 
\setlength{\leftmargin}{1.5em}
\setlength{\parsep}{1pt}
\setlength{\parskip}{0em}
\setlength{\partopsep}{0pt}
\setlength{\topsep}{5pt}}}
{\end{list}}

\def\JGchange#1{{\color{blue}{#1}}}

\newcounter{pfcounter}
\newenvironment{Proof}[1]{\refstepcounter{pfcounter}
\begin{proof}[Proof number
    {\thepfcounter}:\ #1]\color{black}}{\end{proof}}
\newenvironment{ProofI}[1]{\begin{proof}}{\end{proof}}
\newenvironment{ProofA}[1]{\begin{proof}[#1]}{\end{proof}}

\def\SeeProof#1{{{See proof number \ref{#1} on page \pageref{#1} in the
      appendix.}}}

\def\SeeProof#1{{{See proof number \ref{#1} in the
      appendix.}}}

\def\Interval{{\cal I}}
\def\Mman{{\cal M}}

\def\Cdot{{\dot C}}

\def\DixVec{{N}}

\def\deltaFour{\delta^{(4)}}
\def\deltaThree{\delta^{(3)}}

\def\zetaDP{\zeta_{\textup{D}}}

\def\zetaDP{\zeta}

\def\pregeo{{\textup{pre}}}

\def\Jcurr{{\mathcal{J}}}
\def\Mman{{\cal M}}

\def\xhat{{\hat{x}}}
\def\zetahat{{\hat{\zeta}}}
\def\zhat{{\hat{z}}}

\def\THat{{\hat T}}

\def\Ahat{{\hat A}}

\def\zerohat{{\hat 0}}
\def\Ohat{{\hat 0}}
\def\Cdothat{{\hat{\Cdot}}}

\def\gammahat{{\hat\gamma}}
\def\PF{{\varsigma}}

\def\Vz{{\boldsymbol z}}
\def\Vw{{\boldsymbol w}}
\def\Vzero{{\boldsymbol 0}}
\def\Jaabb{{\cal J}^{\iMahat\iMbhat}_{\iMa\iMb}}
\def\Jdotaabb{\dot{\cal J}^{\iMahat\iMbhat}_{\iMa\iMb}}
\def\Jddotaabb{\ddot{\cal J}^{\iMahat\iMbhat}_{\iMa\iMb}}
\def\JJ{{J}}

\def\zetaMultiEllis{\zeta_{\textup{Multi\ Ellis}}}
\def\zetaMultiDixon{\zeta_{\textup{Multi\ Dixon}}}
\def\zetaMultiEllis{\zeta}
\def\zetaMultiDixon{\xi}

\def\lround{(}
\def\rround{)}
\def\lsquare{[}
\def\rsquare{]}

\def\Conserv{{\cal Q}}
\def\TReg{{\cal T}}

\def\VAct#1{{\left\langle#1\right\rangle}}
\def\bump{\psi_1}
\def\bumpf{\psi}

\def\nablaDG{{\boldsymbol\nabla}}
\def\nablaInd{{\overline\nabla}}

\def\TtwoTen{{\color{red}{\cal S}}}
\def\TtwoTenhat{{\color{red}\hat{\cal S}}}
\def\ToneTen{{\color{red}{\cal P}}}
\def\TzeroTen{{\color{red}{\cal W}}}
\def\TarbTen{{\color{red}{\cal U}}}
\def\TtwoTen{{{\cal \phi}}}
\def\TtwoTenhat{{\hat{\cal \phi}}}
\def\ToneTen{{{\cal \theta}}}
\def\TzeroTen{{{\cal \varphi}}}
\def\TarbTen{{{\cal \varphi}}}

\def\Kill{{K}}

\def\iMa{{{\mu}}}
\def\iMb{{{\nu}}}
\def\iMc{{{\rho}}}
\def\iMd{{{\kappa}}}
\def\iMe{{{\alpha}}}
\def\iMf{{{\lambda}}}

\def\iMahat{{{\hat\mu}}}
\def\iMbhat{{{\hat\nu}}}
\def\iMchat{{{\hat\rho}}}
\def\iMdhat{{{\hat\kappa}}}

\def\iSa{{{a}}}
\def\iSb{{{b}}}
\def\iSc{{{c}}}
\def\iSd{{{d}}}
\def\iSe{{{e}}}

\def\iSahat{{{\hat{a}}}}
\def\iSbhat{{{\hat{b}}}}
\def\iSchat{{{\hat{c}}}}

\def\partx{\partial^{(x)}}
\def\partz{\partial^{(z)}}
\def\partx{\partial}
\def\partz{\partial}

\def\Rootg{{\omega}}
\def\ArbTen{{Y}}
\def\GRparam{{\kappa}}
\def\Mone{M_2}
\def\Mtwo{M_3}

\def\jgDchange#1{{\color{brown}{#1}}}
\def\jgPchange#1{{\color{red}{#1}}}
\def\jgNewCh#1{{\color{purple}{#1}}}

\def\jgDchange{}
\def\jgPchange{}
\def\JGchange{}
\def\jgNewCh{}

\def\DEfullstop{\,.}
\def\DEcomma{\,,}
\def\DEnone{}

\def\jgnchange#1{{\color{red}{#1}}}
\def\jgnchange{}

\begin{document}
\jgpapersize
\title{The Distributional Stress-Energy Quadrupole}

\author{Jonathan Gratus$^{1,2,3,*}$, Paolo Pinto$^{1,2,4}$, 
Spyridon Talaganis$^{1,5}$}

\maketitle

\noindent
$^1$ Physics department, Lancaster University, Lancaster LA1 4YB,
\\
$^2$ The Cockcroft Institute Daresbury Laboratory,
Daresbury, Warrington
WA4 4AD UK.
\\
$^3$ \texttt{j.gratus@lancaster.ac.uk}\qquad https://orcid.org/0000-0003-1597-6084
\\
$^4$ \texttt{p.pinto@lancaster.ac.uk}
\\
$^5$ \texttt{s.talaganis@lancaster.ac.uk}\qquad https://orcid.org/0000-0003-0113-7546
\\
$^*$ Corresponding author.

\begin{abstract}
\jgDchange{
We investigate stress-energy tensors constructed from the delta
function on a worldline. We concentrate on quadrupoles as they make
an excellent model for the dominant source of gravitational waves and
have significant novel features.
Unlike the dipole, we show
that the quadrupole has 20 free components which are not determined by
the properties of the stress-energy tensor. These need to be derived
from an underlying model and we give an example motivated from a  
divergent-free dust. We show that the components corresponding to the
partial derivatives representation of the quadrupole, have a gauge
like freedom. We give the change of coordinate formula which involves
second derivatives and two integrals. We also show how to define the
quadrupole without reference to a coordinate systems or a metric. For
the representation using covariant
derivatives, we show how to split a quadrupole into a pure monopole,
pure dipole and pure quadrupole in a coordinate free way.
}
\end{abstract}

\tableofcontents


\section{Introduction}
\label{ch_Intro}

\vspace{2em}

Gravitational wave astronomy \jgDchange{will} give rise to major
developments in gravitational physics and astrophysics. The LIGO and
VIRGO detectors \cite{LIGOScientific:2018mvr,Abbott:2016blz} have
\JGchange{detected} relativistic gravitational two-body systems.  The
existing network of gravitational wave interferometers is expanding
both on Earth (for instance, via KAGRA and
LIGO-India \cite{Unnikrishnan:2013qwa,Akutsu:2017thy}) and in space.

In this article we model the compact source, using a distribution, in
which all the mass is concentrated in one point in space and hence a
worldline in spacetime, but has an extended structure encoded as a
multipole expansion. The zeroth order is the monopole, followed by the
dipole and then the quadrupole. Here we consider the
\jgDchange{quadrupole in detail}. 
\jgPchange{For a Minkowski background,} it is well known
\cite{sathyaprakash2009physics,flanagan2005basics}
 that gravitational
radiation is dominated by the quadrupole moment.

\tikzsetnextfilename{DQ-CQG-figure1}
\begin{figure}[t]
\centering
\begin{tikzpicture}[scale=0.3]
\draw [->,very thick](-8,-21)  -- node[below,pos=0.8] {$x$} +(16,0)  ;
\draw [->,very thick](-8,-21)  -- node[left,pos=0.8] {$t$} +(0,30)  ;
\begin{scope}[shift={(0,6)}]
\filldraw [very thick,draw=black,fill=gray] (0,0) ellipse (3 and 2.5) ;
\end{scope}
\begin{scope}
\filldraw [very thick,draw=black,fill=gray] (0,0.5) to[out=0,in=-135] (2,1.5) 
 arc (135:-135:2) 
 to[out=135,in=0] (0,-.5) 
 to [out=180,in=45] (-2,-1.5) 
 arc (-45:-360+45:2) 
 to [out=-45,in=180] cycle;
\end{scope}
\begin{scope}[shift={(0,-6)}]
\fill [very thick,draw=black,fill=gray]  (0,0.2) to[out=0,in=-135] (4,1.5) 
 arc (135:-135:2) 
 to[out=135,in=0] (0,-.2) 
 to [out=180,in=45] (-4,-1.5) 
 arc (-45:-360+45:2) 
 to [out=-45,in=180] cycle;
\end{scope}
\begin{scope}[shift={(0,-12)}]
\filldraw [very thick,draw=black,fill=gray] (0,0.5) to[out=0,in=-135] (2,1.5) 
 arc (135:-135:2) 
 to[out=135,in=0] (0,-.5) 
 to [out=180,in=45] (-2,-1.5) 
 arc (-45:-360+45:2) 
 to [out=-45,in=180] cycle;
\end{scope}
\begin{scope}[shift={(0,-18)}]
\filldraw [very thick,draw=black,fill=gray] (0,0) ellipse (3 and 2.5) ;
\end{scope}
\end{tikzpicture}
\caption{Schematic showing a blob of matter separating and then
  recombining. Such internal dynamics can take place solely within
  the free components, without affecting the divergencelessness of the
stress-energy tensor (\ref{Intro_Tab_Div_zero}).}
\label{fig_intro_grav_quad_free}
\end{figure}
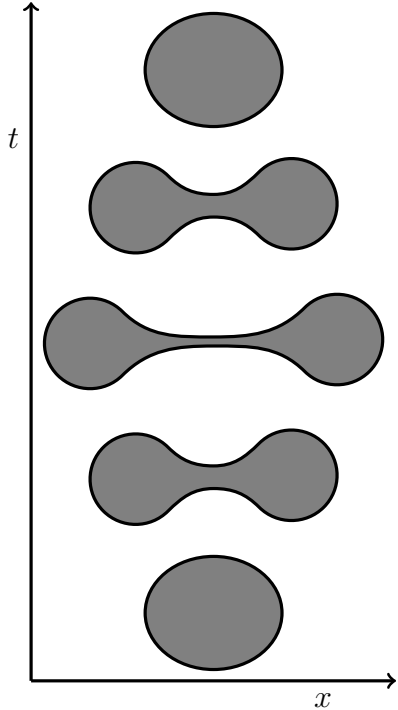

When considering sources of gravitational waves, there are multiple
approaches. \jgDchange{For a pair of simple orbiting masses}, where
relativistic effects can be ignored, one can find analytic
solutions. By contrast the final stages of coalescing black holes
require detailed numerical simulations. Once the stress-energy tensor
is constructed one can evaluate the \jgDchange{perturbation to the
  metric representing the gravitational waves.} Between these two
extremes the post Newtonian approximation \cite{blanchet2014gravitational}
can be used. 

Our approach is different.  In this article we examine the dynamics of
the moments of the distributional quadrupole stress-energy tensor.  
This has a major advantage that
the dynamics \jgnchange{are} encoded as ODEs for the components, as opposed to the
coupled nonlinear PDEs which one is required to solve to model a
general relativistic source.  \jgDchange{In a Minkowski background,
  the gravitational waves can be directly calculated from these
  moments.}  The only constraints we put on the source is that it
obeys the rules of a \JGchange{total} stress-energy tensor, namely
symmetry of its indices and the divergenceless condition. For the
monopole and the dipole it is well known that these conditions
constrain the dynamics so much that they prescribe the ODEs: the
geodesic equation for the monopole and the
Mathisson-Papapetrou-Tulczyjew-Dixon equations for the dipole
\cite{mathisson1937neue,tulczyjew1959motion}.  \jgDchange{In the
  dipole case we assume that the worldline has been given.}  One may
therefore ask if these two conditions also constrain the quadrupole
sufficiently to prescribe the ODEs for its components. In this article
we show that, whereas 40 of the components are prescribed by ODEs, a
further 20 are arbitrary, for example, a quadrupole can expand and
contract as depicted in figure \ref{fig_intro_grav_quad_free}.  Thus
by itself this approach cannot completely prescribe the dynamics of a
quadrupole and one must add additional ODEs, or algebraic equations,
which one can consider to be \defn{constitutive relations} for the
quadrupole. These should arise from an underlying model of the source,
i.e. coalescing black holes will have different constitutive relations
to a rotating ``rigid'' body held together by non gravitational,
e.g. electromagnetic or quantum forces.  Once the constitutive
relations are decided on, the ODEs can be solved and compared to
\jgDchange{numerical simulations or those derived from other models}.

Approximating a distribution of matter with an object at a single point
is a well established method in many branches of physics. Such
approximations are valid if the size of the system is small compared
to other distances involved. For example when considering coalescing
black holes as a source of gravitational waves, the distance between
the black holes is orders of magnitude smaller than their distance to
Earth. Knowing the
dynamics of multipoles may also shed light on \jgDchange{the problem of radiation reaction in the context of dipole and quadrupole dynamics.}

There are many important articles which consider multipole
expansions. These date back to at least the 1950s where Tulczyjew
\cite{tulczyjew1959motion} considered a multipole expansions to derive
the Mathisson-Papapetrou-Tulczyjew-Dixon equations for the dipole. Then in the 
1960s and 1970s Dixon \cite{dixon1964covariant,dixon1967description,DixonII,DixonIII} and
Ellis \cite{Ellis:1975rp} 
considered both charge and mass distributions. They  used 
two different general
formalisms which we compare here, denoting them the Dixon and Ellis
representations.

Recently Steinhoff and Puetzfeld
\cite{steinhoff2010multipolar,steinhoff2012influence,Steinhoff:2014kwa}
calculate the dynamic equation for the components of the
quadrupole. In addition they consider the monopole-dipole and
monopole-dipole-quadrupole system. In all cases the worldline of the
multipole \jgDchange{affects} the dynamics of the components. However
in the above the authors consider if and how the dynamics of the
worldline is \jgDchange{affected} by the higher order moments. They conclude that
one \jgDchange{needs} supplementary conditions in order to determine the
worldline dynamics. We note that these supplementary conditions are
distinct from the constitutive relations described here for the
quadrupole. In this article, excluding section \ref{sch_MD_mono} on
the monopole, the worldline is arbitrary but prescribed. Thus at the
dipole order no supplementary conditions \jgDchange{are} required. However
as stated there are 20 constitutive relations required at the
quadrupole order.

\vspace{1em}

Let $\Mman$ be \jgDchange{a} spacetime with metric $g_{\iMa\iMb}$ and the
Levi-Civita\footnote{It turns out that in most \jgDchange{of our calculations in this article},
  the metric plays no \jgDchange{role} and an arbitrary linear connection can be
  used. See section \ref{ch_CoFree}.}  connection $\nabla_{\iMa}$ with
Christoffel symbol $\Gamma^{\iMa}_{\iMb\iMc}$. Here Greek indices run
$\iMa,\iMb=0,1,2,3$ and Latin indices $\iSa,\iSb=1,2,3$.  Let
$C:\Interval\to\Mman$ where $\Interval\subset\Real$ \jgDchange{is} the worldline
of the source\footnote{Even using proper time in Minkowski space, one
  cannot assume that $\Interval=\Real$ since it is possible to
  accelerate to lightlike infinity in finite proper time.}  with
components $C^{\iMa}(\sigma)$. At this point we do not assume that
$\sigma$ is proper time. Here we consider stress-energy tensors
$T^{\iMa\iMb}$ which are non zero only on the worldline
$C^{\iMa}(\sigma)$, where it has Dirac--$\delta$ like
properties. Such stress-energy tensors are called
\defn{distributional}. 

Being a non linear theory, one cannot simply apply the theory of
distributions to general relativity. It is not meaningful to write
Einstein's equations
\begin{align}
R_{\iMa\iMb}-\tfrac12 g_{\iMa\iMb} R=8\pi\,T_{\iMa \iMb}
\label{Intro_Ein_eqn}
\DEcomma
\end{align}
where the right hand side is a distribution.  This contrasts with
electromagnetism, \jgDchange{which is a linear theory and so one
  often uses distributional sources.} For example an arbitrary moving
point charge gives rise to the Liénard-Wiechard fields.

\JGchange{
There is a large quantity of literature discussing the nature of
distributional stress-energy tensors. Often this links $T^{\iMa\iMb}$
to one or a set of regular stress-energy tensors
$\TReg^{\iMa\iMb}$, for which (\ref{Intro_Ein_eqn}) is valid. Dixon \cite{DixonII}
uses an exponential map, which connects points off the worldline to
points on the worldline, to relate regular and distributional
stress-energy tensors. 
Geroch and  Weatherall
\cite{geroch2018motion} consider an infinite set of stress-energy
tensors satisfying the dominate energy condition and find \jgDchange{conditions}
such that there exists a sequence which tends to the monopole
stress-energy tensor. In section \ref{ch_Ellis_Squz} we show how the
distributional stress-energy tensor $T^{\iMa\iMb}$ is the weak limit
of a set of regular tensors $\TReg^{\iMa\iMb}_\varepsilon$. }

Another approach is to consider
$T_{\iMa\iMb}$ as a source \jgDchange{within the context of} linearised gravity.
 Perturbatively expanding the gravitational metric, $g_{\iMa \iMb}$,
about a background  $\bar{g}_{\iMa \iMb}$,
$g_{\iMa \iMb}
=
\bar{g}_{\iMa \iMb}+\GRparam\,h_{\iMa \iMb}^{(1)}+\cdots
$
where $\GRparam\ll1$ is the perturbation parameter,
and plugging the expansion into the Einstein equation
(\ref{Intro_Ein_eqn}) one has 
\begin{align}
G_{\iMa \iMb}
=
\bar{G}_{\iMa \iMb}
+
\GRparam G^{(1)}_{\iMa \iMb}
+
\cdots
\qquadand
T_{\iMa \iMb}
=
\bar{T}_{\iMa \iMb}
+
\GRparam T^{(1)}_{\iMa \iMb}
+
\cdots
\label{Intro_G_T_expan}
\DEfullstop
\end{align}
Hence the background metric $\bar{g}_{\iMa \iMb}$ satisfies 
$\bar{G}_{\iMa \iMb}=8\pi\,\bar{T}_{\iMa \iMb}$. The linearised
equations are then given by
\begin{align}
G^{(1)}_{\iMa \iMb} = 8\pi\,T^{(1)}_{\iMa \iMb}
\label{Intro_lin_Ein}
\DEfullstop
\end{align}
\jgDchange{In the case when the background metric $\bar{g}_{\mu\mu}$
  is the Minkowski metric $\eta_{\mu\nu}$ then}
(\ref{Intro_lin_Ein})
becomes \cite{flanagan2005basics}
\begin{align}
{
{\Box} \mathcal{H}_{\iMa \iMb}^{(1)}
=-16\pi T_{\iMa \iMb}^{(1)}
}
\label{Intro_lin_Ein_res2}
\DEcomma
\end{align}
where $\mathcal{H}_{\iMa \iMb}^{(1)}=h_{\iMa
  \iMb}^{(1)}-\frac{1}{2}\eta_{\iMa\iMb}(h^{(1)})^\iMc{}_\iMc$,
${\Box}=\partial_\iMa\partial^\iMa$ and we have used the 
\jgnchange{Lorenz gauge, also called the de Donder gauge,} 
($\partial^{\iMa}\mathcal{H}_{\iMa
  \iMb}^{(1)}=0$).
We can give $\mathcal{H}^{(1)}_{\iMa \iMb}$ in terms of an integral over
the retarded Greens functions.
\begin{align}
 \mathcal{H}_{\iMa \iMb}^{(1)}(t,\vec{x})
=
4 \int 
\frac{T_{\iMa \iMb}^{(1)}
(t-\lvert \vec{x}-\vec{x}^{'} 
\rvert,\vec{x}^{'})}{\lvert \vec{x}-\vec{x}^{'} \rvert}
d^{3} \vec{x}^{'} 
\label{Intro_lin_Ein_ret_Green}
\DEfullstop
\end{align}
{One should be careful as there is clearly a contradiction
  between the statement that the perturbation to the background
  stress-energy tensor is small, and the statement that it is
  distributional, and therefore infinite.}  
\JGchange{However as long as one is sufficiently far from the source,
  one can make meaningful statements.}
\JGchange{One result is
  that if $\TReg^{\iMa\iMb}_\varepsilon\to (T^{(1)})^{\iMa\iMb}$
  weakly then, for a point off the
  worldline, the gravitational waves emanating from
  $(T^{(1)})^{\iMa\iMb}$ are the limit of the gravitational waves
  emanating from $\TReg_\varepsilon^{\iMa\iMb}$, i.e.
\begin{align}
\lim_{\epsilon\to 0} \int 
\frac{(\TReg_\varepsilon{})_{\iMa \iMb}
(t-\lvert \vec{x}-\vec{x}^{'} 
\rvert,\vec{x}^{'})}{\lvert \vec{x}-\vec{x}^{'} \rvert}
d^{3} \vec{x}^{'} 
=
\int 
\frac{T^{(1)}_{\iMa \iMb}
(t-\lvert \vec{x}-\vec{x}^{'} 
\rvert,\vec{x}^{'})}{\lvert \vec{x}-\vec{x}^{'} \rvert}
d^{3} \vec{x}^{'} 
\label{Intro_lin_Ein_limit}
\DEfullstop
\end{align}
This requires 
  that \jgDchange{the intersection of the backward}  light cone of the point $(t,\vec{x})$
  with the support of $\TReg_\varepsilon^{\iMa\iMb}$ is compact. 
\jgPchange{This is proved in the appendix, proof number \ref{pf_intro_GW_limit}.}}

\JGchange{
For sources such as coalescing black holes, one cannot use a
regular stress-energy tensor. In this case one can only interpret the
perturbation expansion (\ref{Intro_G_T_expan}) away from the
source. It may then be possible to reinterpret the gravitational waves
as though they were arising from a distributional quadrupole source.}

\JGchange{An alternative approach to interpreting (\ref{Intro_Ein_eqn}) with
distributional $T^{\iMa\iMb}$ is to
extend the theory of distributions  to include
products, for example by using Colombeau algebra
\cite{Steinbauer_2006}.}
\vspace{1em}

In this article we are concerned \textit{only} with the structure of the distributional
stress-energy \jgDchange{tensor}, which we write as $T^{\iMa\iMb}$, and avoid questions
of how it should be applied. 
\JGchange{Since we are dealing with distributions it is most
  convenient to consider $T^{\iMa\iMb}$ as a tensor density of weight
  1. Thus $\Rootg^{-1} T^{\iMa\iMb}$  is a
  tensor, where 
\begin{align}
\Rootg=\sqrt{-\det(g_{\iMa\iMb})}
\DEfullstop
\label{Intro_def_rootg}
\end{align}
  The definition of the covariant derivative of a tensor
  $\ArbTen^{\iMa\iMb\cdots}$
  density of weight 1 is given by
\begin{align}
\nabla_\iMa \ArbTen^{\iMb\iMc\cdots}
=
\Rootg \nabla_\iMa (\Rootg^{-1} \ArbTen^{\iMb\iMc\cdots})
=
- 
\Gamma^\iMd_{\iMa\iMd}\, \ArbTen^{\iMb\iMc\cdots} 
+
\partial_\iMa \ArbTen^{\iMb\iMc\cdots} 
+
\Gamma^\iMb_{\iMa\iMd}\ArbTen^{\iMd\iMc\cdots} 
+
\Gamma^\iMc_{\iMa\iMd}\ArbTen^{\iMb\iMd\cdots} 
+\cdots
\label{Intro_ten_den}
\DEfullstop
\end{align}
where $\Gamma^{\iMb}_{\iMa\iMc}$ are the Christoffel symbols.
In this article the term stress-energy tensor, always refers to a
stress-energy tensor density of weight 1, even if not explicitly
stated. In addition the symbol $T^{\iMa\iMb}$ always refers to a
distributional stress-energy tensor density of weight 1 over the
worldline $C$.}

\JGchange{Since $T^{\iMa\iMb}$ is a total stress-energy
  tensor\footnote{\JGchange{\jgDchange{Non-symmetric stress-energy
        tensors, such as the Minkowski electromagnetic stress-energy
        tensor, are only part of the total stress-energy tensor.}
      However \jgDchange{we} assume that $T^{\iMa\iMb}$ satisfies
      (\ref{Intro_Tab_Sym}) and (\ref{Intro_Tab_Div_zero})
      \jgDchange{because all of the fields are dynamical (there are no
        background fields\cite{gratus2012conservation}) and
        $T^{\mu\nu}$ is the total stress-energy tensor.} } },
  \jgDchange{it satisfies}
\begin{align}
T^{\iMa\iMb}=T^{\iMb\iMa}
\label{Intro_Tab_Sym}
\DEnone
\end{align}
and is divergenceless,
also known as covariantly conserved
\begin{align}
\nabla_{\iMa}  T^{\iMa\iMb}=0
\label{Intro_Tab_Div_zero}
\DEfullstop
\end{align}
which from (\ref{Intro_ten_den}) becomes
\begin{align}
0 &= \nabla_{\iMa} T^{\iMa\iMb}
= \partial_\iMa T^{\iMa\iMb} + \Gamma^{\iMb}_{\iMa\iMc} T^{\iMa\iMc} 
\label{Intro_Tab_Div_zero_density}
\DEfullstop
\end{align}
}

There are several ways of representing a multipole. However we
consider multipoles to be distributions which are integrated with a
symmetric test tensor $\TtwoTen_{\iMa\iMb}=\TtwoTen_{\iMb\iMa}$, so
that\footnote{\JGchange{An integral over $\Mman$ must contain the
    measure $\Rootg$. There is therefore \jgDchange{the following} choice: one can choose
    $T^{\iMa\iMb}$ or $\TtwoTen_{\iMa\iMb}$ to be a density of weight
    1, or put $\Rootg$ 
    explicitly in the integrand.
     Here we have chosen to make $T^{\iMa\iMb}$ a density.}  }
\begin{align}
\int_\Mman T^{\iMa\iMb}\,\TtwoTen_{\iMa\iMb}\, d^4x
\qquad
\text{is a real number.}
\label{Intro_distribution}
\end{align}
\jgDchange{Equation (\ref{Intro_distribution})} can be
written as an integral over the worldline with a number of derivatives
of the Dirac δ-function. \jgDchange{In other words} a multipole of order $k$ is 
\begin{align}
T^{\iMa\iMb}
=
\sum_{r=0}^k \int_\Interval \zeta^{\iMa\iMb\ldots}(\sigma)
\ 
{\cal D}^{(r)}_{\ldots}  \ \deltaFour\big(x-C(\sigma)\big)
\
d\sigma
\label{Intro_general_k_pole}
\DEcomma
\end{align}
where there are $r$ additional indices on $\zeta^{\iMa\iMb\ldots}$ and
${\cal D}^{(r)}_{\ldots}$. The subscript dots on ${\cal
  D}^{(r)}_{\ldots}$ contract with the superscript dots on
$\zeta^{\iMa\iMb\ldots}$. Here ${\cal D}^{(r)}_{\ldots}$ represents
$r$ derivatives of the δ-function. The familiar cases are the
\defn{monopole} when $k=0$, the \defn{dipole} when $k=1$ and the
\defn{quadrupole} when $k=2$. As can be seen from
(\ref{Intro_general_k_pole}) the general dipole contains the monopole
term and the general quadrupole contains both the monopole and dipole
terms. In general, it is not possible to extract the monopole and
dipole terms from the quadrupole, without additional structure such as
a preferred vector field or a coordinate system.  
For the monopole (\ref{Intro_Tab_Sym}) and (\ref{Intro_Tab_Div_zero})
lead to the pre-geodesic equation. By \jgDchange{contrast}, for the dipole and quadrupole
there is no need to assume the worldline $C$ is a geodesic. Therefore,
unless otherwise stated,
we present all the \jgDchange{results} for an arbitrary but prescribed worldline.

There are two important, equivalent representations of
multipoles. One uses the partial derivatives, which we call the
\defn{Ellis} representation. The other uses the covariant derivative
and we call the \defn{Dixon} representation. Both have their
advantages and disadvantages and these are outlined in section
\ref{ch_EllisDixon} below.
The Ellis formulation is greatly simplified when using a coordinate system
$(\sigma,z^1,z^2,z^3)$ which is adapted to the worldline, i.e. where 
\begin{align}
C^0(\sigma)=\sigma
\qquadand
C^\iSa(\sigma)=0
\label{Intro_adapt_coords}
\DEcomma
\end{align}
for $\iSa=1,2,3$. In this coordinate system the integral in
(\ref{Intro_general_k_pole}) can be \JGchange{evaluated}. Observe that
(\ref{Intro_adapt_coords}) implies $\Cdot^0=1$ and $\Cdot^\iSa=0$.

The monopole and dipole have been extensively studied in the
literature, \cite{Han:2016cdh,Kopeikin:2018zro,Blanchet:2006sc}. In this article we concentrate mainly on the
quadrupole \jgDchange{because it has interesting properties that do not appear to have been emphasised previously.} Not only is it the
natural source of gravitational waves, but it has several unusual
properties not seen in the case of the monopole or the \JGchange{dipole}. 
\jgDchange{These properties are given below.}
\begin{bulletlist}
\item
The quadrupole contains free components.
\item
In the Ellis representation, the 
components $\zeta^{\iMa\iMb\iMc\iMd}$ to not transform as tensors
but instead involve second derivatives and double integrals.
\item
\jgPchange{In the Ellis representation, there is no natural way to endow a quadrupole with mass.}
Instead one can only talk about the
energy of a quadrupole and only in the case where there is a
timelike Killing symmetry.
\end{bulletlist}

The $\zeta^{\iMa\iMb\ldots}=\zeta^{\iMa\iMb\ldots}(\sigma)$ are called the
components of $T^{\iMa\iMb}$ and are functions only of the position on the
worldline $C$.
Clearly
from (\ref{Intro_Tab_Sym}) they have the symmetry
\begin{align}
\zeta^{\iMa\iMb\ldots} = \zeta^{\iMb\iMa\ldots}
\label{Intro_zeta_sym}
\DEfullstop
\end{align}
Depending on the representation, we may also choose to impose additional
symmetries for uniqueness. We then apply the divergenceless condition
(\ref{Intro_Tab_Div_zero}) to establish further condition on the
$\zeta^{\iMa\iMb\ldots}$.
We can place the components $\zeta^{\iMa\iMb\ldots}$ into three
categories.
\begin{bulletlist}
\item
Some components are algebraically related to other components and can
therefore be removed.
\item
Some components are determined by a first order ODE. These are
\jgDchange{the} result of the differential equation (\ref{Intro_Tab_Div_zero}). In
order to specify these components it is only necessary to specify
their initial value at some point along the worldline.
\item
This leaves the components we call \defn{free}.  These are not
constrained by (\ref{Intro_Tab_Sym}) and (\ref{Intro_Tab_Div_zero})
and are allowed to take on any
value\footnote{\label{fn_20freeComp}\jgPchange{These 20 free
    components are not the same as the 20 independent components of
    the (reduced) quadrupole stress-energy tensor as described by
    Dixon \cite[Equation
      (1.37)]{DixonIII},\cite{dixon1973definition,bini2009dixon}. See
    appendix \ref{ch_Appx_Dixon} for a discussion.}}. These free
components can however influence the ODE components.

In order to completely specify the dynamics of a quadrupole, 
these free components need to be \jgDchange{determined} by 
constitutive equations. The choice of constitutive equations
depends on a choice of a model for the material. 
In section \ref{ch_Dust} we consider the dust stress-energy tensor
density \cite{Timofeev_2019} and
use it to suggest corresponding constitutive equations.


\end{bulletlist}

In table \ref{tab_number_components} the number of \jgDchange{ODEs} and free
components is given, \jgDchange{and} compared to the electromagnetic dipoles and
quadrupoles. 

\JGchange{In addition, some components may have a freedom in that
  different $\zeta^{\iMa\iMb\ldots}$ correspond to the same
  stress-energy tensor. Equivalently a given stress-energy
  tensor does
  not completely specify the components
  $\zeta^{\iMa\iMb\ldots}$. Examples of this freedom for the dipole
  and quadrupole are given in equations (\ref{MD_diploe_Zeta_Freedom})
  and (\ref{QP_Zeta_Freedom}) below.  In this article we will call
  this a \defn{gauge-like freedom} since \jgDchange{it is} similar to other gauge freedoms
  in that it arises from integrating a physically observable
  tensor. In this case however,
  the components $\zeta^{\iMa\iMb\ldots}$ are not themselves
  tensors.}

\begin{table}[t]
\begin{center}
\begin{tabular}{|c||c|c||c|c|}
\hline
& \multicolumn{2}{c||}{Electromagnetic}
& \multicolumn{2}{c|}{Gravitational}
\\\cline{2-5}
& ODE & free & ODE & free
\\\hline
Monopole & 1 & 0 & 1 & 0 
\\\hline
Semi-dipole & 1 & 3 & 7 & 0
\\\hline
full dipole & 1 & 6 & 10 & 0
\\\hline
semi-quadrupole & 1 & 12 & 22 & 6
\\\hline
full quadrupole & 1 & 20 & 40 & 20
\\\hline
\end{tabular}
\end{center}
\caption{List of the number of components which are determined by
  ODEs and the number which are free, for monopoles, dipoles and
  quadrupoles. The electromagnetic sources refer to a current
  $\Jcurr^{\iMa}$ which is conserved and a source for Maxwell's
  equations. The gravitational source refers to a stress-energy tensor
   $T^{\iMa\iMb}$ which are sources for (linearised) Einstein's
  equations. Each order includes all the lower orders. \jgDchange{The} 10
  components in the full stress-energy dipole includes 1 monopole
  component, while the $(40+20)$ components in the full quadrupole
  includes both dipole and monopole components.  The definitions of the
  semi-dipole and semi-quadrupole are given in section
  \ref{ch_Semi_Q}.}
\label{tab_number_components}
\end{table}

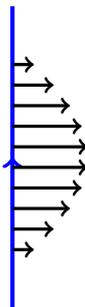
\begin{figure}[t]
\centering
\tikzsetnextfilename{DQ-CQG-figure2}
\begin{tikzpicture}
\draw [ultra thick,->-=0.5,blue] (0,-.5) -- node[black,pos=0.9,left]{} (0,3.5) ;
\foreach \i in {1,2,...,10} 
{ \draw [very thick,->] (0,\i*3/11) -- +({sin(\i*180/11)},0) ;} ;
\end{tikzpicture}
\caption{An electric dipole appears for a finite period of time
  and then disappears. This does not break charge conservation. }
\label{fig_intro_electric_dipole}
\end{figure}

For comparison, the electromagnetic dipole has one ODE component, which is
simply the total charge and satisfies ${dq/d\sigma}=0$, and 
six free components corresponding to the three electric and three
magnetic components \cite{gratus2018correct}. These can be anything
without \jgnchange{breaking} charge
conservation, as seen in figure \ref{fig_intro_electric_dipole}.
For the stress-energy tensor, the free components can correspond to 
the internal matter separating and coalescing, as in figure 
\ref{fig_intro_grav_quad_free}.
One therefore needs additional constitutive
relations which encode the matter one is modelling. In this article we
give an example of constitutive relations which correspond to non
divergent dust.

Given a regular stress-energy tensor $T^{\iMa\iMb}$ and a Killing
vector field $K^{\iMa}$ we can find a conserved quantity $T^{\iMa\iMb}
K_\iMb$ such that $\nabla_{\iMa} (T^{\iMa\iMb} K_\iMb)=0$. The same is
true for the distributional stress-energy tensor. Here $K_{\iMa}$
gives rise to a conserved \JGchange{scalar $\Conserv_\Kill$} along the
worldline $C$. If $K^{\iMa}$ is a timelike Killing vector field it is
natural to associate $\Conserv_\Kill$ as the conserved energy of the
multipole. \jgDchange{However,} the relationship between the energy and mass is 
subtle. In the monopole and dipole case there is a natural definition
of the mass, but the same is not true in the quadrupole case. Even when a
mass can be defined, it is not conserved in general.

\subsubsection*{Outline of article}

As stated above there are two established methods of representing the
stress-energy distribution: one using partial derivatives in
(\ref{Intro_general_k_pole}), which we call the \defn{Ellis}
representation, and the other using covariant derivatives, which we
call the \defn{Dixon} representation. The \jgnchange{pros} and cons of these
two approaches is discussed in section \ref{ch_EllisDixon} and
summarised in table \ref{tab_Ellis_Dixon}. 
\jgNewCh{We show that the Ellis components with respect to adapted
  coordinates are unique can be
  extracted by applying them to particular test tensors.}
In section \ref{ch_MD} we
summarise the key results of the monopole and dipole stress-energy
tensors.
We highlight the Ellis and Dixon representations of the
dipole.  

In section \ref{ch_QP} we examine the quadrupole in detail. In this
section we use the Ellis approach. We give the 
\JGchange{gauge-like} freedom of the
components and the {complicated} change of coordinates which involve second
derivatives and integrals over the worldline, similar to
\cite{gratus2018correct}.
We use the adapted coordinates (\ref{Intro_adapt_coords}) and give the
differential equations arising from the symmetry (\ref{Intro_Tab_Sym})
and divergencelessness (\ref{Intro_Tab_Div_zero}) of
$T^{\iMa\iMb}$. 
We can now identify which components are 
algebraic, which satisfy ODEs and which are free. 
In
subsection \ref{ch_Q_free} we give an example of the free components
in Minkowski spacetime as depicted in figure
\ref{fig_intro_grav_quad_free}.
As stated above, if there is a Killing vector field, there exists a
corresponding conserved quantity. These are given in section
\ref{ch_Q_Cons}. This includes a new interpretation of the conserved
quantities corresponding the three Lorentz boosts.

In section \ref{ch_Dust} we use the limit of the dust stress-energy
tensor as it is squeezed onto the worldline to construct a choice of
constitutive relations to replace the free components with ODEs.

It
is interesting to observe that, using deRham currents, multipoles can
be defined without any additional structure on a
manifold. \jgDchange{In other words,} it is
not necessary to prescribe either a metric or a connection to define a
general multipole.  
This is particularly useful if we wish to extend
the notion of a general multipole tensor distribution to manifolds
such as the tangent bundle which does not possess either metric or
connection. 
However a connection is of course needed to define the
divergencelessness condition (\ref{Intro_Tab_Div_zero}). 
\jgnchange{Although
standard general relativity only considers four dimensional
manifolds with a Lorentzian metric and the Levi-Civita connection,
there is much interest
\cite{dereli1996non,pagani2015quantum,adria2020minimal} in
non-metric compatible connections. Since the approach in the section
does not specify the metric, all the results apply to a non-metric
compatible connection.
Other circumstances where it is advantageous not to prescribe the metric
include transformation optics where one has two
metrics, the gravitational metric and the optical metric. Another case
is when considering varying the metric to derive Einstein's equations
where it is necessary to know precisely the dependency of the various
object on the metric}.

\jgnchange{Up to section \ref{ch_Dust},} we have defined everything in
terms of a coordinate system. However, it is useful to define the multipoles in
a coordinate free manner. \JGchange{When we refer to ``coordinate
  free'', the goal is to define all the mathematical objects and give
  the statement in all the theorems, without reference to a coordinate
  system. Coordinates may, however, be used in proofs.  Thus although
  a vector $V^\iMa$ is invariant under transformations of the
  coordinates, it is defined by how the components change under
  coordinate transformations.  By contrast defining a vector by its
  action on scalar fields follows this coordinate free goal.}
\jgnchange{ There are a number of advantages of this approach.  First,
  complicated coordinate transformations are avoided.  Second, when
  needed, one can derive the coordinate transformations more easily.
  Third, it is easier to present metric free calculations.  Four and
  most importantly, it makes manifest which objects are physical and
  which are merely ``coordinate objects''. This is especially relevant
  in the case of moments where expressions such as $\int_\Mman f(x)
  x^{\iMa_1}\cdots x^{\iMa_k}\,d^4x$ are so dependent on the
  coordinates that there is no coordinate transformation expression.
}

In section
\ref{ch_CoFree} we detail this \jgnchange{metric and coordinate free approach}. 
By contrast to the Ellis approach, the Dixon
approach contains more information about a multipole, namely how it
splits into a monopole term, a dipole term, a quadrupole term and so
on. This split, called here the \defn{Dixon split}, 
is actually \jgnchange{metric and coordinate} independent and the details are
given in section \ref{ch_CoFree_Dixon}.

As noted in \cite{gratus2018correct}, 
without a metric, connection or coordinate
system, it is still
possible to define a pure electric dipole. In this article we call
such a dipole a \defn{semi-dipole}. We observe that the
semi-dipole stress-energy consists of the displacement vector but not
the spin. In section \ref{ch_Semi_Q} we define the semi-dipole and
semi-quadrupole stress-energy tensor.

We conclude, in section \ref{ch_Conclusion}, \jgDchange{and give some of the longer proofs in the appendix.}

\subsubsection*{Notation regarding derivatives}
Given a coordinate system $(x^0,\ldots,x^3)$ then Greek indices
\jgDchange{range over the values}
$\iMa,\iMb=0,\ldots,3$. We write the partial derivatives
\begin{align}
\partx_{\iMa} = \pfrac{}{x^{\iMa}}
\label{Intro_def_partial}
\DEfullstop
\end{align}
In the case of the adapted coordinates $(\sigma,z^1,z^2,z^3)$ obeying
(\ref{Intro_adapt_coords}) we use both Greek indices
$\iMa,\iMb=0,\ldots,3$ and Latin indices $\iSa,\iSb=1,2,3$. In this
case we have
\begin{align}
\partz_{0} = \pfrac{}{\sigma}
\qquadand
\partz_{\iSa} = \pfrac{}{z^\iSa}
\label{Intro_def_partial_adap}
\DEfullstop
\end{align}
Thus, even if not stated explicitly, writing $\partz_{\iSa}$ implies
we are referring to an adapted coordinates system.

Note that in both the adapted and non adapted case we use overdot to
represent differentiation with respect to $\sigma$. In the non adapted
coordinates this is only used for quantities, such as $C^\iMa(\sigma)$
and $\Cdot^\iMa(\sigma)$ which are only defined on the worldline
\jgDchange{of the multipole}. In
the adapted coordinate cases \jgNewCh{it is} synonymous with $\partz_0$.

When we have two non adapted coordinate systems $(x^0,\ldots,x^3)$ and
$(\xhat^\Ohat,\ldots,\xhat^{\hat{3}})$ we use the hat on the index to
indicate the hatted coordinate system. Thus
\begin{align}
\partx_{\iMahat} = \pfrac{}{\xhat^{\iMahat}}
\label{Intro_def_partial_hat}
\DEfullstop
\end{align}
Likewise for the adapted coordinate system
$(\hat{\sigma},\zhat^{\hat{1}},\zhat^{\hat{2}},\zhat^{\hat{3}})$ we
have
\begin{align}
\partz_{\Ohat} = \pfrac{}{\hat\sigma}
\qquadand
\partz_{\iSahat} = \pfrac{}{\zhat^\iSahat}
\label{Intro_def_partial_adap_hat}
\DEfullstop
\end{align}

\section{Dixon's versus Ellis's approaches to multipoles}
\label{ch_EllisDixon}

As stated in the introduction there are two standard approaches to
writing down distributional multipoles, \jgPchange{the Ellis and Dixon
  representations. These are equivalent as any multipoles of order $k$
  can be represented in both notations. The proof that \jgnchange{a} Dixon
  multipole is also an Ellis multipole is given in section
  \ref{sch_Dixon}. The converse is more difficult since it is
  necessary to extract the Dixon components. This involves the Dixon
  split and is given, for quadrupoles, in section \ref{ch_CoFree_Dixon}.}

Both the Ellis and Dixon approaches have advantages and disadvantages
and these are listed in table \ref{tab_Ellis_Dixon}.

\begin{table}
\begin{center}
\begin{tabular}{|p{0.45\textwidth}|p{0.45\textwidth}|}
\hline
\defn{Ellis}
&
\defn{Dixon}
\\\hline
Can be defined using coordinates.
&
Can be defined using coordinates.
\\\hline
Components are unique for adapted coordinates. 
{In a general coordinate system they have a gauge-like freedom.}
&
Components are unique.
\\\hline
For
general coordinate transformation the components require higher derivatives and
integrals.
& 
Components transform as a tensor.
\\\hline
Do not require any additional structure. 
\newline
These can be defined without
referring to a metric or additional vector field.
&
Requires the connection and the Dixon vector $\DixVec_{\iMa}(\sigma)$ for the definition.
\\\hline 
Contains all multipoles up to specific order.
&
Contains all multipoles up to specific order.
\\\hline
It is not possible to extract a multipole of a specific order without
additional structure, for example an adapted coordinate system.
&
Easy to extract a multipoles of any order.
\\\hline
Can be easily defined in a coordinate free way using DeRham push
forward.
&
The Dixon split can be defined in a coordinate free way, but this definition is
complicated. \jgnchange{It} requires the DeRham push forward plus 
a non intuitive additional axiom, given in section
\ref{ch_CoFree_Dixon}.
\\\hline
The dipole can be written in the Ellis representation, which is
consistent with the Mathisson-Papapetrou-Tulczyjew-Dixon equations.
&
The dipole can be written in the Dixon representation, which is
consistent with the Mathisson-Papapetrou-Tulczyjew-Dixon equations.
\\\hline
There is no concept of the mass of the multipole
&
The \jgPchange{monopole term may be used to define the mass, but in general it is not conserved.}
\\\hline
There \JGchange{is no need of a Dixon vector.}
&
There is a complicated formula for the components with respect to
different $\DixVec^{\iMa}(\sigma)$. This will mix
in multipoles of different orders.
\\\hline
One can construct a regular tensor field whose moments, up to $k$, are the
components of the distribution. 
The best method is using squeezed tensors that employ an adapted
coordinate system.
\newline\rule{1.5em}{0em}
In principle \JGchange{it should be possible to} reconstruct the original
regular tensor using the Fourier transform  but this has not been
investigated. 
&
One can construct a tensor field whose moments, up to $k$, are the
components of the distribution. This is by considering the fields on
the transverse hyperspace constructed from the geodesic map of vectors
orthogonal to $\DixVec^{\iMa}(\sigma)$.
\newline\rule{1.5em}{0em}
If all the moments are \jgDchange{known then} one can reconstruct an original
distribution. 
This also requires certain assumptions about analyticity of Fourier
transform.
\\\hline
There is a  formula for extracting the components using test tensors
in adapted coordinate.
&
In principle the components can be extracted using test tensors.
\\\hline
\end{tabular}
\end{center}
\caption{Comparison between the Ellis and Dixon representations.}
\label{tab_Ellis_Dixon}
\end{table}

\subsection{The Ellis approach}

One method \cite{Ellis:1975rp} uses partial derivatives of the Dirac-δ
function. Although Ellis principally defines it for the electric
current $\Jcurr^{\iMa}$ it is easy to extend this for the
stress-energy tensor.
So a multipole of order $k$ is given by 
\begin{align}
T^{\iMa\iMb} = {\frac{1}{k!}}
\int_\Interval \zetaMultiEllis^{\iMa \iMb \iMc_1\ldots\iMc_k}(\sigma)
\ \partx_{\iMc_1} \cdots \partx_{\iMc_k}
\deltaFour\big(x-C(\sigma)\big)\,d\sigma
\label{Intro_Tab_Ellis_Multi}
\DEcomma
\end{align}
where $\zetaMultiEllis^{\iMa \iMb \iMc_1\ldots\iMc_k}(\sigma)$ are smooth
functions of $\sigma$. 
Thus when acting on the test tensor $\TtwoTen_{\iMa\iMb}$
\begin{align}
\int_\Mman T^{\iMa\iMb} \ \TtwoTen_{\iMa\iMb} \ d^4x 
&=
(-1)^{k}  {\frac{1}{k!}}\int_\Interval  
\zetaMultiEllis^{\iMa \iMb \iMc_1\ldots\iMc_k}(\sigma)\,
\big(\partx_{\iMc_1} \cdots \partx_{\iMc_k} 
\TtwoTen_{\iMa\iMb} \big)\big|_{C(\sigma)}
\ d\sigma
\label{Intro_Tab_Ellis_Multi_action}
\DEfullstop
\end{align}
In this article we will refer to this representation of a multipole as
the \defn{Ellis} representation.

The symmetry of $T^{\iMa\iMb}$ leads to
\begin{align}
\zetaMultiEllis^{\iMa \iMb \iMc_1\ldots\iMc_k}
=
\zetaMultiEllis^{\iMb \iMa \iMc_1\ldots\iMc_k}
\label{EllisDixon_ab_symm}
\DEfullstop
\end{align}
In addition the partial derivatives commute
and it is natural to demand that the components  $\zetaMultiEllis^{\iMb \iMa \iMc_1\ldots\iMc_k}$ are
symmetric. Thus we set
\begin{align}
\zetaMultiEllis^{\iMa \iMb \iMc_1\ldots\iMc_k} 
=
\zetaMultiEllis^{\iMa \iMb (\iMc_1\ldots\iMc_k)}
\label{Intro_Zeta_Ellis_sym}
\DEnone
\end{align}
where the round brackets mean the \JGchange{symmetrisation}
of the indices,
\begin{align}
\zetaMultiEllis^{\iMa \iMb (\iMc_1\ldots\iMc_k)}
=
\frac{1}{k!}
\sumdoubleind{\text{All permutations}}{{i_1\ldots i_k}}
\zetaMultiEllis^{\iMa \iMb \iMc_{i_1}\ldots\iMc_{i_k}}
\label{Intro_def_round_brakets}
\DEfullstop
\end{align}

\jgNewCh{
We observe that every multipole of order $k$ is also a multipole of
order $k+1$. This follows since
\begin{align}
\int_\Interval  
\zetaMultiEllis^{\iMa \iMb \iMc_1\ldots\iMc_k}\,
\big(\partx_{\iMc_1} \cdots \partx_{\iMc_k} 
\TtwoTen_{\iMa\iMb} \big)\big|_{C(\sigma)}
\ d\sigma
=
\int_\Interval  
\zetaMultiEllis_{\text{k+1}}^{\iMa \iMb \iMc_1\ldots\iMc_k}
\Cdot^\iMd\big(\partial_\iMd
\partx_{\iMc_1} \cdots \partx_{\iMc_k} 
\TtwoTen_{\iMa\iMb} \big)\big|_{C(\sigma)}
\ d\sigma
\label{Intro_Tab_Ellis_k+1}
\DEnone
\end{align}
where
\begin{align}
\dot{\zetaMultiEllis}_{\text{k+1}}^{\iMa \iMb \iMc_1\ldots\iMc_k}
=
-\zetaMultiEllis^{\iMa \iMb \iMc_1\ldots\iMc_k}
\label{Intro_Tab_Ellis_k+1_def}
\DEfullstop
\end{align}
}

One problem with the Ellis representation is that the
$\zetaMultiEllis^{\iMa \iMb \iMc_1\ldots\iMc_k}$ are not
unique. Examples of the \JGchange{gauge-like} freedom possessed by
$\zetaMultiEllis^{\iMa \iMb \iMc_1\ldots\iMc_k}$ are given in
(\ref{MD_diploe_Zeta_Freedom}) and (\ref{QP_Zeta_Freedom}). This
contrasts with the case when one chooses an adapted coordinate system
below.

\subsection{Adapted coordinates}
\label{ch_Ellis_adap}

In general expressions for multipoles in the Ellis representation are
complicated. They greatly simplify if one chooses an adapted coordinate
system as given by (\ref{Intro_adapt_coords}). In this coordinate
system the integral over $\Interval$ is \jgNewCh{evaluated} and 
\jgDchange{(\ref{Intro_Tab_Ellis_Multi}) becomes}
\begin{align}
T^{\iMa\iMb} = 
\sum_{r=0}^k 
 {\frac{1}{r!}}
\gamma^{\iMa\iMb \iSa_1 \ldots\iSa_r 0\ldots 0}(\sigma)
\ \partz_{\iSa_1} \cdots \partz_{\iSa_r}\,
\deltaThree(\Vz)
\label{Ellis_adap_Ellis_Multi}
\DEcomma
\end{align}
where $\Vz=(z^1,z^2,z^3)$.
The component 
$\gamma^{\iMa\iMb \iSa_1 \ldots\iSa_r 0\ldots 0}$ has $(k-r)$
zero indices, so that $\gamma^{\iMa\iMb \iSa_1 \ldots\iSa_r 0\ldots 0}$ has $2+k$ 
indices. \jgPchange{The proof of (\ref{Ellis_adap_Ellis_Multi})
  is given below after (\ref{Ellis_adap_gam_zeta}) once we have
  calculated $T^{\iMa\iMb}$ on a test function and given the
  relationship between the $\gamma^{\iMa\iMb \iSa_1 \ldots\iSa_r
    0\ldots 0}$ and $\zetaMultiEllis^{\iMa \iMb \iMc_1\ldots\iMc_k}$.}
Since we only differentiate $\deltaThree(\Vz)$
in the $z^\iSa$ direction, when acting on a test tensor
\begin{equation}
\begin{aligned}
\int_\Mman T^{\iMa\iMb}\,\TtwoTen_{\iMa\iMb}\,d^4 x
&=
\sum_{r=0}^k 
{\frac{(-1)^r}{r!}}
\int_\Interval d\sigma \,
\gamma^{\iMa\iMb \iSa_1 \ldots\iSa_r 0\ldots 0}(\sigma)
\,
(\partz_{\iSa_1} \cdots \partz_{\iSa_r}\,\TtwoTen_{\iMa\iMb})
\DEfullstop
\end{aligned}
\label{Ellis_adap_Ellis_Multi_action}
\end{equation}

\begin{ProofI}{Proof of \eqref{Ellis_adap_Ellis_Multi_action} and
    \eqref{Ellis_adap_gam_zeta}} 
\label{pf_Tab_gamma_eval}
\begin{align*}
\int_\Mman T^{\iMa\iMb}\,\TtwoTen_{\iMa\iMb}\,d^4 x
&=
\int_\Interval d\sigma \int_{\textup{space}} d^3\Vz 
\Big(
\sum_{r=0}^k 
{\frac{1}{r!}}
\gamma^{\iMa\iMb \iSa_1 \ldots\iSa_r 0\ldots 0}
\ \partz_{\iSa_1} \cdots \partz_{\iSa_r}\,
\deltaThree(\Vz)
\Big)
\TtwoTen_{\iMa\iMb}
\\&=
\sum_{r=0}^k 
{\frac{(-1)^r}{r!}}
\int_\Interval d\sigma \int_{\textup{space}} d^3\Vz \
\gamma^{\iMa\iMb \iSa_1 \ldots\iSa_r 0\ldots 0}
\,
(\partz_{\iSa_1} \cdots \partz_{\iSa_r}\,\TtwoTen_{\iMa\iMb})
\,\deltaThree(\Vz)
\\&=
\sum_{r=0}^k 
{\frac{(-1)^r}{r!}}
\int_\Interval d\sigma \,
\gamma^{\iMa\iMb \iSa_1 \ldots\iSa_r 0\ldots 0}
\,
(\partz_{\iSa_1} \cdots \partz_{\iSa_r}\,\TtwoTen_{\iMa\iMb})
\DEfullstop
\end{align*}
\end{ProofI}

We still impose the symmetry conditions (\ref{EllisDixon_ab_symm}) and
(\ref{Intro_Zeta_Ellis_sym}) on the $\gamma$'s so that
\begin{align}
\gamma^{\iMa \iMb \iMc_1\ldots\iMc_k}
=
\gamma^{\iMb \iMa \iMc_1\ldots\iMc_k}
=
\gamma^{\iMa\iMb (\iMc_1\ldots\iMc_k)}
\label{Ellis_adap_symms}
\DEfullstop
\end{align}
The relationship between the $\gamma^{\iMa\iMb \iSa_1\ldots\iSa_r 0\ldots 0}$
and $\zetaMultiEllis^{\iMa\iMb \iMc_1\ldots\iMc_k}$ is given
by comparing (\ref{Intro_Tab_Ellis_Multi_action}) and
(\ref{Ellis_adap_Ellis_Multi_action}) for an adapted coordinate system
\begin{align}
\gamma^{\iMa\iMb \iSa_1\ldots\iSa_r 0\ldots 0}
=
\frac{1}{(k-r)!}
{\partial_0^{k-r}} 
\zetaMultiEllis^{\iMa\iMb \iSa_1\ldots\iSa_r0\ldots  0}
\label{Ellis_adap_gam_zeta}
\DEfullstop
\end{align}

\begin{ProofA}{Proof of  \eqref{Ellis_adap_Ellis_Multi} and 
\eqref{Ellis_adap_gam_zeta}. }
\label{pf_zeta_gamma}
\begin{align*}
\int_\Mman T^{\iMa\iMb} \ \TtwoTen_{\iMa\iMb} \ d^4x 
&=
(-1)^{k}  {\frac{1}{k!}}\int_\Interval  
\zetaMultiEllis^{\iMa \iMb \iMc_1\ldots\iMc_k}\,
\big(\partx_{\iMc_1} \cdots \partx_{\iMc_k} 
\TtwoTen_{\iMa\iMb} \big)
\\&=
\sum_{r=0}^k
(-1)^{k}  {\frac{1}{k!}}\ \frac{k!}{r!(k-r)!}\int_\Interval  
\zetaMultiEllis^{\iMa \iMb \iSa_1\ldots\iSa_r0\ldots0}\,
\big(\partx_{\iSa_1} \cdots \partx_{\iSa_r} \partial_0^{k-r}
\TtwoTen_{\iMa\iMb} \big)
\\&=
\sum_{r=0}^k
(-1)^{r} \frac{1}{r!(k-r)!}\int_\Interval  
(\partial_0^{k-r}\zetaMultiEllis^{\iMa \iMb \iSa_1\ldots\iSa_r,0\ldots0})\,
\big(\partx_{\iSa_1} \cdots \partx_{\iSa_r} 
\TtwoTen_{\iMa\iMb} \big)
\DEfullstop
\end{align*}
Hence comparing with \eqref{Ellis_adap_Ellis_Multi_action} gives
\eqref{Ellis_adap_gam_zeta}. \jgnchange{From
(\ref{Ellis_adap_Ellis_Multi_action}) we have (\ref{Ellis_adap_Ellis_Multi}).}
\end{ProofA}

In an adapted coordinate system, the $\gamma^{\iMa\iMb
  \iSa_1\ldots\iSa_r 0\ldots 0}$ are uniquely determined by the
distribution. \jgPchange{This is because we can extract
  $\gamma^{\iMa\iMb \iSa_1\ldots\iSa_r 0\ldots 0}$ by acting on particular
  test functions 
\begin{align}
\gamma^{\iMa\iMb \iSa_1\ldots\iSa_r 0\ldots 0}
&=
(-1)^r\lim_{\epsilon\to 0} 
\int_\Mman T^{\iMa\iMb}\,z^{\iSa_1}\cdots z^{\iSa_r}\,
\bumpf_{\epsilon,\sigma}(\sigma',\Vz)
\,d\sigma'\,d^3\Vz
\DEcomma
\label{Ellis_adap_extract_comp}
\end{align}
where
\begin{align}
\bumpf_{\epsilon,\sigma}(\sigma',\Vz)
=
\epsilon^{-1}\, 
\bump\big((\sigma-\sigma')/\epsilon\big)
\,
\bump\big((z^1)^2+(z^2)^2+(z^3)^2\big)
\DEnone
\label{Ellis_adap_def_bump}
\end{align}
and $\bump:\Real\to\Real$ is a test function with $\bump(0)=1$,
is flat about $0$ and $\int\bump(\sigma)\,d\sigma=1$.
\begin{proof}
From (\ref{Ellis_adap_Ellis_Multi_action}) we have
\begin{align*}
(-1)^r\lim_{\epsilon\to 0} 
&
\int_\Mman T^{\iMa\iMb}\,z^{\iSa_1}\cdots z^{\iSa_r}\,
\bumpf_{\epsilon,\sigma}(\sigma',\Vz)
\,d\sigma'\,d^3\Vz
\\&=
\lim_{\epsilon\to 0} 
\sum_{s=0}^k 
{\frac{(-1)^{s+r}}{s!}}
\int_\Interval d\sigma' \,
\gamma^{\iMa\iMb \iSb_1 \ldots\iSb_s 0\ldots 0}(\sigma')
\,
\partz_{\iSb_1} \cdots \partz_{\iSb_s}
\Big(z^{\iSa_1}\cdots z^{\iSa_r}\,
\bumpf_{\epsilon,\sigma}(\sigma',\Vz)
\Big)
\\&=
\lim_{\epsilon\to 0} 
\sum_{s=0}^k 
{\frac{(-1)^{s+r}}{s!}}
\int_\Interval d\sigma' \,
\gamma^{\iMa\iMb \iSb_1 \ldots\iSb_s 0\ldots 0}(\sigma')
\,
(s!\,\delta^r_s)
\,
\bumpf_{\epsilon,\sigma}(\sigma',\Vz)
\lim_{\epsilon\to 0} 
\\&=
\frac{1}{\epsilon}
\lim_{\epsilon\to 0} 
\int_\Interval d\sigma' \,
\gamma^{\iMa\iMb \iSb_1 \ldots\iSb_r 0\ldots 0}(\sigma')
\,
\bump((\sigma-\sigma')/\epsilon)
\,
\bump(0)
=
\lim_{\epsilon\to 0} 
\int_\Interval d\sigma'' \,
\gamma^{\iMa\iMb \iSb_1 \ldots\iSb_r 0\ldots 0}(\sigma-\epsilon \sigma'')
\,
\bump(\sigma'')
\\&=
\gamma^{\iMa\iMb \iSb_1 \ldots\iSb_r 0\ldots 0}(\sigma)
\,
\int_\Interval d\sigma'' \,
\bump(\sigma'')
=
\gamma^{\iMa\iMb \iSb_1 \ldots\iSb_r 0\ldots 0}(\sigma)
\DEfullstop
\end{align*}
\end{proof}
}
\noindent
The \JGchange{gauge-like} freedom of the
$\zetaMultiEllis^{\iMa \iMb \iSa_1,\ldots\iSa_r0\ldots0}$ in this case
\jgDchange{arises} from the arbitrary constants when integrating
(\ref{Ellis_adap_gam_zeta}) with respect to $\sigma$.

With respect to this coordinate system, one can partition the
multipoles into a monopole, a pure dipole, a pure quadrupole and so
on. However this is a coordinate dependent splitting and these terms
will mix when changing the coordinate system. The coordinate
transformation for quadrupoles is given  in
(\ref{QP_gamma_chage_coords_munu})-(\ref{QP_gamma_chage_coords_00}).
Although they involve up to $k$ derivatives of the coordinate
transformation, they do not require any integrals.

\subsection{Squeezed tensors}
\label{ch_Ellis_Squz}

In an adapted coordinate system, one can construct a one parameter
family of regular stress-energy tensor \JGchange{densities}
$\TReg^{\iMa\iMb}_\varepsilon$ from a given stress-energy tensor
\jgNewCh{$\TReg^{\iMa\iMb}$}, such that in the weak limit
$\TReg^{\iMa\iMb}_\varepsilon\to T^{\iMa\iMb}$ at $\varepsilon\to0$ to
order $k$.  Since we are using adapted coordinates, we write
$(\sigma,\Vz)=(\sigma,z^1,z^2,z^3)$.  We set
\begin{align}
\TReg^{\iMa\iMb}_\varepsilon(\sigma,\Vz)
&=
\frac{1}{\varepsilon^3}
\ \TReg^{\iMa\iMb}\Big(\sigma,\frac{\Vz}{\varepsilon}\Big)
\label{Ellis_Squz_def_T_eps}
\DEfullstop
\end{align}
We assume that $\TReg^{\iMa\iMb}$ has compact support in the transverse
planes, \jgDchange{i.e.} for each $\sigma$, there is a function $R(\sigma)$ such that 
\begin{align}
\TReg^{\iMa\iMb}(\sigma,\Vz)
=
0
\qquadtext{for} 
g_{\iSa\iSb} \,z^\iSa\,z^\iSb > R(\sigma)
\label{Ellis_Squz_compact_supp}
\DEfullstop
\end{align}
This guarantees that all the moments, are finite.

This leads to
\begin{equation}
\begin{aligned}
\TReg^{\iMa\iMb}_\varepsilon(\sigma,\Vz) 
=
\gamma^{\iMa\iMb0\ldots0} \ \deltaThree(\Vz)
+
\varepsilon\,\gamma^{\iMa\iMb\iSa0\ldots0}\,\partz_\iSa\deltaThree(\Vz) 
+
\varepsilon^2\,\gamma^{\iMa\iMb\iSa\iSb0\ldots0} 
\,\partz_\iSa\partz_\iSb\deltaThree(\Vz) 
+\cdots
\DEfullstop
\end{aligned}
\label{Ellis_Squz_Expansion}
\end{equation}
where
\begin{equation}
\begin{aligned}
\gamma^{\iMa\iMb0\ldots0} (\sigma)
&=
\int_{\Real^3} d^3\Vz\ 
\TReg^{\iMa\iMb}\big(\sigma,\Vz\big)
,\qquad
\gamma^{\iMa\iMb\iSa0\ldots0} (\sigma)
=
-\int_{\Real^3} d^3\Vz\ 
z^\iSa\,\TReg^{\iMa\iMb}\big(\sigma,\Vz\big)
,
\\
\gamma^{\iMa\iMb\iSa\iSb0\ldots0} (\sigma)
&=
\int_{\Real^3} d^3\Vz\ 
z^\iSa\,z^\iSb\,\TReg^{\iMa\iMb}\big(\sigma,\Vz\big)
\qquad\text{etc.}
\DEfullstop
\end{aligned}
\label{Ellis_Squz_moments}
\end{equation}
\begin{ProofI}{Proof of \eqref{Ellis_Squz_Expansion} and
    \eqref{Ellis_Squz_moments}}
\label{pf_MomentsOfT}
This follows from setting
$w^\iSa=z^\iSa/\varepsilon$ and Taylor expanding around $\varepsilon=0$
we have
\begin{align*}
\int_{\Real^4}\TReg^{\iMa\iMb}_\varepsilon(&\sigma,\Vz) \,
\TtwoTen_{\iMa\iMb}(\sigma,\Vz)\, 
d\sigma\,d^3z
\\&=
\int_{\Real} d\sigma \int_{\Real^3} d^3z\ 
\TReg^{\iMa\iMb}_\varepsilon(\sigma,\Vz) \,\TtwoTen_{\iMa\iMb}(\sigma,\Vz)\, 
\\&=
\int_{\Real} d\sigma \int_{\Real^3} d^3z\ 
\frac{1}{\varepsilon^3}
\TReg^{\iMa\iMb}\Big(\sigma,\frac{\Vz}{\varepsilon}\Big) \,\TtwoTen_{\iMa\iMb}(\sigma,\Vz)
\\&=
\int_{\Real} d\sigma \int_{\Real^3} d^3\Vw\ 
\TReg^{\iMa\iMb}\big(\sigma,\Vw\big)
\,\TtwoTen_{\iMa\iMb}(\sigma,\varepsilon \Vw)
\\&=
\int_{\Real} d\sigma \int_{\Real^3} d^3\Vw\ 
\TReg^{\iMa\iMb}\big(\sigma,\Vw\big)
\,\TtwoTen_{\iMa\iMb}(\sigma,\Vzero)
+
\varepsilon\int_{\Real} d\sigma \int_{\Real^3} d^3\Vw\ 
\TReg^{\iMa\iMb}\big(\sigma,\Vw\big)
\,w^\iSa\,\big(\partz_{\iSa}\TtwoTen_{\iMa\iMb}\big)(\sigma,\Vzero)
\\&\qquad+
\varepsilon^2\int_{\Real} d\sigma \int_{\Real^3} d^3\Vw\ 
\TReg^{\iMa\iMb}\big(\sigma,\Vw\big)
\,w^\iSa\,w^\iSb\,
\big(\partz_{\iSa}\,\partz_{\iSb}\TtwoTen_{\iMa\iMb}\big)(\sigma,\Vzero)
+\cdots
\\&=
\int_{\Real} d\sigma 
\gamma^{\iMa\iMb0\ldots0} 
\,\TtwoTen_{\iMa\iMb}|_{C(\sigma)}
-
\varepsilon\int_{\Real}\gamma^{\iMa\iMb\iSa0\ldots0} 
d\sigma\ \big(\partz_{\iSa}\TtwoTen_{\iMa\iMb}\big)\big|_{C(\sigma)}
+
\varepsilon^2\int_{\Real}\gamma^{\iMa\iMb\iSa\iSb0\ldots0} 
d\sigma\ \big(\partz_{\iSa}\TtwoTen_{\iMa\iMb}\big)\big|_{C(\sigma)}
+\cdots
\DEfullstop
\end{align*}
\end{ProofI}
Thus there is an intimate relationship between the components of a
distribution and the moments of a regular stress-energy tensor
\JGchange{density}. Here the zeroth order gives the monopole, the
first order the dipole and so on. \jgNewCh{Again}, \jgDchange{the split between
  the different orders} is with respect to the chosen adapted
coordinate system and \jgDchange{the different order terms} will mix
under a coordinate transformation.

\subsection{The Dixon approach}
\label{sch_Dixon}

The alternative approach, largely developed by Dixon uses the covariant derivative and a
choice of a vector field $\DixVec^{\iMa}(\sigma)$ along the
worldline \jgnchange{$C$}. This we will call the \defn{Dixon vector}.
This vector is required to be not orthogonal to the worldline
\jgnchange{$C$}, i.e.
\begin{align}
\DixVec_{\iMa} \, \Cdot^{\iMa} \ne 0
\label{Intro_Tab_Dixon_N_NonOrth}
\DEfullstop
\end{align}
As long as the worldline $C$ is timelike, a natural choice of the
Dixon vector is $\Cdot^{\iMa}$, i.e. 
$\DixVec_{\iMa}
=
g_{\iMa\iMb}\, \Cdot^{\iMa}
$
but this is not the only choice.
\jgNewCh{Having chosen $\DixVec_{\iMa}$, the \defn{Dixon} representation of a
multipole is given 
\cite[Equation (1.9)]{dixon1967description}\cite[Equation  (4.18),
  (7.4), (7.5)]{DixonII}} by
\begin{align}
T^{\iMa\iMb} = \sum_{r=0}^k {\frac{1}{r!}} \nabla_{\iMc_1} \cdots \nabla_{\iMc_r}
\int_\Interval \zetaMultiDixon^{\iMa\iMb \iMc_1\ldots\iMc_r}(\sigma)
\,
\deltaFour\big(x-C(\sigma)\big)\,d\sigma
\label{Intro_Tab_Dixon_Multi}
\DEcomma
\end{align}
where we demand that the components 
$\zetaMultiDixon^{\iMa \iMb \iMc_1\ldots\iMc_k}$ are orthogonal to 
the vector $\DixVec^{\iMa}$,
\begin{align}
\DixVec_{\iMc_j}\
\zetaMultiDixon^{\iMa \iMb \iMc_1\ldots\iMc_k}
= 0
\label{Intro_Tab_Dixon_orthog}
\DEnone
\end{align}
for $j=1,\ldots,k$.
\jgDchange{The covariant derivatives do not commute.
However, they give rise to curvature terms and lower the number of
derivatives. We therefore make the minimal choice and impose 
$\zetaMultiDixon^{\iMa \iMb \iMc_1\ldots\iMc_k}$ are symmetric in the
relevant indices.}
\begin{align}
\zetaMultiDixon^{\iMa \iMb \iMc_1\ldots\iMc_k} 
=
\zetaMultiDixon^{\iMa\iMb (\iMc_1\ldots\iMc_k)}
\label{Intro_Zeta_Dixon_sym}
\DEfullstop
\end{align}
\jgNewCh{Since $T^{\iMa\iMb}$ is a tensor density this enables us to throw the
covariant derivative over onto the test tensor $\TtwoTen_{\iMa\iMb}$, giving 
\begin{align}
\int_\Mman T^{\iMa\iMb} \ \TtwoTen_{\iMa\iMb} \ d^4x 
&=
\sum_{r=0}^k (-1)^{r} {\frac{1}{r!}} \int_\Interval  
\zetaMultiDixon^{\iMa\iMb \iMc_1\ldots\iMc_r}(\sigma)\,
\big(\nabla_{\iMc_1} \cdots \nabla_{\iMc_r} \TtwoTen_{\iMa\iMb} \big)\big|_{C(\sigma)}
\ d\sigma
\label{Intro_Tab_Dixon_Multi_action}
\DEcomma
\end{align}
This follow since if
$v^\iMa$ is a vector density of weight 1 then from (\ref{Intro_ten_den})
$\nabla_\iMa\,v^\iMa=\partial_\iMa\,v^\iMa$. 
}

\jgPchange{All Dixon multipoles of order $k$ are also Ellis multipoles
  of order $k$.
\begin{proof}
Expanding out the left hand side of 
(\ref{Intro_Tab_Dixon_Multi_action}) replaces the covariant
derivatives with partial derivatives and Christoffel symbols. 
The resulting expression is the integral of sum of up to $k$ 
partial derivatives of
$\TtwoTen_{\iMa\iMb}$. Using
(\ref{Intro_Tab_Ellis_k+1}) and (\ref{Intro_Tab_Ellis_k+1_def}) one
can express all the terms to have precisely $k$ derivatives, and hence
one has (\ref{Intro_Tab_Ellis_Multi_action}) for the appropriate
$\zetaMultiEllis^{\iMa \iMb \iMc_1\ldots\iMc_k}$.
\end{proof}
}

From (\ref{Intro_Tab_Dixon_Multi}) we can use the Dixon vector to
perform the \defn{Dixon split} in order to take an arbitrary $k$th
order multipole and split it into a monopole part, a dipole part and
so on. Thus we set
\begin{align}
T^{\iMa\iMb} = \sum_{r=0}^k T^{\iMa\iMb}_{(r)}
\qquadtext{where}
T^{\iMa\iMb}_{(r)} = {\frac{1}{r!}}\nabla_{\iMc_1} \cdots \nabla_{\iMc_r}
\int_\Interval \zetaMultiDixon^{\iMa\iMb \iMc_1\ldots\iMc_r}(\sigma)
\,
\deltaFour\big(x-C(\sigma)\big)\,d\sigma
\label{Intro_Tab_Dixon_Split}
\DEfullstop
\end{align}
In section \ref{ch_CoFree_Dixon} {we present a coordinate free approach
to performing this split. This is presented for quadrupoles, although the procedure
can be extended.} \jgPchange{This is necessary to show that
  all Ellis multipole are also Dixon multipoles.} It also gives a
method to derived the relationship between the Dixon components with
respect to different Dixon vectors.

\begin{table}
\begin{equation*}
\begin{array}{|l|l|}
\hline
\text{speed of light} & [1]
\\\hline
dx^{\iMa} & [L]
\\\hline
g_{\iMa\iMb} & [1]
\\\hline
\Cdot & [L^{-1}]
\\\hline
\Cdot^{\iMa} & [1]
\\\hline
\partx_{\iMa} & [L^{-1}] 
\\\hline
\deltaFour\big(x-C(\sigma)\big) & [L^{-4}]
\\\hline
\text{mass }m & [M]
\\\hline
\end{array}
\qquad\qquad
\begin{array}{|l|l|}
\hline
T^{\iMa\iMb} & [M\,L^{-3}]
\\\hline
\text{test tensor } \TtwoTen_{\iMa\iMb} & [L^{-1}]
\\\hline
\text{dipole displacement }X^{\iMa} & [M L]
\\\hline
\text{dipole 3--momentum }P^{\iMa} & [M]
\\\hline
\text{dipole spin }S^{\iMa\iMb} & [M L]
\\\hline
\zetaMultiEllis^{\iMa\iMb \iMc_{i_1}\ldots \iMc_{i_k}} & [M\,L^{k}]
\\\hline
\zetaMultiDixon^{\iMa\iMb \iMc_{i_1} \ldots \iMc_{i_k}} & [M\,L^{k}]
\\\hline
\gamma^{\iMa\iMb \iSa_{i_1}\ldots\iSa_{i_k}0\ldots 0} & [M\,L^{k}]
\\\hline
\end{array}
\label{Intro_convension_units}
\end{equation*}
\caption{List of units for quantities, in terms of mass $M$ and length
  $L$. The speed of light is 1.}
\label{tab_List_units}
\end{table}


\section{Summary of the monopole and dipole stress-energy tensors.}
\label{ch_MD}

\subsection{The monopole}
\label{sch_MD_mono}

From (\ref{Intro_Tab_Ellis_Multi}) with $k=0$ we have the 
gravitational monopole
\begin{align}
T^{\iMa\iMb}=\int_\Interval \zeta^{\iMa\iMb} \delta\big(x-C(\sigma)\big)\,d\sigma
\label{MD_monpole}
\DEfullstop
\end{align}
The requirement to be a stress-energy tensor 
(\ref{Intro_Tab_Sym}),(\ref{Intro_Tab_Div_zero}) implies that $C$ satisfies
the pre-geodesic equation 
\begin{align}
\Dfrac{\Cdot^\iMa}{\sigma} = \kappa_\pregeo(\sigma)\,\Cdot^\iMa
\label{MD_pre_geodesic}
\DEnone
\end{align}
and
\begin{align}
T^{\iMa\iMb} 
= 
\int_\Interval m_\pregeo(\sigma)\,
\Cdot^{\iMa}\,\Cdot^\iMb\,\delta\big(x-C(\sigma)\big)
\ d\sigma
\label{MD_pre_geodesic_T}
\DEnone
\end{align}
where
\begin{align}
\dot{m}_\pregeo + \kappa_\pregeo\,m_\pregeo=0
\label{MD_pre_geodesic_m_eqn}
\DEfullstop
\end{align}
Here $\Dfrac{}{\sigma}$ represents the covariant derivative along the
worldline, i.e.
\begin{align}
\Dfrac{X^{\iMa}}{\sigma}
=
\dot{X}^{\iMa} + \Gamma^{\iMa}_{\iMb\iMc}\, X^\iMb\,\Cdot^\iMc
\label{MD_def_Dfrac}
\DEnone
\end{align}
and the overdot refers to differentiation with respect to
differentiation with respect to $\sigma$. If $\sigma$ is proper times
so that 
\begin{align}
g_{\iMa\iMb}\,\Cdot^\iMa\,\Cdot^\iMb=-1
\label{MD_propertime}
\DEfullstop
\end{align}
then $\kappa_\pregeo=0$
and (\ref{MD_pre_geodesic}) gives the geodesic equation
\begin{align}
\Dfrac{\Cdot^\iMa}{\sigma} = 0
\label{MD_geodesic}
\DEfullstop
\end{align}
In this case we replace $m_\pregeo$ with $m$ in
(\ref{MD_pre_geodesic_T}). If $m>0$ then we can associate it with the
mass of the source. Thus (\ref{MD_pre_geodesic_T}) becomes

\begin{align}
T^{\iMa\iMb} 
= 
m \int_\Interval 
\Cdot^{\iMa}\,\Cdot^\iMb\,\delta\big(x-C(\sigma)\big)
\,d\sigma
\label{MD_geodesic_T}
\DEfullstop
\end{align}
Thus there \jgDchange{remains} just
one ODE for the remaining component, namely $\dot{m}=0$. There are no
additional free components. See table
\ref{tab_number_components}. However as stated in the introduction, 
we do not impose the geodesic equation for the subsequent analysis of 
the \jgPchange{dipole and
quadrupoles stress-energy tensors.}

\subsection{The dipole}

Setting $k=1$ in (\ref{Intro_Tab_Ellis_Multi}) gives the dipole
\begin{align}
T^{\iMa\iMb}
=
\int_\Interval \zetaDP^{\iMa\iMb\iMc}\, \partx_\iMc
\delta\big(x-C(\sigma)\big)
\,d\sigma
\label{MD_diploe_Tab}
\DEcomma
\end{align}
where the symmetry condition (\ref{Intro_Tab_Sym}) implies
$\zetaDP^{\iMa\iMb\iMc}=\zetaDP^{\iMb\iMa\iMc}$.
We observe that, whereas the components $\zetaDP^{\iMa\iMb\iMc}$ uniquely
specify $T^{\iMa\iMb}$, the converse is not true. That is, given $T^{\iMa\iMb}$
the gauge-like freedom in $\zetaDP^{\iMa\iMb\iMc}$ \jgDchange{is given} by
\begin{align}
\zetaDP^{\iMa\iMb\iMc} \to \zetaDP^{\iMa\iMb\iMc} + M^{\iMa\iMb}\Cdot^\iMc
\label{MD_diploe_Zeta_Freedom}
\DEcomma
\end{align}
where $M^{\iMa\iMb}=M^{\iMb\iMa}$ are any set of constants,
i.e. independent of $\sigma$.

\begin{ProofI}{Proof of \eqref{MD_diploe_Zeta_Freedom}}
\label{pf_Dipole_zeta_freedom}
Substituting \eqref{MD_diploe_Zeta_Freedom} into
\eqref{MD_diploe_Tab} {we have}
\begin{align*}
T^{\iMa\iMb}
\to
T^{\iMa\iMb}
+
\int_\Interval  M^{\iMa\iMb}\Cdot^\iMc \partx_\iMc \JGchange{\delta(x-C(\sigma))}\,d\sigma
&=
T^{\iMa\iMb}
+
\int_\Interval  M^{\iMa\iMb} {\tfrac{d}{d\sigma}} \delta(x-C(\sigma))\,d\sigma
\\&=
T^{\iMa\iMb}
+
\int_\Interval  {\tfrac{d}{d\sigma}} \big(M^{\iMa\iMb} \delta(x-C(\sigma))\big)\,d\sigma
=
T^{\iMa\iMb}
\DEfullstop
\end{align*}
Thus \eqref{MD_diploe_Zeta_Freedom} is a gauge-like freedom. To show it is
the maximum freedom, \jgDchange{we work} in adaptive coordinates. It is
clear that the freedom \eqref{MD_diploe_Zeta_Freedom} is precisely
equivalent to the freedom to choose $\zetaDP^{\iMa\iMb0}$ given
$\gamma^{\iMa\iMb0}$. For details of why this is the maximum gauge-like
freedom see proofs number \ref{pf_ChangCoordZeta} and
\ref{pf_ChangCoordZeta_Gauge} in the appendix about the gauge-like
freedom of the quadrupole
(\ref{QP_Zeta_Freedom}). 
\end{ProofI}


In addition the $\zetaDP^{\iMa\iMb\iMc}$ are not tensorial quantities but
have a coordinate transformation which involves derivatives of the
Jacobian matrix and an integral. Given two coordinate systems
$(x^0,\ldots,x^3)$ and $(\xhat^0,\ldots,\xhat^3)$ then
\begin{align}
\zetahat^{\iMahat\iMbhat\iMchat}
=
J^{\iMahat}_{\iMa} J^{\iMbhat}_\iMb J^{\iMchat}_\iMc  \zeta^{\iMa\iMb\iMc}
-
\Cdothat{}^\iMchat 
\int^\sigma \partial_\iMc(J^{\iMahat}_{\iMa} J^\iMbhat_\iMb)\, 
\zeta^{\iMa\iMb\iMc}\,d\sigma'
\label{MD_diploe_Change_coords}
\DEcomma
\end{align}
where
\begin{align}
J^{\iMahat}_{\iMa} = \pfrac{\xhat^{\iMahat}}{x^{\iMa}}
\label{QP_def_J}
\DEfullstop
\end{align}
\jgnchange{
\begin{ProofA}{Proof of \eqref{MD_diploe_Change_coords}}
Observe that
\begin{align*}
\int_\Interval \zeta^{\iMa\iMb\iMc} \,
&\partx_\iMc \big({J^{\iMahat}_{\iMa} J^{\iMbhat}_\iMb}\big)\, 
\TtwoTenhat_{\iMahat\iMbhat}
\ d\sigma
=
\int_\Interval \zeta^{\iMa\iMb\iMc} \,
\partx_\iMc \big({J^{\iMahat}_{\iMa} J^{\iMbhat}_\iMb}\big)\, 
\Big( \int^\sigma \dfrac{\TtwoTenhat_{\iMahat\iMbhat}\,}{\sigma'} d\sigma'\Big)
\ d\sigma
\\&=
\int_\Interval \zeta^{\iMa\iMb\iMc} \,
\partx_\iMc \big({J^{\iMahat}_{\iMa} J^{\iMbhat}_\iMb}\big)\, 
\bigg( \int^\sigma \Cdothat^\iMchat \partx_\iMchat
\TtwoTenhat_{\iMahat\iMbhat} 
d\sigma'\bigg)
\ d\sigma
=
\int_\Interval
\bigg( \int^\sigma 
 \zeta^{\iMa\iMb\iMc} \,\partx_\iMc \big({J^{\iMahat}_{\iMa} J^{\iMbhat}_\iMb}\big)\, 
d\sigma'\bigg)
\Cdothat^\iMchat \partx_\iMchat
\TtwoTenhat_{\iMahat\iMbhat} \ d\sigma
\DEfullstop
\end{align*}
Hence using (\ref{MD_diploe_Tab}) we have 
\begin{align*}
\int_\Interval \zetahat^{\iMahat\iMbhat\iMchat} \,
&\big(\partx_\iMchat\,\TtwoTenhat_{\iMahat\iMbhat}\big)
\ d\sigma
=
\int_{\Real^4} \THat^{\iMahat\iMbhat}\,\TtwoTenhat_{\iMahat\iMbhat}\,d^4\xhat
=
\int_{\Real^4} T^{\iMa\iMb}\,\TtwoTen_{\iMa\iMb}\,d^4x
=
\int_\Interval \zeta^{\iMa\iMb\iMc} \,
\big(\partx_\iMc\,\TtwoTen_{\iMa\iMb}\big) d\sigma
\\&=
\int_\Interval \zeta^{\iMa\iMb\iMc} \,
\partx_\iMc\,
\big({J^{\iMahat}_{\iMa} J^{\iMbhat}_\iMb}\, \TtwoTenhat_{\iMahat\iMbhat}\big)
\ d\sigma
=
\int_\Interval \zeta^{\iMa\iMb\iMc}\,
\Big(
\partx_\iMc\,\big({J^{\iMahat}_{\iMa} J^{\iMbhat}_\iMb}\big)\, \TtwoTenhat_{\iMahat\iMbhat}
+
{J^{\iMahat}_{\iMa} J^{\iMbhat}_\iMb}\ \partx_\iMc\,\TtwoTenhat_{\iMahat\iMbhat}
\Big)
\ d\sigma
\\&=
\int_\Interval 
\Bigg(
\bigg( \Cdothat^\iMchat 
\int^\sigma 
 \zeta^{\iMa\iMb\iMc} \,\partx_\iMc \big({J^{\iMahat}_{\iMa} J^{\iMbhat}_\iMb}\big)\, 
d\sigma'\bigg)
+
{J^{\iMahat}_{\iMa} J^{\iMbhat}_\iMb J^{\iMchat}_\iMc}\ 
\zeta^{\iMa\iMb\iMc}\,\Bigg)
\partx_{\iMchat}\,\TtwoTenhat_{\iMahat\iMbhat}
\ d\sigma
\DEfullstop
\end{align*}
Hence (\ref{MD_diploe_Change_coords}).
\end{ProofA}
}
Here the freedom to choose the arbitrary constant of integration in
(\ref{MD_diploe_Change_coords}) {is equivalent to the \JGchange{gauge-like}
freedom (\ref{MD_diploe_Zeta_Freedom}).}
\jgPchange{
\begin{ProofI}{Gauge-like}
Consider the cases where the limits of the integral in
(\ref{MD_diploe_Change_coords}) are $\int^\sigma_{\sigma_0}$ and
$\int^\sigma_{\sigma_1}$. \jgDchange{Then} the difference between two expressions
for $\zetahat^{\iMahat\iMbhat\iMchat}$ is 
\begin{align*}
\zetahat^{\iMahat\iMbhat\iMchat} 
\to 
\zetahat^{\iMahat\iMbhat\iMchat}+ 
\Mtwo^{\iMahat\iMbhat}\,\Cdothat{}^\iMchat 
\qquadtext{where}
\Mtwo^{\iMahat\iMbhat}
=
\int_{\sigma_1}^{\sigma_0} 
\partial_\iMc(J^{\iMahat}_{\iMa} J^\iMbhat_\iMb)\, 
\zeta^{\iMa\iMb\iMc}\,d\sigma'
\DEcomma
\end{align*}
hence the gauge-like freedom in (\ref{MD_diploe_Zeta_Freedom}).
\end{ProofI}
}

In adapted coordinates (\ref{Intro_adapt_coords}) then
(\ref{Ellis_adap_Ellis_Multi}) and
(\ref{Ellis_adap_Ellis_Multi_action})
become
\begin{align}
T^{\iMa\iMb} = 
\gamma^{\iMa\iMb0}
\deltaThree(\Vz)
+
\gamma^{\iMa\iMb \iSa}
\ \partz_{\iSa} 
\deltaThree(\Vz)
\quadtext{where}
\gamma^{\iMa\iMb 0}
=
\dot\zeta^{\iMa\iMb0}
\quadand
\gamma^{\iMa\iMb \iSa}
=
\zeta^{\iMa\iMb\iSa}
\label{MD_adap_Ellis}
\DEfullstop
\end{align}

Fortunately for the dipole the requirements (\ref{Intro_Tab_Sym}) and
(\ref{Intro_Tab_Div_zero}) restrict the components
$\zetaDP^{\iMa\iMb\iMc}$ so much 
that $T^{\iMa\iMb}$ can be written solely in terms of tensor
quantities
\begin{align}
T^{\iMa\iMb} 
&=
\int_\Interval 
\hat{P}^{\lround\iMa}\,\Cdot^{\iMb\rround}
\,\delta\big(x-C(\sigma)\big)
d\sigma
+
\nabla_\iMc \int_\Interval 
\hat{S}^{\iMc\lround\iMa}\,\Cdot^{\iMb\rround}
\,\delta\big(x-C(\sigma)\big)
d\sigma
\DEcomma
\label{MD_diploe_Dixon_Tab_NoGeo}
\end{align}
where $\hat{P}^{\iMa}$ and $\hat{S}^{\iMa\iMb}+\hat{S}^{\iMb\iMa}=0$ 
satisfy the
Mathisson-Papapetrou-Tulczyjew-Dixon equations
\begin{align}
\Dfrac{\hat{S}^{\iMa\iMb}}{\sigma}
&= 
\hat{P}^\iMb\Cdot^{\iMa}
-
\hat{P}^{\iMa}\Cdot^\iMb 
\qquadand
\Dfrac{\hat{P}^{\iMa}}{\sigma} 
=
\tfrac12 R^{\iMa}{}_{\iMb\iMc\iMd}\,\Cdot^\iMb\, \hat{S}^{\iMd\iMc} 
\label{MD_diploe_DVa_NoGeo}
\DEfullstop
\end{align}
\JGchange{Given a vector $\DixVec^\iMc$
such that
$\DixVec_\iMc\,\hat{S}^{\iMc\lround\iMa}\,\Cdot^{\iMb\rround}=0$, then
from (\ref{Intro_Tab_Dixon_orthog}) we can interpret
(\ref{MD_diploe_Dixon_Tab_NoGeo}) as the  Dixon
representation of a \jgDchange{dipole} with Dixon vector $\DixVec^\iMc$.}

Clearly we can replace the covariant derivatives with partial
derivatives and Christoffel symbols to give the representation of
the dipole
\begin{equation}
\begin{aligned}
T^{\iMa\iMb} 
&=
\int_\Interval 
\Big(
\hat{P}^{\lround\iMa}\,\Cdot^{\iMb\rround}
+
\hat{S}^{\iMc\lround\iMb}\,\Gamma^{\iMa\rround}{}_{\iMc\iMd} \,\Cdot^\iMd
\Big)\,\delta\big(x-C(\sigma)\big)
d\sigma
+
\int_\Interval 
\hat{S}^{\iMc\lround\iMa}\,\Cdot^{\iMb\rround}
\,\partial_\iMc \,\delta\big(x-C(\sigma)\big)
d\sigma
\DEfullstop
\end{aligned}
\label{MD_diploe_Ellis_Tab_NoGeo}
\end{equation}
However this is not the Ellis representation. \jgnchange{To translate
   (\ref{MD_diploe_Ellis_Tab_NoGeo}) into the Ellis representation 
(\ref{MD_diploe_Tab}) we set}
\begin{align}
\zetaDP^{\iMa\iMb\iMc} 
=
\hat{S}^{\iMc\lround\iMa}\,\Cdot^{\iMb\rround}
+
\Cdot^{\iMc}
\int^\sigma
\Big(
\hat{P}^{\lround\iMa}\,\Cdot^{\iMb\rround}
+
\JGchange{\hat{S}^{\iMe\lround\iMb}\,\Gamma^{\iMa\rround}{}_{\iMe\iMd} \,\Cdot^\iMd}
\Big)\,
d\sigma'
\label{MD_diploe_Ellis_zeta}
\DEfullstop
\end{align}
In the adapted coordinates (\ref{MD_adap_Ellis}) we have
\begin{align}
\gamma^{\iMa\iMb 0}
=
\hat{P}^{\lround\iMa}\,\delta_0^{\iMb\rround}
+
\hat{S}^{\iMc\lround\iMb}\,\Gamma^{\iMa\rround}{}_{\iMc0} 
+
\partial_0(\hat{S}^{0\lround\iMa}\,\delta_0^{\iMb\rround})
\qquadand
\gamma^{\iMa\iMb \iSa}
=
\hat{S}^{\iSa\lround\iMa}\,\delta_0^{\iMb\rround}
\label{MD_gammas}
\DEfullstop
\end{align}

\jgPchange{Let} $\Kill^\iMa$ be Killing vector 
\begin{align}
\nabla_{\iMa}\Kill_\iMb + \nabla_\iMb\Kill_{\iMa} = 0
\label{Q_Cons_symm_Kill}
\DEnone
\end{align}
and let
\begin{align}
\Conserv_\Kill
&=
\gamma^{\iMa00} \,\Kill_{\iMa} - 
\gamma^{\iMa0\iSa} \, \partz_\iSa \,\Kill_{\iMa}
\label{MD_Cons_Q}
\DEfullstop
\end{align}
\jgPchange{We show below that $\Conserv_\Kill$ is a conserved
  quantity.}
From (\ref{MD_gammas}) we have
\begin{align}
\Conserv_\Kill
=
\hat{P}^\iMa\,K_\iMa
+
\tfrac12\hat{S}^{\iMa\iMb}\,\nabla_{\iMb} K_\iMa
\label{MD_Q_K}
\DEfullstop
\end{align}
\begin{ProofA}{Proof that \eqref{MD_Cons_Q} and \eqref{MD_Q_K} are equivalent.}
\label{pf_QK_def}
From \eqref{MD_diploe_DVa_NoGeo} we have
\begin{align*}
\partial_0 \hat{S}^{\iMa\iMb}
&=
\Dfrac{\hat{S}^{\iMa\iMb}}{\sigma}
-
\Gamma^\iMa_{0\iMc}\, \hat{S}^{\iMc\iMb}
-
\Gamma^\iMb_{0\iMc}\, \hat{S}^{\iMa\iMc}
=
\hat{P}^\iMb\Cdot^{\iMa} - \hat{P}^{\iMa}\Cdot^\iMb 
-
\Gamma^\iMa_{0\iMc}\, \hat{S}^{\iMc\iMb}
-
\Gamma^\iMb_{0\iMc}\, \hat{S}^{\iMa\iMc}
\\&=
\hat{P}^\iMb\delta^{\iMa}_0
-
\hat{P}^{\iMa}\delta^\iMb_0 
-
\Gamma^\iMa_{0\iMc}\, \hat{S}^{\iMc\iMb}
-
\Gamma^\iMb_{0\iMc}\, \hat{S}^{\iMa\iMc}
\DEnone
\end{align*}
so
\begin{align*}
\partial_0 \hat{S}^{0\iMa}
&=
\hat{P}^\iMa
-
\hat{P}^{0}\delta^\iMa_0 
-
\Gamma^0_{0\iMc}\, \hat{S}^{\iMc\iMa}
-
\Gamma^\iMa_{0\iMc}\, \hat{S}^{0\iMc}
\DEfullstop
\end{align*}
From \eqref{MD_gammas} we have
\begin{align*}
\gamma^{\iMa00}
&=
\hat{P}^{\lround\iMa}\,\Cdot^{0\rround}
+
\hat{S}^{\iMc\lround 0}\,\Gamma^{\iMa\rround}{}_{\iMc\iMd} \,\Cdot^\iMd
+
\partial_0(\hat{S}^{0\lround\iMa}\,\Cdot^{0\rround})
\\&=
\tfrac12 
\big(
\hat{P}^{\iMa}
+
\hat{P}^0 \delta^{\iMa}_0
+
\hat{S}^{\iMc 0}\,\Gamma^{\iMa}{}_{\iMc0} 
+
\hat{S}^{\iMc\iMa}\,\Gamma^{0}{}_{\iMc0}
+
\partial_0(\hat{S}^{0\iMa})
\big)
\\&=
\tfrac12 
\big(
\hat{P}^{\iMa}
+
\hat{P}^0 \delta^{\iMa}_0
+
\hat{S}^{\iMc 0}\,\Gamma^{\iMa}{}_{\iMc0} 
+
\hat{S}^{\iMc\iMa}\,\Gamma^{0}{}_{\iMc0}
+ 
\hat{P}^\iMa
-
\hat{P}^{0}\delta^\iMa_0 
-
\Gamma^0_{0\iMc}\, \hat{S}^{\iMc\iMa}
-
\Gamma^\iMa_{0\iMc}\, \hat{S}^{0\iMc}
\big)
\\&=
\hat{P}^{\iMa}
+
\hat{S}^{\iMc 0}\,\Gamma^{\iMa}{}_{\iMc0} 
\DEnone
\end{align*}
and
\begin{align*}
\gamma^{\iMa0 \iSa}
=
\tfrac12
\hat{S}^{\iSa\iMa}
+
\tfrac12
\hat{S}^{\iSa0}\delta^\iMa_0
\DEfullstop
\end{align*}
From \eqref{Q_Cons_symm_Kill} we have
\begin{align*}
0
&=
\nabla_{\iSa}K_0 + \nabla_{0}K_\iSa
=
\partial_{\iSa}K_0 + \partial_{0}K_\iSa - 2\Gamma^\iMa_{\iSa 0} K_\iMa
\DEfullstop
\end{align*}
Hence from \eqref{MD_Cons_Q} we have
\begin{align*}
\Conserv_\Kill
&=
\gamma^{\iMa00} \,\Kill_{\iMa} - 
\gamma^{\iMa0\iSa} \, \partz_\iSa \,\Kill_{\iMa}
=
\big(\hat{P}^{\iMa}
+
\hat{S}^{\iMc 0}\,\Gamma^{\iMa}{}_{\iMc0} 
\big)K_\iMa
- 
\tfrac12
\big(\hat{S}^{\iSa\iMa}
+
\hat{S}^{\iSa0}\delta^\iMa_0
\big)\, \partz_\iSa \,\Kill_{\iMa}
\\&=
\hat{P}^{\iMa}K_\iMa
+
\hat{S}^{\iMc 0}\,\Gamma^{\iMa}{}_{\iMc0} 
K_\iMa
- 
\tfrac12
\hat{S}^{\iSa\iMa}
\,\partz_\iSa\Kill_{\iMa}
-
\tfrac12
\hat{S}^{\iSa0}
\,\partz_\iSa\Kill_{0}
\\&=
\hat{P}^{\iMa}K_\iMa
+
\hat{S}^{\iMc 0}\,\Gamma^{\iMa}{}_{\iMc0} 
K_\iMa
- 
\tfrac12
\hat{S}^{\iSa\iMa}
\,\partz_\iSa\Kill_{\iMa}
+
\tfrac12
\hat{S}^{\iSa0}
\,\partz_0\Kill_{\iSa}
-
\hat{S}^{\iSa0}
\Gamma^\iMa_{\iSa 0} K_\iMa
\\&=
\hat{P}^{\iMa}K_\iMa
+ 
\tfrac12
\hat{S}^{\iMa\iSa}
\,\partz_\iSa\Kill_{\iMa}
+
\tfrac12
\hat{S}^{\iMa0}
\,\partz_0\Kill_{\iMa}
=
\hat{P}^{\iMa}K_\iMa
+ 
\tfrac12
\hat{S}^{\iMa\iMb}
\,\partz_\iMb\Kill_{\iMa}
=
\hat{P}^{\iMa}K_\iMa
+ 
\tfrac12
\hat{S}^{\iMa\iMb}
\,\nabla_\iMb\Kill_{\iMa}
\DEfullstop
\end{align*}
\end{ProofA}
\begin{ProofA}{Proof that $\Conserv_\Kill$ in \eqref{MD_Q_K} is conserved.}
\label{pf_QK_consved}
Since $\Kill_{\iMa}$ is \jgnchange{Killing} we have
\begin{align*}
\nabla_\iMa\nabla_\iMb\Kill_\iMc
=
R^\iMd{}_{\iMa\iMb\iMc} \Kill_\iMd
\DEfullstop
\end{align*}
From \eqref{MD_Q_K} and (\ref{MD_diploe_DVa_NoGeo}) we have 
\begin{align*}
\dot\Conserv_\Kill
&=
\Dfrac{\Conserv_\Kill}{\sigma}
=
\Dfrac{\hat{P}^\iMa}{\sigma}\,K_\iMa
+
\hat{P}^\iMa\,\Cdot^\iMb\nabla_\iMb K_\iMa
+
\tfrac12\Dfrac{\hat{S}^{\iMa\iMb}}{\sigma}\,\nabla_{\iMb}\ K_\iMa
+
\tfrac12\hat{S}^{\iMa\iMb}\,\Cdot^\iMc\nabla_{\iMc}\nabla_{\iMb}\ K_\iMa
\\&=
\tfrac12 R^{\iMa}{}_{\iMb\iMc\iMd}\,\Cdot^\iMb\, \hat{S}^{\iMd\iMc} 
\,K_\iMa
+
\hat{P}^\iMa\,\Cdot^\iMb\nabla_\iMb K_\iMa
+
\tfrac12
\Big(\hat{P}^\iMb\Cdot^{\iMa}
-
\hat{P}^{\iMa}\Cdot^\iMb\Big)
\,\nabla_{\iMb}K_\iMa
+
\tfrac12\hat{S}^{\iMa\iMb}\,\Cdot^\iMc\nabla_{\iMc}\nabla_{\iMb}K_\iMa
\\&=
\tfrac12 R^{\iMa}{}_{\iMb\iMc\iMd}\,\Cdot^\iMb\, \hat{S}^{\iMd\iMc} 
\,K_\iMa
+
\tfrac12\hat{S}^{\iMa\iMb}\,\Cdot^\iMc\,R^\iMd{}_{\iMc\iMb\iMa} \Kill_\iMd
=0
\DEfullstop
\end{align*}
\end{ProofA}

\vspace{1em}

The situation is simplified in the case when $C$ is a geodesic. In
this case we can  
use the Dixon representation with $\DixVec^{\iMa}=\Cdot^{\iMa}$. 
\begin{equation}
\begin{aligned}
T^{\iMa\iMb} 
&=
\int_\Interval 
\Big(
m \Cdot^{\iMa}\,\Cdot^\iMb
+
P^{\lround\iMa}\,\Cdot^{\iMb\rround}
\Big)\,\delta\big(x-C(\sigma)\big)
d\sigma
+
\nabla_\iMc \int_\Interval 
\Big(
X^\iMc \Cdot^{\iMa}\,\Cdot^\iMb
+
S^{\iMc\lround\iMa}\,\Cdot^{\iMb\rround}
\Big)\,\delta\big(x-C(\sigma)\big)
d\sigma
\DEcomma
\end{aligned}
\label{MD_diploe_Dixon_Tab}
\end{equation}
where
\begin{align}
\hat{S}^{\iMa\iMb} = S^{\iMa\iMb} - X^{\iMa}\Cdot^\iMb + X^\iMb\Cdot^{\iMa}
\qquadand
\hat{P}^{\iMa}=P^{\iMa}+m\Cdot^\iMa
\label{MD_geodesic_Deqns}
\DEfullstop
\end{align}

\begin{ProofA}{Proof of the relationship between
    \eqref{MD_diploe_DSab} and \eqref{MD_diploe_DVa_NoGeo}}
\label{pf_MattPap}
In this proof we refer to the two equations in
\eqref{MD_diploe_DVa_NoGeo} as (\ref{MD_diploe_DVa_NoGeo}.1) and
(\ref{MD_diploe_DVa_NoGeo}.2) and
likewise for (\ref{MD_diploe_DSab}.1) to (\ref{MD_diploe_DSab}.4).
From \eqref{MD_geodesic_Deqns} and (\ref{MD_geodesic}) we have
\begin{align*}
\Dfrac{\hat{S}^{\iMa\iMb}}{\sigma} 
&-
\hat{P}^\iMb\Cdot^{\iMa}
+
\hat{P}^{\iMa}\Cdot^\iMb 
\\&= 
\Dfrac{S^{\iMa\iMb}}{\sigma} 
- \Dfrac{X^{\iMa}}{\sigma}\Cdot^\iMb 
- X^{\iMa}\Dfrac{\Cdot^\iMb}{\sigma}  
+ \Dfrac{X^\iMb}{\sigma}\Cdot^{\iMa}
+ X^\iMb\Dfrac{\Cdot^{\iMa}}{\sigma}
-
({P}^\iMb+m\Cdot^\iMb)\Cdot^{\iMa}
+
({P}^{\iMa}+m\Cdot^{\iMa})\Cdot^\iMb 
\\&= 
\Dfrac{S^{\iMa\iMb}}{\sigma} 
- \Big(\Dfrac{X^{\iMa}}{\sigma} - P^{\iMa}\Big)\Cdot^\iMb 
+ \Big(\Dfrac{X^\iMb}{\sigma} - P^{\iMb}\Big)\Cdot^{\iMa}
\DEfullstop
\end{align*}
Hence (\ref{MD_diploe_DSab}.2) and (\ref{MD_diploe_DSab}.4) imply
(\ref{MD_diploe_DVa_NoGeo}.1). By contrast from
(\ref{MD_diploe_DVa_NoGeo}.1) we can project out
(\ref{MD_diploe_DSab}.2) and (\ref{MD_diploe_DSab}.4) using
$\Cdot_\iMa$.

Likewise from \eqref{MD_geodesic_Deqns} we have
\begin{align*}
\Dfrac{\hat{P}^{\iMa}}{\sigma} 
-
\tfrac12 R^{\iMa}{}_{\iMb\iMc\iMd}\,\Cdot^\iMb\, \hat{S}^{\iMd\iMc} 
&=
\Dfrac{{P}^{\iMa}}{\sigma} + \Dfrac{m}{\sigma}\Cdot^\iMa +
m\Dfrac{\Cdot^\iMa}{\sigma} 
-
\tfrac12 R^{\iMa}{}_{\iMb\iMc\iMd}\,\Cdot^\iMb\, 
\big({S}^{\iMd\iMc} - X^{\iMd}\Cdot^\iMc + X^\iMc\Cdot^{\iMd}\big)
\\&=
\Dfrac{{P}^{\iMa}}{\sigma} + \dot{m}\Cdot^\iMa 
-
\tfrac12 R^{\iMa}{}_{\iMb\iMc\iMd}\,\Cdot^\iMb\,{S}^{\iMd\iMc} 
-
R^{\iMa}{}_{\iMb\iMc\iMd}\,\Cdot^\iMb\, 
X^{\iMd}\Cdot^\iMc
\DEfullstop
\end{align*}
Thus (\ref{MD_diploe_DSab}.1) and (\ref{MD_diploe_DSab}.3) imply
(\ref{MD_diploe_DVa_NoGeo}.2). By contrast from
(\ref{MD_diploe_DVa_NoGeo}.2) we can project out
(\ref{MD_diploe_DSab}.1) and (\ref{MD_diploe_DSab}.3) using
$\Cdot_\iMa$.
\end{ProofA}

These quantities have intuitive meaning. See Table
\ref{tab_List_units}
for the units associated with each component.
\begin{bulletlist}
\item
The rest mass $m$.
\item
A displacement vector  $X^{\iMa}$ with $X_{\iMa}\,\Cdot^{\iMa}=0$.
\item
The rate of change of the displacement vector $P^{\iMa}$ with
$P_{\iMa}\,\Cdot^{\iMa}=0$.
\item
A spin tensor $S^{\iMa\iMb}$ with $S^{\iMa\iMb}+S^{\iMb\iMa}=0$ and
$\Cdot_{\iMa}\,S^{\iMa\iMb}=0$.
\end{bulletlist}
These satisfy
\begin{align}
\dot{m}=0
\,,\qquad
\Dfrac{X^{\iMa}}{\sigma} 
=
P^{\iMa}
\,,\qquad
\Dfrac{P^{\iMa}}{\sigma} 
= 
\tfrac12 R^{\iMa}{}_{\iMb\iMc\iMd}\,\Cdot^\iMb\, S^{\iMd\iMc} 
+R^{\iMa}{}_{\iMb\iMc\iMd}\,\Cdot^\iMb\, \Cdot^\iMc\,X^\iMd
\,,\qquad
\Dfrac{S^{\iMa\iMb}}{\sigma}
&=
0
\label{MD_diploe_DSab}
\DEfullstop
\end{align}

Counting the number of components we see there are 10 ODEs, which
completely \jgPchange{determine the dynamical evolution of the dipole
  components on the prescribed worldline.} 
There are no additional free components.

As we see below, the same situation does not occur for the
quadrupoles. The conditions (\ref{Intro_Tab_Sym}) and
(\ref{Intro_Tab_Div_zero}) do not completely determine the dynamics of
all the components, it is not possible to write all the components in
terms of tensors, and \jgDchange{it is not possible to associate the concept of mass with the quadrupole.}

A particular case of the dipole is when $S^{\iMa\iMb}=0$, which is
compatible with its dynamic equation (\ref{MD_diploe_DSab}). We call
this case a \defn{semi-dipole}. The notion of semi-dipoles and
semi-quadrupoles is purely geometric and is addressed in section
\ref{ch_Semi_Q}.


\section{The quadrupole stress-energy tensor.}
\label{ch_QP}

Setting $k=2$ in (\ref{Intro_Tab_Ellis_Multi}) gives the formula for a
quadrupole,
\begin{align}
T^{\iMa\iMb}
=
\frac12\int_\Interval \zeta^{\iMa\iMb\iMc\iMd}(\sigma) 
\,\partx_\iMc\partx_\iMd 
\delta\big(x-C(\sigma)\big)\,d\sigma
\label{QP_Tab}
\DEcomma
\end{align}
so that the action on the test tensor $\TtwoTen_{\iMa\iMb}$ is given by
\begin{align}
\int_{\Real^4} T^{\iMa\iMb}\,\TtwoTen_{\iMa\iMb}\,d^4x
=
\frac12\int_\Interval \zeta^{\iMa\iMb\iMc\iMd}(\sigma) \,
\big(\partx_\iMc\partx_\iMd \TtwoTen_{\iMa\iMb}\big)\big|_{C(\sigma)}\ d\sigma
\label{QP_Tab_action}
\DEfullstop
\end{align}
From (\ref{Intro_Tab_Sym}) we impose 
\begin{align}
\zeta^{\iMa\iMb\iMc\iMd} = \zeta^{\iMb\iMa\iMc\iMd}
\label{QP_zeta_sym_ab}
\DEcomma
\end{align}
and due to the commutation of partial derivatives we also set
\begin{align}
\zeta^{\iMa\iMb\iMc\iMd} = \zeta^{\iMa\iMb\iMd\iMc}
\label{QP_zeta_sym_cd}
\DEfullstop
\end{align}
\jgnchange{As in the dipole case}, the $\zeta^{\iMa\iMb\iMc\iMd}$ are not uniquely
specified by the $T^{\iMa\iMb}$, with the 
\JGchange{gauge-like} freedom
\begin{align}
\zeta^{\iMa\iMb\iMc\iMd} \to 
\zeta^{\iMa\iMb\iMc\iMd} + 
\Mone^{\iMb\iMa}\, \Cdot^{\lround\iMc}\, C^{\iMd\rround}
+
\Mtwo^{\iMa\iMb\lround\iMc}\,\Cdot^{\iMd\rround}
\label{QP_Zeta_Freedom}
\DEnone
\end{align}
where $\Mone^{\iMb\iMa}$ and $\Mtwo^{\iMa\iMb\iMc}$ are arbitrary
constants.
\begin{ProofI}{Proof of \eqref{QP_Zeta_Freedom}}
\label{pf_Quad_zeta_freedom}
Similar to the proof of \eqref{MD_diploe_Zeta_Freedom}, we have
\begin{align*}
\int_\Interval  \Mone^{\iMa\iMb}\,\Cdot^{\lround\iMc}\, C^{\iMd\rround}
&\partial_\iMc\partial_\iMd\delta\big(x-C(\sigma)\big)\,d\sigma 
=
\int_\Interval  \Mone^{\iMa\iMb}\,C^{\iMd}\Cdot^{\iMc}\, 
\partial_\iMc\partial_\iMd\delta\big(x-C(\sigma)\big)\,d\sigma 
\\&=
\Mone^{\iMa\iMb} \int_\Interval  C^{\iMd} \dfrac{}{\sigma}
\Big(\partial_\iMd\delta\big(x-C(\sigma)\big)\Big)\,d\sigma 
\\&=
\Mone^{\iMa\iMb} \int_\Interval  \dfrac{}{\sigma} \Big( C^{\iMd} 
\partial_\iMd\delta\big(x-C(\sigma)\big)\Big)\,d\sigma 
-
\Mone^{\iMa\iMb}\int_\Interval  \Cdot^{\iMd} 
\partial_\iMd\delta\big(x-C(\sigma)\big)\,d\sigma 
\\&=
-
\Mone^{\iMa\iMb}\int_\Interval  \dfrac{}{\sigma}\delta\big(x-C(\sigma)\big)\,d\sigma 
=
0
\DEfullstop
\end{align*}
and
\begin{align*}
\int_\Interval  \Mtwo^{\iMa\iMb\lround\iMd} 
\Cdot^{\iMc\rround}\, 
\partial_\iMc\partial_\iMd\delta\big(x-C(\sigma)\big)\,d\sigma 
&=
\int_\Interval  \Mtwo^{\iMa\iMb\iMd} 
\Cdot^{\iMc}\, 
\partial_\iMc\partial_\iMd\delta\big(x-C(\sigma)\big)\,d\sigma 
\\&=
\Mtwo^{\iMa\iMb\iMd} 
\int_\Interval  
\dfrac{}{\sigma}\Big(\partial_\iMd\delta\big(x-C(\sigma)\big)\Big)
\,d\sigma 
=
0
\DEfullstop
\end{align*}
To see why this incorporates all the gauge-like freedom we use the adapted
coordinates system. Therefore this proof (proof number
\ref{pf_Quad_zeta_all_gauge}) is given in the
appendix. 
\end{ProofI}

As in \cite{gratus2018correct},
under change of coordinate $(x^0,\ldots,x^3)$ to 
$(\xhat^{\hat 0},\ldots,\xhat^{\hat 3})$ we have have a complicated
transformation involving derivatives and integrals
\begin{equation}
\begin{aligned}
\zetahat^{\iMahat\iMbhat\iMchat\iMdhat} 
&=
\zeta^{\iMa\iMb\iMc\iMd}\,
{\Jaabb}\,
J^{\iMchat}_\iMc\, J^{\iMdhat}_\iMd
-\tfrac12
\Cdothat^\iMchat\int^\sigma \zeta^{\iMa\iMb\iMc\iMd}\,
\Big({\Jaabb} (\partx_\iMc\,J^{\iMdhat}_\iMd)+
2\,\partx_\iMc\,({\Jaabb})\,J^{\iMdhat}_\iMd\Big) \,d\sigma'
\\&\qquad
-\tfrac12
\Cdothat^\iMdhat\int^\sigma \zeta^{\iMa\iMb\iMc\iMd}\,
\Big({\Jaabb} (\partx_\iMc\,J^{\iMchat}_\iMd)+
2\,\partx_\iMc\,({\Jaabb})\,J^{\iMchat}_\iMd\Big) \,d\sigma'
\\&\quad
+\tfrac12
\Cdothat^{\iMdhat}\int^{\sigma} 
\Cdothat^{\iMchat}\int^{\sigma'} \zeta^{\iMa\iMb\iMc\iMd} \,
\partx_\iMc\,\partx_\iMd \,\big({\Jaabb}\big)\,d\sigma''\,
d\sigma'
+\tfrac12
\Cdothat^{\iMchat}\int^{\sigma} 
\Cdothat^{\iMdhat}\int^{\sigma'} \zeta^{\iMa\iMb\iMc\iMd} \,
\partx_\iMc\,\partx_\iMd \,\big({\Jaabb}\big)\,d\sigma''\,
d\sigma'
\DEcomma
\end{aligned}
\label{QP_Zeta_Change_Coords}
\end{equation}
where $J^\iMahat_{\iMa}$ is given by
\begin{align}
J^\iMahat_{\iMa}
=
\pfrac{\xhat^\iMahat}{x^\iMa}
\label{QP_def_Jaa}
\DEnone
\end{align}
and
\begin{align}
{\Jaabb}=J^{\iMahat}_{\iMa}\,J^{\iMbhat}_\iMb\, 
\label{QP_def_Jaabb}
\DEfullstop
\end{align}
This is proved in the appendix, proof number \ref{pf_ChangCoordZeta}.
It is not necessary to give the lower limits of the integrals as these
are incorporate in \JGchange{gauge-like} freedom
(\ref{QP_Zeta_Freedom}).
\begin{ProofI}{}
\JGchange{
Similar to the proof following (\ref{QP_def_J}), consider 
$\int^\sigma_{\sigma_0}$ and
$\int^\sigma_{\sigma_1}$.
Let $\Ahat^{\iMahat\iMbhat}(\sigma)=\zeta^{\iMa\iMb\iMc\iMd} \,
\partx_\iMc\,\partx_\iMd \,\big({\Jaabb}\big)$ then taking the last
term
\begin{align*}
\int^{\sigma}_{\sigma_0} 
\Cdothat^{\iMdhat}\int^{\sigma'}_{\sigma_0} \Ahat^{\iMahat\iMbhat}(\sigma'')\,d\sigma''\,
d\sigma'
&-
\int^{\sigma}_{\sigma_1} 
\Cdothat^{\iMdhat}\int^{\sigma'}_{\sigma_1} \Ahat^{\iMahat\iMbhat}(\sigma'')\,d\sigma''\,
d\sigma'
\\&=
\int^{\sigma}_{\sigma_0} 
\Cdothat^{\iMdhat}\int^{\sigma'}_{\sigma_0} \Ahat^{\iMahat\iMbhat}\,d\sigma''\,
d\sigma'
-
\int^{\sigma}_{\sigma_1} 
\Cdothat^{\iMdhat}\int^{\sigma'}_{\sigma_0} \Ahat^{\iMahat\iMbhat}\,d\sigma''\,
d\sigma'
\jgnchange{-}
\int^{\sigma}_{\sigma_0} 
\Cdothat^{\iMdhat}\int^{\sigma_0}_{\sigma_1} \Ahat^{\iMahat\iMbhat}\,d\sigma''\,
d\sigma'
\\&=
\jgnchange{\int^{\sigma_1}_{\sigma_0}} 
\Cdothat^{\iMdhat}\int^{\sigma'}_{\sigma_0} \Ahat^{\iMahat\iMbhat}\,d\sigma''\,
d\sigma'
\jgnchange{\,-\,}
\hat{C}^{\iMdhat}\int^{\sigma_0}_{\sigma_1} \Ahat^{\iMahat\iMbhat}\,d\sigma''
\jgnchange{\,+\,}
\hat{C}^{\iMdhat}(\sigma_0)\int^{\sigma_0}_{\sigma_1} \Ahat^{\iMahat\iMbhat}\,d\sigma''
\DEfullstop
\end{align*}
Hence the difference between two expressions
for $\zetahat^{\iMahat\iMbhat\iMchat\iMdhat}$ is 
$\zetahat^{\iMahat\iMbhat\iMchat\iMdhat} \to 
\zetahat^{\iMahat\iMbhat\iMchat\iMdhat} + 
\Mone^{\iMbhat\iMahat}\, \Cdothat{}^{\lround\iMchat}\, \hat{C}^{\iMdhat\rround}
+
\Mtwo^{\iMahat\iMbhat\lround\iMchat}\,\Cdothat{}^{\iMdhat\rround}$
where
\begin{align*}
\Mone^{\iMbhat\iMahat}
=
\jgnchange{\int^{\sigma_1}_{\sigma_0}} \Ahat^{\iMahat\iMbhat}\,d\sigma
\DEnone
\end{align*}
and
\begin{align*}
\Mtwo^{\iMahat\iMbhat\iMchat}
=
\jgnchange{\int^{\sigma_1}_{\sigma_0}} 
\Cdothat^{\iMdhat}\int^{\sigma'}_{\sigma_0} \Ahat^{\iMahat\iMbhat}\,d\sigma''\,
d\sigma'
+
\hat{C}^{\iMdhat}(\sigma_0)\int^{\sigma_0}_{\sigma_1} \Ahat^{\iMahat\iMbhat}\,d\sigma''
- 
\int^{\sigma_1}_{\sigma_0} \zeta^{\iMa\iMb\iMc\iMd}\,
\Big({\Jaabb} (\partx_\iMc\,J^{\iMchat}_\iMd)+
2\,\partx_\iMc\,({\Jaabb})\,J^{\iMchat}_\iMd\Big) \,d\sigma'
\DEfullstop
\end{align*}
}
\end{ProofI}
It also
is necessary to check that (\ref{QP_Zeta_Change_Coords}) is consistent with the
gauge-like freedom (\ref{QP_Zeta_Freedom}). This is given in proof
number \ref{pf_ChangCoordZeta_Gauge} in the appendix.

\vspace{1em}

As stated in the introduction the quadrupole is greatly simplified if
we choose adapted coordinates given in (\ref{Intro_adapt_coords}),
so that $\Cdot^{\iMa}=\delta^{\iMa}_0$. Equation (\ref{QP_Tab})
can now be written in terms of components $\gamma^{\iMa\iMb\iMc\iMd}$
\begin{align}
T^{\iMa\iMb}(\sigma,\Vz)
&=
\gamma^{\iMa\iMb00}(\sigma) \,\deltaThree(\Vz)
+
\gamma^{\iMa\iMb0\iSa}(\sigma)\, \partz_\iSa \deltaThree(\Vz)
+
\tfrac12
\gamma^{\iMa\iMb\iSa\iSb}(\sigma)\,
\partz_\iSa\partz_\iSb 
\deltaThree(\Vz)
\label{QP_Tab_gamma}
\DEnone
\end{align}
so that from (\ref{Ellis_adap_Ellis_Multi_action}) becomes
\begin{align}
\int_\Mman T^{\iMa\iMb}\,\TtwoTen_{\iMa\iMb}\,d^4 x
&=
\int_\Interval \Big(\gamma^{\iMa\iMb00}\,\TtwoTen_{\iMa\iMb}
-
\gamma^{\iMa\iMb0\iSa}\,(\partz_\iSa
\TtwoTen_{\iMa\iMb}) 
+
\tfrac12 
\gamma^{\iMa\iMb\iSa\iSb}
(\partz_\iSa \partz_\iSb\,\TtwoTen_{\iMa\iMb})
\Big)\,d\sigma
\label{QP_Tab_gamma_action}
\DEfullstop
\end{align}
Here again we impose
\begin{align}
\gamma^{\iMa\iMb\iMc\iMd} = \gamma^{\iMb\iMa\iMc\iMd}
\qquadand
\gamma^{\iMa\iMb\iMc\iMd} = \gamma^{\iMa\iMb\iMd\iMc}
\label{QP_gamma_sym}
\DEfullstop
\end{align}
In adapted coordinates, the components $\gamma^{\iMa\iMb\iMc\iMd}$ are uniquely
determined from $T^{\iMa\iMb}$, so there is no \JGchange{gauge-like} freedom, as in
(\ref{QP_Zeta_Freedom}).
In this coordinate system we can still express $T^{\iMa\iMb}$ in terms of
(\ref{QP_Tab}), and the relationship between
$\gamma^{\iMa\iMb\iMc\iMd}$ and $\zeta^{\iMa\iMb\iMc\iMd}$ is given by
\begin{align}
\gamma^{\iMa\iMb00}=\tfrac12 \ddot \zeta^{\iMa\iMb00}, \quad
\gamma^{\iMa\iMb \iSa 0}=\dot \zeta^{\iMa\iMb \iSa 0} \quadand
\gamma^{\iMa\iMb \iSa \iSb}= \zeta^{\iMa\iMb \iSa \iSb}
\label{QP_gamma_zeta}
\DEnone
\end{align}
which is consistent with (\ref{QP_Zeta_Freedom}). This follows from
(\ref{Ellis_adap_gam_zeta}).

It is now much easier to express the differential and algebraic
equations on the components arising from the divergenceless conditions
(\ref{Intro_Tab_Div_zero}), \jgPchange{as proved below}. 
\begin{align}
\dot\gamma^{\iMa000}
&=
- \Gamma^{\iMa}_{\iMb\iMc}\, \gamma^{\iMc\iMb00}
+(\partz_{\iSa}\Gamma^0_{\iMb\iMc})\, \gamma^{\iMc \iMb 0 \iSa}
-\tfrac12\big(\partz_{\iSb}\partz_{\iSa}\Gamma^0_{\iMb\iMc}\big) 
\gamma^{\iMc \iMb \iSa \iSb}
\DEcomma
\label{QP_DTeqn_a000}
\\ 
\dot \gamma^{\iMa00\iSa}
&=
-\gamma^{\iMa\iSa00}
- \Gamma^{\iMa}_{\iMb\iMc}\, \gamma^{\iMc \iMb 0 \iSa}
+ (\partz_{\iSb}\Gamma^{\iMa}_{\iMb\iMc})\, \gamma^{\iMc \iMb \iSb \iSa}
\DEcomma
\label{QP_DTeqn_a00m}
\\ 
\dot \gamma^{\iMa 0 \iSa\iSb}
&=
- 2\gamma^{\iMa (\iSb\iSa) 0}
- \Gamma^{\iMa}_{\iMb\iMc}\, \gamma^{\iMc \iMb \iSa \iSb}
\label{QP_DTeqn_a0mn}
\DEnone
\end{align}
together with the algebraic equation
\begin{align}
\gamma ^{\iMa (\iSa \iSb \iSc)}=0
\label{QP_DTeqn_alg}
\DEfullstop
\end{align}
\begin{ProofA}{Proof of
    \eqref{QP_DTeqn_a000}-\eqref{QP_DTeqn_alg}}
\label{pf_Quad_dynm_eqn}
From \eqref{Intro_Tab_Div_zero_density} we have for any test vector 
$\ToneTen^{\iMb}$
\begin{align*}
0
&=
\int_\Mman (\nabla_\iMa T^{\iMa\iMb} )\, \ToneTen_{\iMb}\,d^4 x
=
\int_\Mman 
\big(\partial_\iMa T^{\iMa\iMb} + 
\Gamma^\iMb_{\iMa\iMc} T^{\iMa\iMc} \big)\, 
\ToneTen_{\iMb}\,d^4 x
=
\int_\Mman 
T^{\iMa\iMb} \big(\Gamma^\iMc_{\iMa\iMb} \,\ToneTen_{\iMc}-\partial_\iMa \ToneTen_{\iMb} \big)\, 
d^4 x
\\&=
\int_\Mman 
\Big(\gamma^{\iMa\iMb00} \,\deltaThree(\Vz)
+
\gamma^{\iMa\iMb0\iSa}\, \partz_\iSa \deltaThree(\Vz)
+
\tfrac12
\gamma^{\iMa\iMb\iSa\iSb}\,
\partz_\iSa\partz_\iSb 
\deltaThree(\Vz)
\Big)
\big(\Gamma^\iMc_{\iMa\iMb} \,\ToneTen_{\iMc}-\partial_\iMa \ToneTen_{\iMb} \big)\, 
d^4 x
\\&=
\int_\Interval d\sigma\Big(
\gamma^{\iMa\iMb00}\,
\big(\Gamma^\iMc_{\iMa\iMb} \,\ToneTen_{\iMc}-\partial_\iMa \ToneTen_{\iMb} \big)
-
\gamma^{\iMa\iMb0\iSa}\,\partz_\iSa
\big(\Gamma^\iMc_{\iMa\iMb} \,\ToneTen_{\iMc}-\partial_\iMa \ToneTen_{\iMb} \big)
+
\tfrac12 
\gamma^{\iMa\iMb\iSa\iSb}
\partz_\iSa \partz_\iSb
\big(\Gamma^\iMc_{\iMa\iMb} \,\ToneTen_{\iMc}-\partial_\iMa \ToneTen_{\iMb} \big)
\Big)
\\&=
\int_\Interval d\sigma\Big(
\gamma^{\iMa\iMb00}\,\Gamma^\iMc_{\iMa\iMb} \,\ToneTen_{\iMc}
-
\gamma^{\iSa\iMb00}\,\partial_\iSa \ToneTen_{\iMb} 
+
\dot\gamma^{0\iMb00}\,\ToneTen_{\iMb} 
\\&\qquad\qquad
-
\gamma^{\iMa\iMb0\iSa}\,\partz_\iSa
\big(\Gamma^\iMc_{\iMa\iMb} \,\ToneTen_{\iMc}\big)
+
\gamma^{\iSb\iMb0\iSa}\,\partz_\iSa
\partial_\iSb \ToneTen_{\iMb} 
-
\dot\gamma^{0\iMb0\iSa}\,\partz_\iSa
\ToneTen_{\iMb} 
\\&\qquad\qquad
+
\tfrac12 
\gamma^{\iMa\iMb\iSa\iSb}
\partz_\iSa \partz_\iSb
\big(\Gamma^\iMc_{\iMa\iMb} \,\ToneTen_{\iMc}\big)
-
\tfrac12 
\gamma^{\iSc\iMb\iSa\iSb}
\partz_\iSa \partz_\iSb
\partial_\iSc \ToneTen_{\iMb} 
+
\tfrac12 
\dot\gamma^{0\iMb\iSa\iSb}
\partz_\iSa \partz_\iSb
\ToneTen_{\iMb} 
\Big)
\\&=
\int_\Interval d\sigma\Big(
\gamma^{\iMa\iMb00}\,\Gamma^\iMc_{\iMa\iMb} \,\ToneTen_{\iMc}
-
\gamma^{\iSa\iMb00}\,\partial_\iSa \ToneTen_{\iMb} 
+
\dot\gamma^{0\iMc00}\,\ToneTen_{\iMc} 
\\&\qquad\qquad
-
\gamma^{\iMa\iMb0\iSa}\,(\partial_\iSa\Gamma^\iMc_{\iMa\iMb}) \,\ToneTen_{\iMc}
-
\gamma^{\iMa\iMb0\iSa}\,
\Gamma^\iMc_{\iMa\iMb} \,\partial_\iSa\ToneTen_{\iMc}
+
\gamma^{\iSb\iMb0\iSa}\,\partial_\iSa
\partial_\iSb \ToneTen_{\iMb} 
-
\dot\gamma^{0\iMb0\iSa}\,\partial_\iSa
\ToneTen_{\iMb} 
\\&\qquad\qquad
+
\tfrac12 
\gamma^{\iMa\iMb\iSa\iSb}
\big(\partial_\iSa \partial_\iSb
\Gamma^\iMc_{\iMa\iMb}\big)\ToneTen_{\iMc}
+
\gamma^{\iMa\iMb\iSa\iSb}
\big(\partial_\iSa 
\Gamma^\iMc_{\iMa\iMb} \big)
\,\big(\partial_\iSb\ToneTen_{\iMc}\big)
+
\tfrac12 
\gamma^{\iMa\iMb\iSa\iSb}
\Gamma^\iMc_{\iMa\iMb} \partial_\iSa \partial_\iSb
\ToneTen_{\iMc}
\\&\qquad\qquad\qquad
-
\tfrac12 
\gamma^{\iSc\iMb\iSa\iSb}
\partial_\iSa \partial_\iSb
\partial_\iSc \ToneTen_{\iMb} 
+
\tfrac12 
\dot\gamma^{0\iMb\iSa\iSb}
\partial_\iSa \partial_\iSb
\ToneTen_{\iMb} 
\Big)
\\&=
\int_\Interval d\sigma\bigg(
\ToneTen_{\iMc}
\Big(\gamma^{\iMa\iMb00}\,\Gamma^\iMc_{\iMa\iMb} 
+
\dot\gamma^{0\iMc00}
-
\gamma^{\iMa\iMb0\iSa}\,(\partial_\iSa\Gamma^\iMc_{\iMa\iMb})
+
\tfrac12 
\gamma^{\iMa\iMb\iSa\iSb}
\big(\partial_\iSa \partial_\iSb
\Gamma^\iMc_{\iMa\iMb}\big)
\Big)
\\&\qquad\qquad
-\partial_\iSa \ToneTen_{\iMc} \Big(
\gamma^{\iSa\iMc00}\,
+
\gamma^{\iMa\iMb0\iSa}\,\Gamma^\iMc_{\iMa\iMb} 
+
\dot\gamma^{0\iMc0\iSa}
-
\gamma^{\iMa\iMb\iSb\iSa}
\big(\partial_\iSb 
\Gamma^\iMc_{\iMa\iMb} \big)
\Big)
\\&\qquad\qquad
+\partial_\iSa\partial_\iSb \ToneTen_{\iMc} 
\Big(\gamma^{\iSb\iMc0\iSa}
+
\tfrac12 
\gamma^{\iMa\iMb\iSa\iSb}
\Gamma^\iMc_{\iMa\iMb}
+
\tfrac12 
\dot\gamma^{0\iMc\iSa\iSb}
\Big)
-
\tfrac12 
\gamma^{\iSc\iMb\iSa\iSb}
\partial_\iSa \partial_\iSb
\partial_\iSc \ToneTen_{\iMb} 
\bigg)
\DEfullstop
\end{align*}
The terms with $\ToneTen_{\iMc}$, $\partial_\iSa\ToneTen_{\iMc}$,
$\partial_\iSa\partial_\iSb\ToneTen_{\iMc}$ and
$\partial_\iSa\partial_\iSb\partial_\iMc\ToneTen_{\iMc}$ are
\jgPchange{independent}. From this we get
\eqref{QP_DTeqn_a000}-\eqref{QP_DTeqn_alg}. Note we must take
the symmetric part with respect to $\iSb,\iSa$.
\end{ProofA}

We can now count the number of components of the quadrupole. From
(\ref{QP_DTeqn_a000})-(\ref{QP_DTeqn_a0mn}) we have 40 first order
ODEs. However not all the components are determined by these
ODEs. From (\ref{QP_gamma_sym}) we start with 100 components. The
algebraic equation (\ref{QP_DTeqn_alg}) gives 40 independent equations
so that there are 60 independent components. Thus 40 are determined by
ODEs and the remaining 20 are free components. As stated in the
introduction these free components need to be replaced by 
constitutive equations. However the choice of constitutive equations
depends on a choice of \jgPchange{an underlying model for the
  stress-energy tensor}. An example of such
constitutive equations is given in section \ref{ch_Dust} below.


Under change of adapted coordinate $(\sigma,z^1,z^2,z^3)$ to
$(\hat\sigma,\zhat^1,\zhat^2,\zhat^3)$ we have
\begin{align}
\gammahat^{\iMahat\iMbhat\iSahat\iSbhat}
&=
{\Jaabb}\, J^\iSahat_\iSa\, J^\iSbhat_\iSb\, \gamma^{\iMa\iMb\iSa \iSb}
\DEcomma
\label{QP_gamma_chage_coords_munu}
\\
\gammahat^{\iMahat\iMbhat\iSahat\zerohat}
&=
{\Jaabb}\, J^\iSahat_\iSa\, \gamma^{\iMa\iMb\iSa0}
+
\big(
{\Jaabb}\,J^\zerohat_\iSb \, J^\iSahat_\iSa\, 
\gamma^{\iMa\iMb\iSa\iSb}\big)\dot{}
-
\tfrac12
\big(
\partial_\iSa J^\iSahat_{\iSb}\,{\Jaabb}
+
J^\iSahat_\iSa\, \partx_\iSb {\Jaabb} 
+
J^\iSahat_\iSb\, \partx_\iSa {\Jaabb} 
\big)
\gamma^{\iMa\iMb\iSa\iSb}
\DEcomma
\label{ChangeCoords_gamma_abmu0}
\\
\gammahat^{\iMahat\iMbhat\Ohat\Ohat}
&=
\Jaabb\,\gamma^{\iMa \iMb 00}
+
\Jaabb\,J^\Ohat_{\iSc} \,\dot\gamma^{\iMa\iMb\iSa0}
+
\big(
(\Jaabb\ J^\Ohat_{\iSc})\dot{} 
-
\partx_{\iSc} {\Jaabb}
\big)
\gamma^{\iMa \iMb \iSc 0}
\nonumber
\\&\quad
+
\tfrac12\big((\Jaabb\ J^\Ohat_{\iSd} \, J^\Ohat_{\iSc})\, 
\gamma^{\iMa  \iMb \iSc \iSd}\big)\ddot{}
-
\big(\big(
\tfrac12
\partial_\iSc J^\Ohat_{\iSd}\,{\Jaabb}+
J^\Ohat_{\iSd}\, \partx_{\iSc} {\Jaabb}
\big)
\gamma^{\iMa \iMb \iSc \iSd}\big)\dot{}
+  (\tfrac12\partx_{\iSc}\partx_{\iSd}{\Jaabb})\, \gamma^{\iMa \iMb \iSc \iSd}
\label{QP_gamma_chage_coords_00}
\DEnone
\end{align}
\SeeProof{pf_gamma_change_coords} Although these may be considered more
complicated than (\ref{QP_Zeta_Change_Coords}) they does not involve any
integrals.
We have assumed that $\sigma$ and
$\hat\sigma$ parameterise the same points on the worldline $C$.
Thus on the worldline
$J^\iMahat_0=\delta^\iMahat_0$. However this does not imply 
$\partx_\iMb J^{\iMahat}_{0}=0$.

\subsection{The static semi-quadrupole and the free components}
\label{ch_Q_free}

To get an intuition about the free components, consider the dynamic
equations (\ref{QP_DTeqn_a000})-(\ref{QP_DTeqn_alg}) on a flat
Minkowski background with Cartesian coordinates
$(t=z^0,z^1,z^2,z^3)=(t,\Vz)$ and with the worldline at
$\Vz=\Vzero$. Thus we can set $t=\sigma$ so that $C^0(t)=t$ and
$C^\iSa(t)=0$. The dynamic equations
(\ref{QP_DTeqn_a000})-(\ref{QP_DTeqn_alg}) become
\begin{align}
\dot\gamma^{\iMa000} &= 0
\DEcomma
\label{Q_free_gdot_a000}
\\
\dot\gamma^{\iMa0\iSa0} &= -\gamma^{\iMa\iSa00}
\DEcomma
\label{Q_free_gdot_a0mu0}
\\
\dot\gamma^{\iMa0\iSb\iSa} &= -2\gamma^{\iMa(\iSa\iSb)0}
\DEcomma
\label{Q_free_gdot_a0munu}
\\
\gamma^{\iMa(\iSa\iSb\jgnchange{\iSc})} &=0
\label{Q_free_gdot_amunurho}
\DEfullstop
\end{align}

As a further simplification, consider only the semi-quadrupole. This is when
\begin{align}
\gamma^{\iMa\iSa\iSb\jgnchange{\iSc}}=0
\label{Q_free_semi}
\DEfullstop
\end{align}
According to table \ref{tab_number_components}, there should be 22 ODE
components and 6 free components.  This arises since
(\ref{Q_free_semi}) implies $\gamma^{\iSa0\iSb\jgnchange{\iSc}}=0$ which
\jgnchange{eliminates} all
but 6 of the ODEs in
(\ref{Q_free_gdot_a0munu}). 
See section
\ref{ch_Semi_Q} below for full details.

The general solution is given by
\begin{equation}
\begin{gathered}
\gamma^{0000} = m,\quad
\gamma^{\iSa000} = P^\iSa,\quad
\gamma^{00\iSa0} = X^\iSa-t\, P^\iSa ,\quad
\\
\gamma^{00\iSb\iSa} = \kappa^{\iSb\iSa}(t),\quad
\gamma^{\iSb0\iSa0} = S^{\iSb\iSa}-\tfrac12\dot{\kappa}^{\iSb\iSa}(t) ,\quad
\gamma^{\iSb\iSa00} = \tfrac12\ddot{\kappa}^{\iSb\iSa}(t),\quad
\gamma^{\jgnchange{\iSc}\iSb\iSa0}=0
\end{gathered}
\label{Q_free_semi_soln}
\end{equation}
where the 10 \jgDchange{quantities} $m,P^\iSa,X^\iSa,S^{\iSa\iSb}$
\jgDchange{are constants, $S^{\iSa\iSb}$ satisfies}
$S^{\iSa\iSb}+S^{\iSb\iSa}=0$ and the six free components
$\kappa^{\iSb\iSa}(t)$ \jgDchange{satisfy}
$\kappa^{\iSb\iSa}(t)=\kappa^{\iSa\iSb}(t)$. Here we interpret
$m$ as the total mass, $P^\iSa$ as the momentum and $S^{\iSb\iSa}$ as the
spin. The six free components $\kappa^{\iSa\iSb}(t)$ are the
moments of inertia. Since there are 22 ODEs there should be 22
constants of integration. As well as the 10 already given, the
remaining 12 are the six initial conditions for $\kappa^{\iSa\iSb}(0)$
and for $\dot\kappa^{\iSa\iSb}(0)$.


\begin{ProofA}{Proof of~\eqref{Q_free_semi_soln}}
\label{pf_SemiQuad_Count}
For the semi-quadrupole \jgnchange{\eqref{Q_free_semi}, then} 
\eqref{Q_free_gdot_amunurho} is automatically satisfied.
Equations \eqref{Q_free_gdot_a000}-\eqref{Q_free_gdot_a0munu} become
\begin{gather*}
\dot\gamma^{0000} = 0,\quad
\dot\gamma^{\iSa000} = 0,\quad
\dot\gamma^{00\iSa0} = -\gamma^{0\iSa00},\quad
\jgnchange{\dot\gamma^{0\iSb\iSa0}} = -\gamma^{\iSb\iSa00},\quad
\dot\gamma^{00\iSb\iSa} = -\gamma^{0(\iSa\iSb)0},\quad
0=\dot\gamma^{0\iSc\iSb\iSa} = -\gamma^{\iSc(\iSa\iSb)0}
\DEfullstop
\end{gather*}
It may appear that we have not stated anything about
$(\gamma^{\iSc\iSa\iSb0}-\gamma^{\iSc\iSb\iSa0})$. 
However due to the symmetry of $\gamma^{\iSc\iSa\iSb0}$ we have
\begin{align*}
\gamma^{\iSc\iSa\iSb0}-\gamma^{\iSc\iSb\iSa0}
=
\gamma^{\iSa\iSc\iSb0}-\gamma^{\iSb\iSc\iSa0}
=
-\gamma^{\iSa\iSb\iSc0}+\gamma^{\iSb\iSa\iSc0}
=
0
\DEfullstop
\end{align*}
Thus from the last equation above we have $\gamma^{\iSc\iSb\iSa0}=0$.
Setting $\gamma^{00\iSb\iSa}=\kappa^{\iSa\iSb}(\sigma)$ we have 
$\gamma^{0(\iSb\iSa)0}=\dot{\kappa}^{\iSa\iSb}$ and
$\gamma^{\iSb\iSa00}=\ddot{\kappa}^{\iSa\iSb}$.
The remaining constants in \eqref{Q_free_semi_soln} are then
\jgDchange{determined}.
\end{ProofA}


Consider the components of $T^{\iMa\iMb}$ as arising from squeezing a regular
stress-energy tensor density $\TReg^{\iMa\iMb}(t,\Vz)$ as in section
\ref{ch_Ellis_Squz}. 
Thus
\begin{align}
\gamma^{\iMa\iMb00} = \int_{\Real^3} \TReg^{\iMa\iMb}(t,\Vz) d^3\Vz,\quad
\gamma^{\iMa\iMb\iSa0} = \int_{\Real^3}
\TReg^{\iMa\iMb}(t,\Vz)\,z^\iSa\ d^3\Vz,
\quad
\gamma^{\iMa\iMb\iSa\iSb} = \int_{\Real^3} \TReg^{\iMa\iMb}(t,\Vz)\,z^\iSa\,z^\iSb\ d^3\Vz
\label{Q_free_Squz_componets}
\DEfullstop
\end{align}
Comparing (\ref{Q_free_semi_soln}) and (\ref{Q_free_Squz_componets})
we see
\begin{equation}
\begin{gathered}
m = \int_{\Real^3} \TReg^{00}(t,\Vz) d^3\Vz,\qquad
P^\iSa = \int_{\Real^3} \TReg^{\iSa0}(t,\Vz) d^3\Vz,\qquad
X^\iSa = t\,P^\iSa + \int_{\Real^3} \TReg^{00}(t,\Vz) z^\iSa\, d^3\Vz,
\\
S^{\iSb\iSa} = \int_{\Real^3} 
z^{\lsquare\iSa}\,\TReg^{\iSb\rsquare 0}(t,\Vz)  d^3\Vz
\qquadand
\kappa^{\iSa\iSb} = 
\int_{\Real^3} 
z^{\iSa}\,z^{\iSb}\,\TReg^{00}(t,\Vz)  d^3\Vz
\label{Q_free_Squz_mPXkS}
\DEfullstop
\end{gathered}
\end{equation}

For example let $P^\iSa=0$ and
$S^{\iSa\iSb}=0$ then
\begin{align}
m = \int_{\Real^3} \TReg^{00}(t,\Vz) \, d^3\Vz,\qquad
\kappa^{\iSa\iSb}(t) = \int_{\Real^3} z^\iSa\,z^\iSb\,\TReg^{00}(t,\Vz) \,d^3\Vz
\label{Q_free_no_p}
\DEfullstop
\end{align}
Since $\kappa^{\iSa\iSb}(t)$ are free components we can choose any
$\TReg^{\iMa\iMb}(t,\Vz)$ we like so long as its total integral is $m$ and
they are 
sufficiently symmetric that $P^\iSa=0$ and
$S^{\iSa\iSb}=0$  hold.  \jgDchange{This can be achieved if, for example}
$\TReg^{\iMa\iMb}(t,\Vz)$ is symmetric about the three
directions $z^\iSa$. This explains why we can choose to have a
distribution of matter which separates and then coalesces as in figure
\ref{fig_intro_grav_quad_free}.

\subsection{Conserved quantities}
\label{ch_Q_Cons}
Recall that a Killing vector (\ref{Q_Cons_symm_Kill}) leads to a
conserved quantity in the dipole case. The same is true for
quadrupole. 
In an adapted coordinate system $(\sigma,z^1,z^2,z^3)$ the
conserved quantity $\Conserv_\Kill$ is given by
\begin{align}
\Conserv_\Kill
&=
\gamma^{\iMa000} \Kill_{\iMa} -
\gamma^{\iMa0\iSa0} \partz_\iSa \Kill_{\iMa} + 
\tfrac12
\gamma^{\iMa0\iSa\iSb} \partz_\iSa\partz_\iSb \Kill_{\iMa}
\label{Q_Cons_Q}
\DEfullstop
\end{align}
\begin{ProofI}{Proof of \eqref{MD_Cons_Q} and \eqref{Q_Cons_Q}}
\label{pf_QK}
Let $\TzeroTen$ be a test function. Thus
\begin{align*}
\int_\Mman  \nabla_\iMa (T^{\iMa\iMb}\,\Kill_\iMb)\, \TzeroTen\, d^4x
=
\int_\Mman  (\nabla_\iMa T^{\iMa\iMb}\,\Kill_\iMb +
T^{\iMa\iMb}\,\nabla_\iMa \,\Kill_\iMb\, \TzeroTen )\,  d^4x
=
0
\DEfullstop
\end{align*}
from \eqref{Intro_Tab_Sym}, (\ref{Intro_Tab_Div_zero}) and
\eqref{Q_Cons_symm_Kill}.
Since $T^{\iMa\iMb}$ is a tensor
density then so is $T^{\iMa\iMb}\Kill_\iMb$. 
Hence
\begin{align*}
0 
&=
\int_\Mman  \nabla_\iMa (T^{\iMa\iMb}\,\Kill_\iMb)\, \TzeroTen\, d^4x
=
\int_\Mman  T^{\iMa\iMb}\,\Kill_\iMb\, \nabla_\iMa \TzeroTen\, d^4x
=
\int_\Mman  T^{\iMa\iMb}\,\Kill_\iMb\, \partial_\iMa \TzeroTen\, d^4x
\\&
=
\int_\Interval \Big(\gamma^{\iMa\iMb00}\,\Kill_\iMb\partial_\iMa\TzeroTen
-
\gamma^{\iMa\iMb0\iSa}\,\partz_\iSa
(\Kill_\iMb\partial_\iMa\TzeroTen)
+
\tfrac12 
\gamma^{\iMa\iMb\iSa\iSb}
\partz_\iSa \partz_\iSb\,(\Kill_\iMb\partial_\iMa\TzeroTen)
\Big)\,d\sigma
\\&=
\int_\Interval \Big(\partial_\iMa\TzeroTen
\big(\gamma^{\iMa\iMb00}\,\Kill_\iMb
-
\gamma^{\iMa\iMb0\iSa}\,\partz_\iSa \Kill_\iMb
+
\tfrac12 
\gamma^{\iMa\iMb\iSa\iSb} \partz_\iSa \partz_\iSb\,\Kill_\iMb\big)
\Big)\,d\sigma
+\text{higher derivatives of $\TzeroTen$.}
\\&=
\int_\Interval \Big(\partial_0\TzeroTen
\big(\gamma^{0\iMb00}\,\Kill_\iMb
-
\gamma^{0\iMb0\iSa}\,\partz_\iSa \Kill_\iMb
+
\tfrac12 
\gamma^{0\iMb\iSa\iSb} \partz_\iSa \partz_\iSb\,\Kill_\iMb\big)
\Big)\,d\sigma
\\&\qquad\qquad
+
\int_\Interval \Big(\partial_\iSc\TzeroTen\ 
\big(\gamma^{\iSc\iMb00}\,\Kill_\iMb
-
\gamma^{\iSc\iMb0\iSa}\,\partz_\iSa \Kill_\iMb
+
\tfrac12 
\gamma^{\iSc\iMb\iSa\iSb} \partz_\iSa \partz_\iSb\,\Kill_\iMb\big)
\Big)\,d\sigma
+\text{higher derivatives of $\TzeroTen$.}
\\&=
-
\int_\Interval \Big(\TzeroTen\ 
\partial_0\big(\gamma^{0\iMb00}\,\Kill_\iMb
-
\gamma^{0\iMb0\iSa}\,\partz_\iSa \Kill_\iMb
+
\tfrac12 
\gamma^{0\iMb\iSa\iSb} \partz_\iSa \partz_\iSb\,\Kill_\iMb\big)
\Big)\,d\sigma
+\text{higher derivatives of $\TzeroTen$.}
\\&=
-
\int_\Interval \big(\TzeroTen\ 
\dot\Conserv_K
\big)\,d\sigma
+\text{higher derivatives of $\TzeroTen$.}
\end{align*}
Thus since we can extract the different derivatives of $\TzeroTen$ we have
$\dot\Conserv_K=0$.
\end{ProofI}

It is worth exploring the conserved quantities on the static
semi-quadrupole given by (\ref{Q_free_semi_soln}). In Minkowski
spacetime there are 10 Killing vectors. 
\begin{bulletlist}
\item
Mass or Energy: for $\Kill_0=1$, $\Kill_\iSa=0$ we have $\Conserv_\Kill=m$.
\item
\jgDchange{Momentum: for $\Kill_0=0$, $\Kill_1=1$, $\Kill_2=0$ and $\Kill_3=0$
then
$\Conserv_\Kill=p_1$. Likewise for the other two cases.}
\item
Angular momentum and spin: let
$\Kill_0=0$, $\Kill_1=z^2$, $\Kill_2=-z^1$ and
$\Kill_3=0$.
We have
\begin{align*}
\Conserv_\Kill
&=
\gamma^{1000} \Kill_1 + 
\gamma^{2000} \Kill_2 + 
\gamma^{2010} \partz_1 \Kill_2 +
\gamma^{1020} \partz_2 \Kill_1
\\&=
p^1\,z^2 - p^2\,z^1
+
\big(S^{12}-\dot{\kappa}^{12}(t)\big)
-
\big(S^{21}-\dot{\kappa}^{21}(t)\big)
=
S^{12}
\DEfullstop
\end{align*}
Likewise for the other two cases.
\item
Boost: Let 
$\Kill_0=z^1$, $\Kill_1=t+t_0$, $\Kill_2=0$ and
$\Kill_3=0$ for some fixed $t_0$. Then
\begin{align*}
\Conserv_\Kill
&=
\gamma^{0000} \Kill_0 + 
\gamma^{1000} \Kill_1 + 
\gamma^{0010} \partz_1 \Kill_0 
=
m\,z^1
+
P^1\,(t+t_0)
+
(X^1-t\,P^1)
=
X^1+t_0\,P^1
\DEfullstop
\end{align*}
Likewise for the other two cases.
\end{bulletlist}
Thus the 10 Killing symmetries of Minkowski spacetime correspond
directly to the 10 \jgDchange{constants} of the solution to static semi-quadrupole.
This also gives a new interpretation to the three somewhat obscure
conserved quantities corresponding to the three boosts.

\section{Non-divergent 
dust model of a quadrupole and the corresponding constitutive relations.}
\label{ch_Dust}

\JGchange{The familiar dust model is given in terms of a scalar 
$\varrho$ and a vector field $U^{\iMa}$ with $g_{\iMa\iMb}\,U^{\iMa}\,U^\iMb=-1$. The
stress-energy tensor density is given by
\begin{align}
\TReg^{\iMa\iMb} = \varrho\, U^{\iMa}\,U^\iMb\,\Rootg
\label{Dust_SE_tensor}
\DEcomma
\end{align}
where $\Rootg=\sqrt{-\det(g_{\iMa\iMb})}$. Then 
the divergenceless condition implies that the $U^{\iMa}$ are
geodesics
\begin{align}
U^{\iMa}\,\nabla_{\iMa}\,U^\iMb = 0
\label{Dust_geodesic}
\DEnone
\end{align}
and the flow $\varrho$ is conserved
\begin{align}
\nabla_{\iMa} (\varrho \,U^{\iMa} ) &= 0
\label{Dust_conserved}
\DEfullstop
\end{align}
}Furthermore let us assume that the dust is non divergent, so that it
preserves the measure, i.e.
\begin{align}
U^{\iMa}\,\partx_{\iMa} \Rootg
=0
\label{Dust_non_div}
\DEfullstop
\end{align}
so that $\partial_{\iMa} (\varrho \,U^{\iMa} ) = 0$.

In order to create a squeezed tensor $\TReg_\varepsilon^{\iMa\iMb}$ from
$\TReg^{\iMa\iMb}$ we need \jgDchange{to} choose a coordinate system. It is natural to
choose the coordinate adapted to $U^{\iMa}$ so that $U^{\iMa}=\delta_0^{\iMa}$. This
gives
$\dot\varrho=0$
so that we can write $\varrho=\varrho(\Vz)$. Likewise we have
$\iSa=\iSa(\Vz)$. Hence
\begin{align}
\TReg^{\iMa\iMb}(\sigma,\Vz) 
&= 
\varrho(\Vz)\, \delta^{\iMa}_0\,\delta^\iMb_0\, \iSa(\Vz)
\label{Dust_Tab_adap}
\DEfullstop
\end{align}
We require that $\varrho(\Vz)=0$ for large $\Vz$. From
(\ref{Ellis_Squz_moments}) we see
\begin{equation}
\begin{aligned}
\gamma^{\iMa\iMb00} (\sigma)
&=
\delta_0^{\iMa}\,\delta_0^\iMb
\int_{\Real^3} d^3\Vz\ 
\varrho(\Vz) \,\iSa(\Vz)
,
\\
\gamma^{\iMa\iMb\iSa0} (\sigma)
&=
-\delta_0^{\iMa}\,\delta_0^\iMb \int_{\Real^3} d^3\Vz\ 
z^\iSa\,\varrho(\Vz)\,\iSa(\Vz)
,\quad
\\
\gamma^{\iMa\iMb\iSa\iSb} (\sigma)
&=
\delta_0^{\iMa}\,\delta_0^\iMb \int_{\Real^3} d^3\Vz\ 
z^\iSa\,z^\iSb\,\varrho(\Vz)\,\iSa(\Vz)
\DEfullstop
\end{aligned}
\label{Dust_Squz_moments}
\end{equation}
Since both $\varrho$ and $\iSa$ are independent of $\sigma$ we have the
dynamic equations
\begin{align}
\dot\gamma^{\iMa\iMb00} = 0,\quad
\dot\gamma^{\iMa\iMb\iSa0} = 0\quadand
\dot\gamma^{\iMa\iMb\iSa\iSb} = 0
\label{Dust_gamma_dot}
\DEfullstop
\end{align}
These are consistent with the dynamic equations
(\ref{QP_DTeqn_a000})-(\ref{QP_DTeqn_a0mn}) since in the adapted
coordinate system the \jgDchange{geodesic} equation becomes $\Gamma^{\iMa}_{00}=0$.

Equation (\ref{Dust_gamma_dot}) completely defines the
dynamics. However, our goal is \jgDchange{to} use (\ref{Dust_gamma_dot}) to
inspire the constitutive relations in the case when we are not
modelling a non-divergent dust, and 
(\ref{QP_DTeqn_a000})-(\ref{QP_DTeqn_a0mn}) hold.
One option is to require that some of the free components are in fact
constants. This is challenging because we need to be consistent with 
(\ref{QP_DTeqn_a000})-(\ref{QP_DTeqn_a0mn}).

As a simple example, consider the static semi-quadrupole \JGchange{in
  Minkowski spacetime}, given by
(\ref{Q_free_semi_soln}). The non-divergent dust constitutive
relations would make $\kappa^{\iSa\iSb}(t)$ a constant. It would also
make $P^\iSa=0$. This
replaces (\ref{Q_free_semi_soln}) with
\begin{equation}
\begin{gathered}
\gamma^{0000} = m,\quad
\gamma^{\iSa000} = 0,\quad
\gamma^{00\iSa0} = X^\iSa ,\quad
\\
\gamma^{00\iSb\iSa} = \kappa^{\iSb\iSa},\quad
\gamma^{\iSb0\iSa0} = S^{\iSb\iSa},\quad
\gamma^{\iSb\iSa00} = 0,\quad
\gamma^{\iSc\iSb\iSa0}=0
\DEfullstop
\end{gathered}
\label{Dust_Q_free_semi_soln}
\end{equation}


\section{The coordinate free and metric free approach to
  quadrupoles.}
\label{ch_CoFree}

\JGchange{As stated in the introduction we can construct the
  \jgDchange{distributional} stress-energy tensor density with only a
  connection. \jgPchange{In this section we do not assume the
    connection is Levi-Civita connection. Indeed there is no mention
    of a metric at all.} \jgnchange{This, as stated, is
    particularly useful in the case of non-metric connections, or when
    there is no metric or multiple metrics.} To simplify the
  mathematics we assume the metric is torsion free. However this too
  can be relaxed with the result of additional torsion terms in many
  expressions.} 

In \cite{gratus2018correct} the authors present a coordinate free
definition of submanifold distributions, also known as deRham
currents, in terms of the deRham push
forward \cite{deRham1984differentiable} and standard operations.

Since we are using coordinate free notation we write a vector field as
$V\in\Gamma T\Mman$. Here $T\Mman$ is the tangent bundle of spacetime
and $\Gamma T\Mman$ refers to sections of the tangent bundle. A
vector \jgDchange{at a} point $p\in\Mman$ \jgDchange{is written} $V\in T_p\Mman$. 
A vector field and vectors at a point are differential operators and
we write the action of a vector on a scalar field \JGchange{using
  angle brackets} as $V\VAct{f}$.
The bundle of
$p$--forms is written $\Lambda^p\Mman$ so a $p$--form field is written
$\alpha\in\Gamma\Lambda^p\Mman$.

\JGchange{Given a coordinate system $(x^0,\ldots,x^3)$ then we write
$V=V^{\iMa}\partx_{\iMa}$. Here $\partx_{\iMa}$ are basis vectors and
$V^{\iMa}$ are indexed scalar fields. Thus
\begin{align}
V\VAct{f} = V^{\iMa} \,\partial_{\iMa} f
\qquadtext{where}
V^{\iMa} = V\VAct{x^\iMa}
\label{CoFree_def_V}
\DEfullstop
\end{align}
}

For 1--forms
$\alpha\in\Gamma\Lambda^1\Mman$ we can write $\alpha=\alpha_{\iMa}\,dx^{\iMa}$
where again $\alpha_{\iMa}$ are indexed scalar fields.

\subsection{The two types of $\nabla$}
\label{ch_Nablas}

In the literature on general relativity and differential geometry,
there are two \jgDchange{conventions} used when referring to the covariant
derivative. One is typically used when using index tensor notation,
the other when one is using coordinate free notation. Usually one has
simply to choose one convention \jgDchange{and present} all the results using
that. We have done this up to now using index notation. However in
this section we wish to present a coordinate free definition of all
the objects. As a result it is necessary to use both definitions of the
covariant derivatives, sometimes in the same expression.  So to avoid
confusion, from now on we introduce two different symbols.

The covariant derivative which we have used up to this point and which
``knows'' about the index of an object we write
$\nablaInd_{\iMa}$. \jgDchange{The action}
on the indexed scalar fields $V^{\iMa}$ \jgDchange{is} then
\begin{align}
\nablaInd_{\iMa} V^\iMb = \partial_{\iMa}(V^\iMb) + V^\iMc\,\Gamma^\iMb_{\iMa\iMc}
\label{Nablas_def_NablaInd}
\DEfullstop
\end{align}
\jgDchange{In other words,} the Christoffel symbols are tied to the \jgDchange{indices}.
By contrast the coordinate free covariant derivative is written
$\nablaDG_V$ where $V\in\Gamma T\Mman$. In this case the Christoffel
symbol satisfies
\begin{align}
\Gamma^{\iMa}_{\iMb\iMc}\, \partial_{\iMa} = \nablaDG_{\partial_\iMb} \partial_\iMc 
\label{Nablas_def_NablaDG_Chris}
\DEfullstop
\end{align}
This covariant derivative knows about the tensor structure, but not
the \jgDchange{indices}. Thus 
\begin{align}
\nablaDG_U V^{\iMa} = U\VAct{V^{\iMa}} = \JGchange{U^{\iMb}\,\partial_\iMb V^\iMa}
\label{Nablas_NablaDG_scalar}
\DEfullstop
\end{align}
The two covariant derivatives are related via the following
\begin{equation}
\begin{aligned}
\nablaDG_U (V)
&=
U^\iMb (\nablaInd_\iMb V^{\iMa})\,\partial_{\iMa}
\DEcomma
\end{aligned}
\label{Nablas_relate_defs}
\end{equation}
since
\begin{align*}
\nablaDG_U (V)
&=
\nablaDG_U (V^{\iMa}\,\partial_{\iMa})
=
U\VAct{V^{\iMa}}\partial_{\iMa} + U^\iMb\,V^{\iMa}
\nablaDG_{\partial_\iMb}\partial^{\iMa}
=
U\VAct{V^{\iMa}}\partial_{\iMa} + U^\iMb\,V^{\iMa}
\Gamma^\iMc_{\iMa\iMb}\partial_\iMc
\\&=
U^\iMb\big(\partial_\iMb\VAct{V^{\iMa}}
+ V^\iMc\Gamma^{\iMa}_{\iMb\iMc}\big)\partial_{\iMa}
=
U^\iMb (\nablaInd_\iMb V^{\iMa})\partial_{\iMa}
\DEfullstop
\end{align*}

\jgDchange{Setting k=2 in (\ref{Intro_Tab_Dixon_Multi}) gives the
  Dixon quadrupole, and we see that it contains the operator
  $\nablaInd_{\iMa}\nablaInd_\iMb$.}
\jgDchange{Thus (\ref{Intro_Tab_Dixon_Multi})} is tensorial with respect to the
indices $\iMa$ and $\iMb$. To give \jgDchange{a} coordinate free definition we define for
any tensor $S$,
\begin{align}
\nablaDG^2_{U,V} S
&=
\nablaDG_U\,\nablaDG_V S
-
\nablaDG_{\nablaDG_U V} S
\label{Nablas_nable^2}
\DEfullstop
\end{align}
This definition can be extended to arbitrary order.
This is clearly tensorial in $U$, but is also tensorial (also known as
f-linear) with respect
to $V$. Thus
\begin{align}
\nablaDG^2_{(fU),V} S
=
\nablaDG^2_{U,(fV)} S
=
f\,\nablaDG^2_{U,V} S
\label{Nablas_nable^2_lin}
\DEfullstop
\end{align}
\begin{ProofI}{Proof of \eqref{Nablas_nable^2_lin}}
\label{pf_nabla^2_flin}
\begin{align*}
\nablaDG^2_{U,fV} S
&=
\nablaDG_U\,\nablaDG_{(fV)} S
-
\nablaDG_{\nablaDG_U (fV)} S
=
\nablaDG_U\,(f\nablaDG_{V} S)
-
\nablaDG_{(f\nablaDG_U V + U\VAct{f} V)} S
\\&=
f\,\nablaDG_U\,\nablaDG_{V} S + U\VAct{f}\,\nablaDG_V S
-
f\nablaDG_{\nablaDG_U V} S - U\VAct{f}\,\nablaDG_V S
= f \nablaDG^2_{U,V} S
\DEfullstop
\end{align*}
\end{ProofI}

The relationship between $\nablaDG^2_{U,V}$ and
$\nablaInd_{\iMa}\nablaInd_\iMb$ is given by
\begin{align}
\nablaDG^2_{U,V} W 
= 
U^\iMb\,V^\iMc\,
\Big(\nablaInd_\iMb\,\nablaInd_\iMc W^{\iMa}\Big) \partial_{\iMa}
\label{Nablas_connenting}
\DEfullstop
\end{align}
for any vector $W^{\iMa}$.
\begin{ProofI}{Proof of \eqref{Nablas_connenting}}
\label{pf_nabla^2_nabla}
\begin{align*}
\nablaDG^2_{U,V} W
&=
\nablaDG_U\,\nablaDG_V W 
-
\nablaDG_{\nablaDG_U V} W
=
U^\iMb \nablaInd_\iMb\,(\nablaDG_V W)^{\iMa} \partial_{\iMa}
-
(\nablaDG_U V)^\iMc (\nablaInd_\iMc W^{\iMa}) \partial_{\iMa}
\\&=
U^\iMb \nablaInd_\iMb\,(V^\iMc \nablaInd_\iMc W^{\iMa})\partial_{\iMa} 
-
U^\iMb (\nablaInd_\iMb V^\iMc) (\nablaInd_\iMc W^{\iMa}) \partial_{\iMa}
\\&=
U^\iMb 
\Big(\nablaInd_\iMb\,(V^\iMc \nablaInd_\iMc W^{\iMa})
-
(\nablaInd_\iMb V^\iMc) (\nablaInd_\iMc W^{\iMa})\Big) \partial_{\iMa}
\\&=
U^\iMb\,V^\iMc\,
\Big(\nablaInd_\iMb\,\nablaInd_\iMc W^{\iMa}\Big) \partial_{\iMa}
\DEfullstop
\end{align*}
\end{ProofI}

\subsection{Defining distributional forms}
\label{ch_CoFree_Forms}

Following Schwartz, we define a distributional $p$--form by
\jgDchange{its action} 
on a test $(4-p)$--form $\TarbTen\in\Gamma\Lambda^{4-p}M$, i.e. \jgDchange{a}
$(4-p)$--form with compact support \cite{gratus2018correct}.
Given $\alpha\in\Gamma\Lambda^pM$ is
a smooth $p-$form, we construct a regular distribution $\alpha^D$
via
\begin{align}
\alpha^D[\TarbTen]=\int_M \TarbTen\wedge \alpha
\label{Defs_def_alpha_D}
\DEfullstop
\end{align}
The definition of the wedge product, Lie derivatives, internal
contraction and exterior derivatives on distributions are defined to
be consistent with (\ref{Defs_def_alpha_D}). Thus for a distribution
$\Psi$ we set
\begin{align}
\begin{gathered}
(\Psi_1+\Psi_2)[\TarbTen]
=
\Psi_1[\TarbTen]+\Psi_2[\TarbTen]
\,,\quad
(\beta\wedge\Psi)[\TarbTen]=\Psi[\TarbTen\wedge\beta]
\,,\quad
(d\Psi)[\TarbTen]=(-1)^{(3-p)}\Psi[d\TarbTen]
\,,
\\
(i_v\Psi)[\TarbTen]=(-1)^{(3-p)}\Psi[i_v\TarbTen]
\quadand
(L_v\Psi)[\TarbTen]=-\Psi[L_v\TarbTen]
\label{Defs_opps_on_Psi}
\end{gathered}
\DEnone
\end{align}
for $v\in\Gamma T\Mman$.
Given \jgDchange{that} $C:\Interval\to\Mman$, is a closed embedding,
\jgDchange{the DeRham} push forward
with respect to $C$ of a $p$--form,
$\alpha\in\Gamma\Lambda^p\Interval$ is given by the distribution
\begin{align}
\big(C_\PF(\alpha)\big)[\TarbTen] = \int_\Interval C^\star(\TarbTen)\wedge\alpha
\label{CoFree_def_C_pf}
\DEfullstop
\end{align}
where $\TarbTen$ is a test form of degree $0$ or $1$ and
$C^\star(\TarbTen)$ is the \JGchange{pullback of
$\TarbTen\in\Gamma\Lambda^q\Mman$ to $\Gamma\Lambda^q\Interval$. 
This has degree $\deg\big(C_\PF(\alpha)\big)=3+p$.}
\JGchange{A general form
distribution is then given by} \jgDchange{applying an arbitrary number
  of wedges, exterior derivatives, etc, to $C_\PF(\alpha)$ using
  the rules given in (\ref{Defs_opps_on_Psi}).}

The \defn{order of a multipole} is defined as follows. If
\begin{equation}
\begin{aligned}
\Psi[\lambda^{k+1} \TarbTen]=0
\quadtext{for all}
&\lambda\in\Gamma\Lambda^0M\text{ and }
\TarbTen\in\Gamma_0\Lambda^1M
\quadtext{such that}
C^\star(\lambda)=0
\DEcomma
\end{aligned}
\label{Defs_order}
\end{equation}
then we say that the order of $\Psi$ is at most $k$. Since we impose that
$\lambda$ vanishes on the image of $C$, \jgDchange{(\ref{Defs_order})} implies that we need to
differentiate the argument $\lambda^{k+1} \TarbTen$ at least $k+1$ times
for $\Psi[\lambda^{k+1} \TarbTen]\ne0$. We say dipoles have order at most one
and quadrupoles have order at most two. Therefore the terms in a
dipole have at most one derivative, and those in a quadrupole at most
two. This is consistent with the fact that the set of quadrupoles include all
dipoles.

The deRham push forward is compatible with the exterior derivative 
\begin{align}
d\,C_\PF(\alpha)=C_\PF(d\alpha)
\label{CoFree_d_C_PF}
\DEcomma
\end{align}
and the internal contraction for \jgDchange{fields tangent to $C$}
\begin{align}
i_w\,C_\PF(\alpha) 
=
C_\PF(i_v\,\alpha)
\qquadtext{where}
w\in\Gamma T\Mman,\ 
v\in\Gamma T\Interval,\
C_\star(v|_\sigma) = w|_{C(\sigma)} 
\quadtext{for all} \sigma\in\Interval
\label{CoFree_iv_C_PF}
\DEfullstop
\end{align}

\subsection{The stress-energy 3--forms}
\label{ch_CoFree_SE}

\jgPchange{Since we wish to work without a metric it is natural to use
  a the stress-energy 3--forms \cite{benn1987introduction}. The
  relationship to $T^{\iMa\iMb}$ is given in (\ref{CoFree_def_Tab}) below.}
We exploit the fact \jgDchange{that} the stress-energy
3--forms have a similar structure to the electromagnetic
current 3--form. \JGchange{This enables us to use the technology
  developed in  \cite{gratus2018correct}.}

\jgPchange{We define} the stress-energy form $\tau$ as a map which
takes a 1--form $\alpha\in\Gamma\Lambda^1\Mman$ and gives a deRham
current 3--form $\tau_\alpha$ over the worldline $C$.
\begin{align}
\alpha \mapsto \tau_\alpha
\label{CoFree_tau_map}
\DEfullstop
\end{align}
The map (\ref{CoFree_tau_map}) is not \jgPchange{`f'-linear} but does satisfy
\begin{align}
\tau_{(\alpha+\beta)} = \tau_\alpha + \tau_\beta
\qquadand
\tau_{(f\alpha)}[\ToneTen] = \tau_{\alpha}[f\ToneTen]
\label{CoFree_tau_lin}
\DEcomma
\end{align}
for any test 1--form $\ToneTen$ and \jgDchange{scalar field $f$}.

Using $\tau_\alpha$ we define a tensor valued distribution $\tau$
which takes a tensor of type (0,2) as an argument. This is
defined as
\begin{align}
\tau[\ToneTen\otimes\alpha]
=
\tau_\alpha[\ToneTen]
\label{CoFree_SE_def_tau}
\DEfullstop
\end{align}
The stress-energy tensor is symmetric (\ref{Intro_Tab_Sym}) 
and divergenceless (\ref{Intro_Tab_Div_zero}). \jgPchange{We show
  below that} the symmetry condition
is given by
\begin{align}
\tau[\beta\otimes\alpha]=\tau[\alpha\otimes\beta]
\label{CoFree_SE_tau_symm}
\DEcomma
\end{align}
and the divergenceless condition is given by
\begin{align}
D\tau=0
\label{CoFree_SE_Dtau=0}
\DEcomma
\end{align}
where
\begin{align}
(D\tau)[\ToneTen] = -\tau[D\ToneTen]
\label{CoFree_SE_def_Dtau}
\DEnone
\end{align}
and
\begin{align}
(D\ToneTen)(U,V) = (\nablaDG_V\ToneTen)(U)
\label{CoFree_SE_def_Dphi}
\DEfullstop
\end{align}

Using a coordinate system, we can convert the map
(\ref{CoFree_tau_map}) into indexed 3--forms via
\begin{align}
\tau^{\iMa} = \tau_{dx^{\iMa}}
\label{CoFree_def_indexed_tau}
\DEfullstop
\end{align}
The relationship between the stress-energy forms and the tensor
density $T^{\iMa\iMb}$ is given by
\begin{align}
\int_\Interval T^{\iMa\iMb}\,\TtwoTen_{\iMa\iMb}\, d^4x 
&= 
\tau^{\iMa}[\TtwoTen_{\iMa\iMb}\,dx^\iMb]
=
\jgnchange{\tau[\TtwoTen_{\iMa\iMb}\,dx^\iMb\otimes dx^\iMa]}
\label{CoFree_def_Tab}
\DEfullstop
\end{align}
Using this coordinate system, (\ref{CoFree_SE_tau_symm}) becomes
\begin{align}
dx^{\iMa}\wedge \tau^\iMb = dx^\iMb\wedge \tau^{\iMa}
\label{CoFree_tau^a_sym}
\DEcomma
\end{align}
and (\ref{CoFree_SE_Dtau=0}) becomes
\begin{align}
d\tau^{\iMa} + \Gamma^{\iMa}_{\iMb\iMc}\,dx^\iMc \wedge \tau^\iMb
=
0
\label{CoFree_def_D_coords}
\DEfullstop
\end{align}
\begin{ProofI}{Proof of \eqref{CoFree_def_D_coords}}
\label{pf_D_coords}
 Let $\ToneTen$ be a test 1--form then
\begin{align*}
(D\ToneTen)(U,V)
&=
(\nablaDG_V\ToneTen)(U)
=
U^\iMb (\nablaDG_V\ToneTen)_\iMb
=
U^\iMb\, V^{\iMa}\,\nablaInd_{\iMa} \ToneTen_{\iMb}
=
(\nablaInd_\iMb \ToneTen_{\iMa})\  
(dx^\iMb\otimes dx^{\iMa}) (U,V)
\DEcomma
\end{align*}
hence
\begin{align*}
D\ToneTen=
(\nablaInd_\iMb \ToneTen_{\iMa})\  
(dx^\iMb\otimes dx^{\iMa}) 
\DEfullstop
\end{align*}
Thus
\begin{align*}
D\tau[\ToneTen]
&=
-\tau[D\ToneTen]
=
-\tau[(\nablaInd_\iMb \ToneTen_{\iMa})\  
(dx^\iMb\otimes dx^{\iMa})]
=
-\tau^{\iMa}[(\nablaInd_\iMb \ToneTen_{\iMa})\  
dx^\iMb]
=
-\tau^{\iMa}[(\partial_\iMb \ToneTen_{\iMa}-\Gamma^\iMc_{\iMb\iMa} \ToneTen_{\iMc})\,dx^\iMb]
\\&=
-\tau^{\iMa}[\partial_\iMb \ToneTen_{\iMa}\,dx^\iMb-\Gamma^\iMc_{\iMb\iMa} \ToneTen_{\iMc}\,dx^\iMb]
=
-\tau^{\iMa}[\partial_\iMb \ToneTen_{\iMa}\,dx^\iMb]
+
\tau^{\iMa}[\Gamma^\iMc_{\iMb\iMa} \ToneTen_{\iMc}\,dx^\iMb]
=
-\tau^{\iMa}[d\ToneTen_{\iMa}]
+
\Gamma^\iMc_{\iMb\iMa} \,dx^\iMb\wedge\tau^{\iMa}[\ToneTen_{\iMc}]
\\&=
d\tau^{\iMa}[\ToneTen_{\iMa}]
+
\Gamma^\iMc_{\iMb\iMa} \,dx^\iMb\wedge\tau^{\iMa}[\ToneTen_{\iMc}]
=
\big(d\tau^\iMc
+
\Gamma^\iMc_{\iMb\iMa} \,dx^\iMb\wedge\tau^{\iMa}\big)[\ToneTen_{\iMc}]
\DEfullstop
\end{align*}
\end{ProofI}

\subsection{Killing forms and conservation}
\label{ch_CoFree_Kill}

Killing forms (\ref{Q_Cons_symm_Kill}) can be written in a coordinate
free way. The 1--form $\alpha\in\Gamma\Lambda^1\Mman$ is Killing if
\begin{align}
(\nablaDG_V\alpha)(V)=0
\label{CoFree_Kill_def_Kill}
\DEcomma
\end{align}
for all vectors $V\in\Gamma T\Mman$.
\jgPchange{If there is a metric $g$, $\nablaDG$ is metric compatible and
  $\alpha$ is the metric dual of $K$ then (\ref{CoFree_Kill_def_Kill})
  is equivalent to $K$ being Killing. This follows since 
$(\nablaDG_V \alpha)(V) = g(\nablaDG_V K,V) $.}
From (\ref{CoFree_def_D_coords}) and (\ref{CoFree_tau^a_sym}) we have
\begin{align*}
d\tau_\alpha 
&=
d(\alpha_\iMa\,\tau^\iMa)
=
d \alpha_\iMa\wedge \tau^\iMa
+ \alpha_\iMa\wedge d\tau^\iMa
=
(\partial_\iMc \alpha_\iMa)\,dx^\iMc\wedge \tau^\iMa
- \Gamma^{\iMa}_{\iMb\iMc} \alpha_\iMa \,dx^\iMc\wedge d\tau^\iMb
=
\nablaInd_\iMc\alpha_\iMb\,dx^\iMc\wedge d\tau^\iMb
\\&=
\tfrac12
(\nablaInd_\iMc\alpha_\iMb-\nablaInd_\iMb\alpha_\iMc)\,dx^\iMc\wedge d\tau^\iMb
\DEfullstop
\end{align*}
Hence if $\alpha\in\Gamma\Lambda^1\Mman$ is a Killing 1--form then
from (\ref{Q_Cons_symm_Kill}) $d\tau_\alpha=0$.
This gives an alternative method of proving (\ref{Q_Cons_Q}).


\subsection{\jgDchange{The definition} of components}
\label{ch_CoFree_Coords}

Using (\ref{QP_Tab}) and (\ref{CoFree_def_Tab}) we deduce in an
arbitrary coordinate system
\begin{align}
\tau^{\iMa} = \tfrac12\, i_\iMb\, L_\iMc\, L_\iMd
C_\PF(\zeta^{\iMa\iMb\iMc\iMd} d\sigma)
\label{CoFree_tau_a_arb_coord}
\DEcomma
\end{align}
where $i_\iMb=i_{\partial_\iMb}$ and $L_\iMc=L_{\partial_\iMc}$.
\begin{ProofI}{Proof of \eqref{CoFree_tau_a_arb_coord}}
\label{pf_tau_Tab_zetas}
From \eqref{CoFree_def_Tab} and (\ref{QP_Tab_action}) we have
\begin{align*}
\tau^{\iMa}[\TtwoTen_{\iMa\iMb}\,dx^\iMb]
&=
\tfrac12\, i_\iMe\, L_\iMc\, L_\iMd
C_\PF(\zeta^{\iMa\iMe\iMc\iMd} d\sigma)
[\TtwoTen_{\iMa\iMb}\,dx^\iMb]
=
\tfrac12\, L_\iMc\, L_\iMd
C_\PF( \zeta^{\iMa\iMe\iMc\iMd} d\sigma)
[i_\iMe\,\TtwoTen_{\iMa\iMb}\,dx^\iMb]
\\&=
\tfrac12\, \delta^{\iMb}_{\iMe} L_\iMc\, L_\iMd
C_\PF(\zeta^{\iMa\iMe\iMc\iMd} d\sigma)
[\TtwoTen_{\iMa\iMb}]
=
\tfrac12\,  L_\iMc\, L_\iMd
C_\PF(\zeta^{\iMa\iMb\iMc\iMd} d\sigma)
[\TtwoTen_{\iMa\iMb}]
\\&=
\tfrac12\,  
C_\PF(\zeta^{\iMa\iMb\iMc\iMd} d\sigma)
[\partx_\iMc\, \partx_\iMd\TtwoTen_{\iMa\iMb}]
=
\tfrac12\,  
\int_\Interval \zeta^{\iMa\iMb\iMc\iMd}\,
(\partx_\iMc\, \partx_\iMd\TtwoTen_{\iMa\iMb})\, d\sigma
=
\int_\Interval T^{\iMa\iMb}\,\TtwoTen_{\iMa\iMb}\, d^4x 
\DEfullstop
\end{align*}
\end{ProofI}

In an adapted coordinate system (\ref{Intro_adapt_coords}) \jgDchange{equation}
(\ref{QP_Tab_gamma}) implies
\begin{align}
\tau^{\iMa}
&=
i_\iMb \, C_\PF(\gamma^{\iMa\iMb00}\,d\sigma)
+
i_\iMb \, L_\iSa\, C_\PF(\gamma^{\iMa\iMb0\iSa}\,d\sigma)
+
\tfrac12 i_\iMb \, L_\iSa\,L_\iSb\,
C_\PF(\gamma^{\iMa\iMb\iSa\iSb}\,d\sigma)
\label{CoFree_tau_a_adp_coord}
\DEfullstop
\end{align}
\begin{ProofI}{Proof of \eqref{CoFree_tau_a_adp_coord}}
\label{pf_tau_Tab_gammas}
\begin{align*}
\tau^{\iMa}[\TtwoTen_{\iMa\iMb}\,dx^\iMb]
&=
i_\iMe \, C_\PF(\gamma^{\iMa\iMe00}\,d\sigma)
[\TtwoTen_{\iMa\iMb}\,dx^\iMb]
+
i_\iMe \, L_\iSa\, C_\PF(\gamma^{\iMa\iMe0\iSa}\,d\sigma)
[\TtwoTen_{\iMa\iMb}\,dx^\iMb]
\\&\qquad\qquad
+
\tfrac12 i_\iMe \, L_\iSa\,L_\iSb\,
C_\PF(\gamma^{\iMa\iMe\iSa\iSb}\,d\sigma)
[\TtwoTen_{\iMa\iMb}\,dx^\iMb]
\\&=
C_\PF(\gamma^{\iMa\iMb00}\,d\sigma)
[\TtwoTen_{\iMa\iMb}]
+
 L_\iSa\, C_\PF(\gamma^{\iMa\iMb0\iSa}\,d\sigma)
[\TtwoTen_{\iMa\iMb}]
+
\tfrac12 L_\iSa\,L_\iSb\,
C_\PF(\gamma^{\iMa\iMb\iSa\iSb}\,d\sigma)
[\TtwoTen_{\iMa\iMb}]
\\&=
C_\PF(\gamma^{\iMa\iMb00}\,d\sigma)
[\TtwoTen_{\iMa\iMb}]
-
 C_\PF(\gamma^{\iMa\iMb0\iSa}\,d\sigma)
[\partz_\iSa \TtwoTen_{\iMa\iMb}]
+
\tfrac12 
C_\PF(\gamma^{\iMa\iMb\iSa\iSb}\,d\sigma)
[\partz_\iSa \partz_\iSb\,\TtwoTen_{\iMa\iMb}]
\\&=
\int_\Interval \gamma^{\iMa\iMb00}\,\TtwoTen_{\iMa\iMb}\,d\sigma
-
\int_\Interval \gamma^{\iMa\iMb0\iSa}\,(\partz_\iSa
\TtwoTen_{\iMa\iMb}) d\sigma
+
\tfrac12 
\int_\Interval \gamma^{\iMa\iMb\iSa\iSb}
(\partz_\iSa \partz_\iSb\,\TtwoTen_{\iMa\iMb})\,d\sigma
\\&=
\int_\Interval \Big(\gamma^{\iMa\iMb00}\,\TtwoTen_{\iMa\iMb}
-
\gamma^{\iMa\iMb0\iSa}\,(\partz_\iSa
\TtwoTen_{\iMa\iMb}) 
+
\tfrac12 
\gamma^{\iMa\iMb\iSa\iSb}
(\partz_\iSa \partz_\iSb\,\TtwoTen_{\iMa\iMb})
\Big)\,d\sigma
\DEfullstop
\end{align*}
\end{ProofI}

As stated the advantage of using an adapted coordinate system is that
the $\gamma^{\iMa\iMb\iMc\iMd}$ are unique, \JGchange{as seen from
  (\ref{Ellis_adap_extract_comp}), (\ref{Ellis_adap_def_bump}).}

\subsection{Semi-dipoles and semi-quadrupoles}
\label{ch_Semi_Q}

Having defined the quadrupoles in a coordinate free manner, one can
identify properties which can be defined without
reference to a coordinate system. In \cite{gratus2018correct} we defined the
semi-dipole and semi-quadrupole electromagnetic 3--form. The
semi-dipole corresponded to the purely electric \jgnchange{dipole}. One can
likewise define the semi-dipole and semi-quadrupole stress-energy
distributions. In this case we say that $\tau_\alpha$ is an semi-multipole of
order at most $\ell$  if
\begin{equation}
\begin{aligned}
\tau_\alpha[\lambda^{\ell} d\mu]=0
\quadtext{for all}&
\lambda,\mu\in\Gamma\Lambda^0M\quadtext{such that}
C^\star(\lambda)=C^\star(\mu)=0
\DEfullstop
\end{aligned}
\label{Defs_Elec_order}
\end{equation}
We observe that the semi-dipole ($\ell=1$) corresponds to the case
when the spin tensor is $S^{\iSb\iSa}=0$. The semi-quadrupole ($\ell=2$), does not
have a natural interpretation, but is used as a quadrupole with fewer
components. 

When we apply this to the quadrupole (\ref{CoFree_tau_a_adp_coord}),
in an adapted coordinate system $(\sigma,z^1,z^2,z^3)$, we see that
the semi-quadrupole is given by
\begin{align}
\tau^{\iMa}
&=
i_\iMb \, C_\PF(\gamma^{\iMa\iMb00}\,d\sigma)
+
i_\iMb \, L_\iSa\, C_\PF(\gamma^{\iMa\iMb0\iSa}\,d\sigma)
+
\tfrac12 L_\iSa\,L_\iSb\,
C_\PF(\gamma^{\iMa0\iSa\iSb})
\label{Semi_Q_tau_a_adp_coord}
\DEfullstop
\end{align}
This gives 22 ODE components and 6 free components as indicated in
table \ref{tab_number_components}. We presented the general solution
for the static
semi-quadrupole in section \ref{ch_Q_free}.

\begin{ProofA}{Proof of \eqref{Semi_Q_tau_a_adp_coord} and
    Semi-quadrupole counting.}
A simple application using $\lambda=\lambda_1+\lambda_2$ and
$\lambda=\lambda_1-\lambda_2$ implies we can replace
\eqref{Defs_Elec_order} for $\ell=2$ with
\eqref{Defs_Elec_order} with $\tau_\alpha[\lambda_1\lambda_2 d\iSa]=0$
where $C^\star(\lambda_1)=C^\star(\lambda_2)=C^\star(\iSa)=0$. 
\jgDchange{Hence}
\begin{align*}
0
=
\lim_{\epsilon\to 0} 
\tau^{\iMa}[z^\iSa\,z^\iSb\,d(z^\iSc\bumpf_{\epsilon,\sigma'})]
=
\gamma^{\iMa\iSc\iSa\iSb}(\sigma')
\end{align*}
\label{pf_Semi_Q_cond}
and hence \eqref{Ellis_adap_gam_zeta}.

We can now count the number and type of components. The dynamic
equation \eqref{QP_DTeqn_a000} and (\ref{QP_DTeqn_a00m}) remain
unchanged but \eqref{QP_DTeqn_a0mn} becomes 
\begin{align*}
\gamma^{\iSc \iSb 0 \iSa}
=
- \Gamma^\iSc{}_{00}\, \gamma^{0 0 \iSa \iSb}
\qquadand
\dot \gamma^{0 0 \iSa\iSb}
&=
- 2\gamma^{0 (\iSb 0) \iSa}
- \Gamma^0{}_{00}\, \gamma^{0 0 \iSa \iSb}
\DEcomma
\end{align*}
since the symmetry condition \eqref{QP_gamma_sym} implies
$\gamma^{\iSc0\iSa\iSb}=\gamma^{0\iSc\iSa\iSb}=0$. Thus we have 4+12+6=22
ODEs. 

Starting with the 100 components given after applying
\eqref{QP_gamma_sym} we have $9\times 6=54$ constraints coming from 
$\gamma^{\iMa\iSc\iSa\iSb}=0$ plus $18$ constraints coming from the
first equation above. This leaves 28 components. Of these 22 are given
by the ODEs and 6 are free.
\end{ProofA}



\subsection{The coordinate free definition of the Dixon 
split only using $\DixVec$  and the connection}
\label{ch_CoFree_Dixon}

We have defined the stress-energy distribution without reference to
a coordinate system. When writing this in terms of coordinates
(\ref{CoFree_tau_a_arb_coord}) and (\ref{Ellis_Squz_moments}) we see
that this corresponds directly to the Ellis representation of the
multipoles. 
Here we show how to perform the Dixon split
(\ref{Intro_Tab_Dixon_Split}) which 
\jgDchange{separates} the multipoles into different
orders with respect to a 1--form $\DixVec$ along the
curve. \jgDchange{Our procedure separates}  the quadrupole into a pure Dixon quadrupole term, a pure
Dixon dipole term and a monopole term. The pattern however is clear.
The Dixon split (\ref{Intro_Tab_Dixon_Split}) requires defining
$\tau_{(0)}$, $\tau_{(1)}$ and $\tau_{(2)}$ such
that an arbitrary quadrupole \jgDchange{has the form} 
\begin{align}
\tau = 
\tau_{(0)} + 
\tau_{(1)} + 
\tau_{(2)} 
\label{CoFree_Dixon_split}
\DEfullstop
\end{align}
Using (\ref{CoFree_def_Tab}) to convert these into
$T^{\iMa\iMb}_{(r)}$ \jgDchange{we find}
that $T^{\iMa\iMb}=T^{\iMa\iMb}_{(0)}+T^{\iMa\iMb}_{(1)}+T^{\iMa\iMb}_{(2)}$ where
\begin{align}
\tau_{(0)} [\TtwoTen]
&=
\int_\Mman T^{\iMa\iMb}_{(0)} \ \TtwoTen_{\iMa\iMb} \ d^4x 
=
\int_\Interval  
\zetaMultiDixon^{\iMa\iMb}(\sigma)\,
\TtwoTen_{\iMa\iMb}(\sigma)\ d\sigma
\,, 
\label{CoFree_Dixon_Tab_Mono}
\\
\tau_{(1)} [\TtwoTen]
&=
\int_\Mman T^{\iMa\iMb}_{(1)} \ \TtwoTen_{\iMa\iMb} \ d^4x 
=
\int_\Interval  
\zetaMultiDixon^{\iMa\iMb\iMc}(\sigma)\,
(\nablaInd_\iMc \TtwoTen_{\iMa\iMb})|_{C(\sigma)}\ d\sigma
\,, 
\label{CoFree_Dixon_Tab_Dip}
\\
\tau_{(2)} [\TtwoTen]
&=
\int_\Mman T^{\iMa\iMb}_{(2)} \ \TtwoTen_{\iMa\iMb} \ d^4x 
=
\int_\Interval  
\zetaMultiDixon^{\iMa\iMb\iMc\iMd}(\sigma)\,
(\nablaInd_\iMc\nablaInd_\iMd \TtwoTen_{\iMa\iMb})|_{C(\sigma)}\ d\sigma
\label{CoFree_Dixon_Tab_Quad}
\DEfullstop
\end{align}
The Dixon split is with respect to a 1--form, as opposed to a vector
along $C$, \jgDchange{in} order to avoid using the metric. The one
requirement is that the 1--form $\DixVec$ combined with the vector
$\Cdot$ is nowhere zero, i.e.
\begin{align}
\DixVec(\Cdot)\ne0
\label{CoFree_Dixon_not_orthog}
\DEfullstop
\end{align}

In order to perform the Dixon split, it is necessary to define a
radial vector fields. We
say that $R\in\Gamma TM$ is \defn{Radial} (to second order) 
with respect to $C$ and
$\DixVec$ if for all $p=C(\sigma)$
\begin{align}
R|_{p} &= 0
,\qquad
(\nablaDG_V R)|_p = V|_p
\qquadand
\big(\nablaDG^2_{U,V} R\big)\big|_p = 0
\label{CoFree_Dixon_def_Rad}
\DEcomma
\end{align}
for all vectors $U,V\in T\Mman$ such that $N(V)=N(U)=0$. 
Below in (\ref{CoFree_Dixon_Radial_expan}) we express the components of $R$ with
respect to a coordinate system, which is adapted both for $C$ and
$\DixVec$.

Using this radial vector, the Dixon split (\ref{CoFree_Dixon_split}) 
is given by
\begin{align}
\tau_{(0)} [\TtwoTen]
&=
\tau[\TtwoTen-\nabla_R\TtwoTen + \tfrac12\nabla^2_{R,R} \TtwoTen] 
\DEcomma
\label{CoFree_Dixon_mono}
\\
\tau_{(1)} [\TtwoTen]
&=
\tau[\nabla_R\TtwoTen-\nabla^2_{R,R} \TtwoTen] 
\DEcomma
\label{CoFree_Dixon_dip}
\\
\tau_{(2)} [\TtwoTen]
&=
\tau[\tfrac12\nabla^2_{R,R} \TtwoTen] 
\label{CoFree_Dixon_quad}
\DEfullstop
\end{align}
where $\TtwoTen$ is an type (0,2) test tensor. 
The proof of (\ref{CoFree_Dixon_mono})-(\ref{CoFree_Dixon_quad}) is
given below.
One advantage of \jgDchange{using
  (\ref{CoFree_Dixon_mono})-(\ref{CoFree_Dixon_quad})} 
is that one can now show how the
Dixon components mix when one changes $\DixVec$, \jgNewCh{and that all
  Ellis multipoles are also Dixon Multipoles.}

\subsubsection*{Proofs of Dixon split}
\label{ch_Apendx_DixonSplit}

In this section we work in a coordinate system $(\sigma,z^1,z^2,z^3)$,
which is adapted both for $C$ and $\DixVec$, so that
$\DixVec=\DixVec_0 d\sigma$ with $\DixVec_0\ne 0$. We see that if
$N(V)=0$ then \jgDchange{$V^0=0$}. Likewise we can
replace $\xi^{\iMa\iMb\iMc\iMd}$ with $\xi^{\iMa\iMb\iSa\iSb}$ since
$\xi^{\iMa\iMb0\iSa}=\xi^{\iMa\iMb\iSa0}=0$.

In this coordinate system a radial
vector $R$ has the properties
\begin{align}
\begin{gathered}
R^{\iMa}|_p=0
,\quad
\partial_{\iMa} R^0|_p = 0
,\quad
\partial_{\iMa} R^\iSa|_p = \delta_{\iMa}^\iSa
,\quad
\partial_0 \partial_{\iMa} R^\iMb|_p = 0
,\\
\partial_\iSb\partial_\iSc R^0|_p = -2\Gamma^0_{\iSb\iSc}
\quadand
\partial_\iSb\partial_\iSc R^\iSa|_p = -\Gamma^\iSa_{\iSb\iSc}
\DEcomma
\end{gathered}
\label{CoFree_Dixon_Radial_diffs_R^mu}
\end{align}
for any $p=C(\sigma)$.
This can be expressed as 
\begin{align}
R^0 = - 
z^\iSb z^\iSc\,\Gamma^0_{\iSb\iSc}\,\partial_0  
+
O(\Vz^3)
\qquadand
R^\iSa = z^\iSa - \tfrac12 z^\iSb z^\iSc\,\Gamma^\iSa_{\iSb\iSc}  + 
O(\Vz^3)
\label{CoFree_Dixon_Radial_expan_alt}
\DEcomma
\end{align}
or alternatively as
\begin{align}
R= z^\iSa\,\partz_\iSa 
- 
z^\iSb z^\iSc\,\Gamma^0_{\iSb\iSc}\,\partial_0  
-
\tfrac12 z^\iSb z^\iSc\,\Gamma^\iSa_{\iSb\iSc}\,\partial_\iSa  
+  O(\Vz^3)
\label{CoFree_Dixon_Radial_expan}
\DEfullstop
\end{align}
where $O(\Vz^3)$ is any function (or vector)
of $(\sigma,z^1,z^2,z^3)$ which is at least
cubic in its $z^\iSa$ arguments. 

\begin{ProofA}{Proof of
    \eqref{CoFree_Dixon_Radial_diffs_R^mu}}
In the adapted coordinate system, assume first that $R^{\iMa}$ satisfies 
\eqref{CoFree_Dixon_Radial_diffs_R^mu} and that $U,V$ satisfy
$N(U)=N(V)=0$, so $U^0=V^0=0$. 

Clearly from either (\ref{CoFree_Dixon_def_Rad}.1) or
(\ref{CoFree_Dixon_Radial_diffs_R^mu}.1) we have $R|_p=0$. Here
(\ref{CoFree_Dixon_def_Rad}.1) refers to the first equation in
\eqref{CoFree_Dixon_def_Rad}. 
\begin{align*}
\big(\nablaDG_V R - V\big)^{\iMa}\big|_p
=
\big(V^\iMb\partial_\iMb(R^{\iMa}) + V^\iMb R^\iMc \Gamma^{\iMa}_{\iMb\iMc} - V^{\iMa}\big)\big|_p
=
\big(V^\iSa (\partial_\iSa(R^{\iMa}) - \delta^{\iMa}_\iSa)\big)\big|_p
\DEfullstop
\end{align*}
Thus (\ref{CoFree_Dixon_def_Rad}.2) is
equivalent to (\ref{CoFree_Dixon_Radial_diffs_R^mu}.2),
(\ref{CoFree_Dixon_Radial_diffs_R^mu}.3).
From (\ref{CoFree_Dixon_Radial_diffs_R^mu}.2) and
(\ref{CoFree_Dixon_Radial_diffs_R^mu}.3)  we have
(\ref{CoFree_Dixon_Radial_diffs_R^mu}.4)

From (\ref{CoFree_Dixon_def_Rad}.2) we have, 
(implicitly evaluating at $p$),
\begin{align*}
\nablaInd_\iSb(\nablaInd_\iSc R^\iSa)
&=
\partial_\iSb(\nablaInd_\iSc R^\iSa)
+
(\nablaInd_\iSc R^\iSd)
\Gamma^\iSa_{\iSb\iSd} 
-
(\nablaInd_\iSd R^\iSa)
\Gamma^\iSd_{\iSb\iSc} 
\\&=
\partial_\iSb\partial_\iSc R^\iSa
+
\partial_\iSb(R^\iSe\,\Gamma^\iSa_{\iSc\iSe})
+
(\partial_\iSc R^\iSd)
\Gamma^\iSa_{\iSb\iSd} 
+
R^\iSe \Gamma^\iSd_{\iSc\iSe}
\Gamma^\iSa_{\iSb\iSd} 
-
(\partial_\iSd R^\iSa)
\Gamma^\iSd_{\iSb\iSc} 
-
R^\iSe \Gamma^\iSa_{\iSd\iSe}
\Gamma^\iSd_{\iSb\iSc} 
\\&=
\partial_\iSb\partial_\iSc R^\iSa
+
\delta^\iSe_\iSb\,\Gamma^\iSa_{\iSc\iSe}
+
\delta^\iSd_\iSc 
\Gamma^\iSa_{\iSb\iSd} 
-
\delta^\iSa_\iSd 
\Gamma^\iSd_{\iSb\iSc} 
=
\partial_\iSb\partial_\iSc R^\iSa
+
\Gamma^\iSa_{\iSc\iSb}
+
\Gamma^\iSa_{\iSb\iSc} 
-
\Gamma^\iSa_{\iSb\iSc} 
\\&=
\partial_\iSb\partial_\iSc R^\iSa
+
\Gamma^\iSa_{\iSc\iSb}
\DEnone
\end{align*}
and
\begin{align*}
\nablaInd_\iSb(\nablaInd_\iSc R^0)
&=
\partial_\iSb(\nablaInd_\iSc R^0)
+
(\nablaInd_\iSc R^\iSd)
\Gamma^0_{\iSb \iSd} 
-
(\nablaInd_\iSd R^0)
\Gamma^\iSd_{\iSb\iSc} 
=
\partial_\iSb\partial_\iSc R^0
+
\partial_\iSb(R^\iSe\,\Gamma^0_{\iSc\iSe})
+
\Gamma^0_{\iSb \iSc} 
\\&=
\partial_\iSb\partial_\iSc R^0
+
\partial_\iSb(R^\iSe)\,\Gamma^0_{\iSc\iSe}
+
\Gamma^0_{\iSb \iSc} 
=
\partial_\iSb\partial_\iSc R^0
+
2\Gamma^0_{\iSb \iSc} 
\DEfullstop
\end{align*}
Thus
\begin{align*}
\nablaDG^2_{U,V} R
&=
V^{\iMa} U^\iMb (\partial_{\iMa}\partial_\iMb R^\iSa+ \Gamma^\iSa_{\iSc\iSb})
\partial_\iSa
+
V^{\iMa} U^\iMb (\partial_{\iMa}\partial_\iMb R^0+ 2\Gamma^\iSa_{\iSc\iSb})
\partial_0
\\&=
V^\iSa U^\iSb (\partial_\iSb\partial_\iSc R^\iSa+ \Gamma^\iSa_{\iSc\iSb})
\partial_\iSa
+
V^\iSa U^\iSb (\partial_\iSb\partial_\iSc R^0+ 2\Gamma^\iSa_{\iSc\iSb})
\partial_0
\DEfullstop
\end{align*}
Hence (\ref{CoFree_Dixon_def_Rad}.3) \jgDchange{holds} if and only if
(\ref{CoFree_Dixon_Radial_diffs_R^mu}.5) and 
(\ref{CoFree_Dixon_Radial_diffs_R^mu}.6) \jgDchange{hold.}

\end{ProofA}


\begin{ProofA}{Proof of \eqref{CoFree_Dixon_quad}}
\label{pf_DixonSplit_quad}
In the adapted coordinate system
and evaluating at $C(\sigma)$ we have
\begin{align*}
\xi^{\iMa\iMb} 
(R^\iMe\, R^\iMf\, \nablaInd_\iMe\nablaInd_\iMf \TtwoTen_{\iMa\iMb})
=
0
\DEfullstop
\end{align*}
Thus the monopole term \eqref{CoFree_Dixon_Tab_Mono} does not
contribute to $\tau_{(2)}$. Likewise
\begin{align*}
\xi^{\iMa\iMb\iMc\iMd} \nablaInd_\iMc
(R^\iMe\, R^\iMf\, \nablaInd_\iMe\nablaInd_\iMf \TtwoTen_{\iMa\iMb})
=
0
\DEcomma
\end{align*}
so the dipole term \eqref{CoFree_Dixon_Tab_Dip} does not
contribute to $\tau_{(2)}$.
Finally we have
\begin{equation}
\begin{aligned}
\xi^{\iMa\iMb\iMc\iMd} \nablaInd_\iMc\nablaInd_\iMd &
(R^\iMe\, R^\iMf\, \nablaInd_\iMe\nablaInd_\iMf \TtwoTen_{\iMa\iMb})
=
\xi^{\iMa\iMb\iSa\iSb} \nablaInd_{\iSa}\nablaInd_{\iSb}
(R^\iMe\, R^\iMf\, \nablaInd_\iMe\nablaInd_\iMf \TtwoTen_{\iMa\iMb})
=
\xi^{\iMa\iMb\iSa\iSb} 
(\partial_{\iSa}\partial_{\iSb}(R^\iMe\, R^\iMf)\, \nablaInd_\iMe\nablaInd_\iMf \TtwoTen_{\iMa\iMb})
\\&=
\xi^{\iMa\iMb\iSa\iSb} (\delta_\iSa^\iMe\delta_\iSb^\iMf+\delta_\iSa^\iMf\delta_\iSb^\iMe)
(\nablaInd_\iMe\nablaInd_\iMf \TtwoTen_{\iMa\iMb})
=
2 \xi^{\iMa\iMb\iSa\iSb} 
(\nablaInd_\iSa\nablaInd_\iSb \TtwoTen_{\iMa\iMb})
=
2 \xi^{\iMa\iMb\iMc\iMd} 
(\nablaInd_\iMc\nablaInd_\iMd \TtwoTen_{\iMa\iMb})
\DEfullstop
\end{aligned}
\label{pf_CoFree_Dixon_xi^abcd_RR}
\end{equation}
Thus $\tau_{(2)}$ is given by \eqref{CoFree_Dixon_quad}.

\end{ProofA}


\begin{ProofA}{Proof of \eqref{CoFree_Dixon_dip}}
\label{pf_DixonSplit_dip}
Since
\begin{align*}
\xi^{\iMa\iMb} 
(R^\iMe\,\nablaInd_\iMe\,\TtwoTen_{\iMa\iMb} - R^\iMe\, R^\iMf\, \nablaInd_\iMe\nablaInd_\iMf 
\TtwoTen_{\iMa\iMb})
=0
\DEcomma
\end{align*}
the monopole term does not contribute to $\tau_{(1)}$.
Also 
\begin{align*}
\nablaInd_\iSa\nablaInd_\iSb\,(R^\iMe\,\nablaInd_\iMe\,\TtwoTen_{\iMa\iMb})
&=
\nablaInd_\iSa\big((\nablaInd_\iSb R^\iMe)\nablaInd_\iMe\,\TtwoTen_{\iMa\iMb}\big)
+
\nablaInd_\iSa\big( R^\iMe \nablaInd_\iSb\nablaInd_\iMe\,\TtwoTen_{\iMa\iMb}\big)
\\&=
(\nablaInd_\iSa \nablaInd_\iSb R^\iMe)\nablaInd_\iMe\,\TtwoTen_{\iMa\iMb}
+
(\nablaInd_\iSb R^\iMe)\nablaInd_\iSa \nablaInd_\iMe\,\TtwoTen_{\iMa\iMb}
+
(\nablaInd_\iSa R^\iMe) \nablaInd_\iSb\nablaInd_\iMe\,\TtwoTen_{\iMa\iMb}
+
R^\iMe \nablaInd_\iSa\nablaInd_\iSb\nablaInd_\iMe\,\TtwoTen_{\iMa\iMb}
\\&=
\delta^\iMe_\iSb \nablaInd_\iSa \nablaInd_\iMe\,\TtwoTen_{\iMa\iMb}
+
\delta^\iMe_\iSa \nablaInd_\iSb\nablaInd_\iMe\,\TtwoTen_{\iMa\iMb}
=
\nablaInd_\iSa \nablaInd_\iSb\,\TtwoTen_{\iMa\iMb}
+
\nablaInd_\iSb\nablaInd_\iSa\,\TtwoTen_{\iMa\iMb}
\DEfullstop
\end{align*}
Hence
\begin{equation}
\begin{aligned}
\xi^{\iMa\iMb\iMc\iMd}\, \nablaInd_\iMc\nablaInd_\iMd\,(R^\iMe\,\nablaInd_\iMe\,\TtwoTen_{\iMa\iMb})
&=
\xi^{\iMa\iMb\iSa\iSb}\, \nablaInd_\iSa\nablaInd_\iSb\,(R^\iMe\,\nablaInd_\iMe\,\TtwoTen_{\iMa\iMb})
=
\xi^{\iMa\iMb\iSa\iSb}\, \big(\nablaInd_\iSa \nablaInd_\iSb\,\TtwoTen_{\iMa\iMb}
+
\nablaInd_\iSb\nablaInd_\iSa\,\TtwoTen_{\iMa\iMb}\big)
\\&=
2 \xi^{\iMa\iMb\iSa\iSb} 
(\nablaInd_\iSa\nablaInd_\iSb \TtwoTen_{\iMa\iMb})
=
2 \xi^{\iMa\iMb\iMc\iMd} 
(\nablaInd_\iMc\nablaInd_\iMd \TtwoTen_{\iMa\iMb})
\DEfullstop
\end{aligned}
\label{pf_CoFree_Dixon_xi^abcd_R}
\end{equation}
Thus using \eqref{pf_CoFree_Dixon_xi^abcd_RR} we see
\begin{align*}
\xi^{\iMa\iMb\iMc\iMd}\, \nablaInd_\iMc\nablaInd_\iMd\,
(R^\iMe\,\nablaInd_\iMe\,\TtwoTen_{\iMa\iMb} - R^\iMe\, R^\iMf\, \nablaInd_\iMe\nablaInd_\iMf 
\TtwoTen_{\iMa\iMb})
&=
0
\DEfullstop
\end{align*}
Thus the quadrupole term \eqref{CoFree_Dixon_Tab_Quad} does not
contribute to $\tau_{(1)}$.
Finally
\begin{equation}
\begin{aligned}
\xi^{\iMa\iMb\iMc}\, \nablaInd_\iMc
(R^\iMe\,\nablaInd_\iMe\,\TtwoTen_{\iMa\iMb})
&=
\xi^{\iMa\iMb\iSa}\, \nablaInd_\iSa
(R^\iMe\,\nablaInd_\iMe\,\TtwoTen_{\iMa\iMb})
=
\xi^{\iMa\iMb\iSa}\, (\nablaInd_\iSa R^\iMe)\,\nablaInd_\iMe\,\TtwoTen_{\iMa\iMb}
=
\xi^{\iMa\iMb\iSa}\, \delta_\iSa^\iMe\,\nablaInd_\iMe\,\TtwoTen_{\iMa\iMb}
\\&=
\xi^{\iMa\iMb\iSa}\, \nablaInd_\iSa\,\TtwoTen_{\iMa\iMb}
=
\xi^{\iMa\iMb\iMc}\, \nablaInd_\iMc\,\TtwoTen_{\iMa\iMb}
\DEfullstop
\end{aligned}
\label{pf_CoFree_Dixon_xi^abc_R}
\end{equation}
Thus $\tau_{(1)}$ is given by \eqref{CoFree_Dixon_dip}.
\end{ProofA}


\begin{ProofA}{Proof of \eqref{CoFree_Dixon_mono}}
\label{pf_DixonSplit_mono}
From \eqref{pf_CoFree_Dixon_xi^abcd_RR} and
\eqref{pf_CoFree_Dixon_xi^abcd_R} we have
\begin{align*}
\xi^{\iMa\iMb\iMc\iMd}\, \nablaInd_\iMc\nablaInd_\iMd\,
(\TtwoTen_{\iMa\iMb} - 
R^\iMe\,\nablaInd_\iMe\,\TtwoTen_{\iMa\iMb} + \tfrac12R^\iMe\, R^\iMf\, \nablaInd_\iMe\nablaInd_\iMf 
\TtwoTen_{\iMa\iMb})
&=
0
\DEfullstop
\end{align*}
Thus the quadrupole term \eqref{CoFree_Dixon_Tab_Quad} does not
contribute to $\tau_{(0)}$. Using \eqref{pf_CoFree_Dixon_xi^abc_R} we
have
\begin{align*}
\xi^{\iMa\iMb\iMc}\, \nablaInd_\iMc
(\TtwoTen_{\iMa\iMb} - 
R^\iMe\,\nablaInd_\iMe\,\TtwoTen_{\iMa\iMb} + \tfrac12R^\iMe\, R^\iMf\, \nablaInd_\iMe\nablaInd_\iMf 
\TtwoTen_{\iMa\iMb})=0
\DEcomma
\end{align*}
so the dipole term \eqref{CoFree_Dixon_Tab_Dip} does not
contribute to $\tau_{(0)}$.
Finally
\begin{align*}
\xi^{\iMa\iMb}\, 
(\TtwoTen_{\iMa\iMb} - 
R^\iMe\,\nablaInd_\iMe\,\TtwoTen_{\iMa\iMb} + \tfrac12R^\iMe\, R^\iMf\, \nablaInd_\iMe\nablaInd_\iMf 
\TtwoTen_{\iMa\iMb})
&=
\xi^{\iMa\iMb}\, \TtwoTen_{\iMa\iMb}
\DEcomma
\end{align*}
so $\tau_{(0)}$ is given by \eqref{CoFree_Dixon_mono}.
\end{ProofA}

\section{Discussion and outlook.}
\label{ch_Conclusion}

We have derived a number of key results about the distributional
quadrupole stress-energy tensor, in particular the existence of the
free components, which require additional constitutive relations to
prescribe. An example of these constitutive relations is given. We
have also given the coordinate transformation of the quadrupole
components, the conserved quantities in the presence of a Killing
vector and a definition of semi-quadrupoles. \jgDchange{We presented a metric and
coordinate free definition of the quadrupole and a way of separating
the quadrupole into the monopole, dipole and quadrupole terms
corresponding to the Dixon representation.}

The understanding of the quadrupole stress-energy tensor distribution
is important for the study of gravitational wave sources, as well as
being interesting in its own right. The existence of free components
imply that it is not possible to know everything about a quadrupole
simply from the initial conditions. There is clearly much research
that needs to be done to find appropriate constitutive relations to
replace the free components with ODEs or algebraic relations.
\jgDchange{It may be possible to calculate these from underlying
  models, such as two orbiting black holes or neutron stars.}  In
section \ref{ch_Dust} we \jgDchange{presented} only a very simple
constitutive relation corresponding to a dust model. With
increasing sensitivity of gravitational wave astronomy one can hope to
test the different constitutive relations using experimental data.

Although the observation of the need for constitutive relations for
the quadrupole on a prescribe worldline is new, there are other cases
where the need for constitutive relations has been observed. For
example \cite{steinhoff2010multipolar}, they are needed to determine
how dipoles or quadrupoles \jgDchange{affect} the worldline. There are other
situations where one can expect constitutive relations will be
needed. In future work we intend to look at the dynamics of
charged multipoles in an electromagnetic field. One would expect in
this case that constitutive relations are also needed, especially since a
dipole has nine components, but the electromagnetic current, which
provides the force and torque, has only six components. These
constitutive relations describe the differences between the charge
distribution and the mass distribution in the dipole. The situation
has an additional challenge in that the electromagnetic field
\jgDchange{diverges}
on the worldline. This poses another question that has been tackled by
many authors: how does a dipole respond to its own electromagnetic
field \cite{Gralla:2009md,gratus2015self,ferris2011origin}.

We have investigated the \jgDchange{representation of the
  quadrupole using partial derivatives (the Ellis representation).} As
well as the differential equations, we have given the gauge-like
freedom, the change of coordinates, the adapted coordinates and the
change of coordinates for adapted coordinates. It is natural to ask
what new features will arise for \jgDchange{sextupoles}. One will
expect that the gauge-like freedom for \jgDchange{sextupoles} will
include a term with $\Cdot^{\iMa}\, C^\iMb\,\int^\sigma
C^\iMc\,d\sigma'$.

Having definitions which are coordinate free can be very useful. They
make it clear which objects are coordinate dependent and which are
truly geometric.  Although the Ellis representation of
multipoles is easy to define in a coordinate free manner, here we have
derived a coordinate free definition of {Dixon's split of the
  quadrupole into the pure monopole, dipole and quadrupole terms.
\jgnchange{In future work, we intend to reproduce these results in
  coordinates. This will enable us to  write the Dixon components
  $\zetaMultiDixon^{\iMa\iMb\ldots}$ in terms of Ellis components
  $\zetaMultiEllis^{\iMa\iMb\ldots}$, and also derive the complicated
  relationship between the Dixon components
  $\zetaMultiDixon^{\iMa\iMb\ldots}$ for different $\DixVec^{\iMa}$.}
The dynamical equations for the Ellis quadrupole components were
derived (\ref{QP_DTeqn_a000})-(\ref{QP_DTeqn_alg}). Using this 
split, one could translate these into dynamical equations for the
Dixon components.} \jgPchange{This will enable a comparison between
  the 20 independent components of (reduced) quadrupole stress-energy
  tensor as described by Dixon, as discussed in appendix \ref{ch_Appx_Dixon}.}

Although spacetime is endowed with both a metric and a connection,
there is much research into which objects can be defined without such
structures. In some cases this \JGchange{questions the underlying
  physics}, asking whether the electromagnetic field is more
fundamental than the gravitational field \cite{hehl2016kottler}. In
other cases it is useful for examining how an object \jgDchange{depends} on a
metric or a connection. This is necessary when
\jgDchange{calculating the result of varying a Lagrangian} with
respect to the metric. It is \jgDchange{useful} therefore, that a general
multipole does not require any additional structures beyond those in a
general manifold. This means that one can define multipoles on other
manifolds such as tangent bundles or jet bundles. Such an approach may
also give an insight into prescribing constitutive relations. 
Of course a connection is required \jgnchange{in order} to demand the
stress-energy distribution \jgnchange{be} divergenceless, but there is no
requirement \jgnchange{that such a connection be Levi-Civita}. All the
coordinate free presentation from section \ref{ch_CoFree} does not
require a metric, so one can choose a metric compatible or a non
metric compatible connection. We have demanded that the connection is
torsion free. On the whole this is to simplify the equations so that
we do not have to write down all the torsion components and their
derivatives. One can reproduce the results with these extra terms and
it would be interesting to see how the results depend on the torsion.

\section*{Acknowledgement}

JG is grateful for the support provided
by STFC (the Cockcroft Institute ST/P002056/1) and
EPSRC (the Alpha-X project EP/N028694/1). ST and PP would like to thank
the Faculty of Science and Technology, Lancaster University for their support.
JG would like to thank the anonymous referee of
\cite{gratus2018correct} for suggesting applying the technique to
gravity.

We would like to thank Dr. David Burton for reading the manuscript and
making suggestions.

\bibliographystyle{unsrt}
\bibliography{bibliography}

\appendix

\section*{Appendix}

\section{Details of the proofs}

\subsection{Proof from introductory sections}

\def\freg{{\mathfrak{f}}}
\def\fdist{{f}}
\def\norm#1{{\lvert#1\rvert}}
\def\manU{{\cal U}}
\def\manV{{\cal V}}

\begin{Proof}{Proof of \eqref{Intro_lin_Ein_limit}: \jgnchange{gravitational
    waves from a distribution.}}
\label{pf_intro_GW_limit}
Fix components $\iMa\iMb$ with respect \jgDchange{to} the global Cartesian
coordinate system and set $\freg_\varepsilon$
and $\fdist$ to be the components
$\TReg_\varepsilon^{\iMa\iMb}$ and $(T^{(1)})^{\iMa\iMb}$.
Then $\freg_\varepsilon$ is a family of regular scalars with
support in a region $\manU$ about the worldline $C$ 
and $\fdist$ is a distribution on the worldline and 
$\freg_\varepsilon\to\fdist$ weakly. Given a point $(t,\vec{x})$ not
on the worldline and outside $\manU$, \jgDchange{let} $\manV$ be the 
intersection of the 
backward light cone of $(t,\vec{x})$ and $\manU$. The assumption
after (\ref{Intro_lin_Ein_limit}) implies  
$\manV$ is compact. 

\jgDchange{Introduce} an adapted coordinate system (\ref{Ellis_adap_Ellis_Multi})
which is adapted for both the worldline and the backward light cone of
the point $(t,\vec{x})$ and such that $z^\iSa=x^\iSa-x'{}^\iSa$.
 
Let $\chi(\Vz)$ be a test
function which coincides with 
${\norm{\Vz}}^{-1}$ on $\manV$ and let $\bumpf(\sigma)$ another test
function such that the support of $\bumpf(\sigma)\chi(\Vz)$ does not
include $(t,\vec{x})$. Then since $\bumpf(\sigma)\chi(\Vz)$ is a test function
\begin{align*}
\lim_{\epsilon\to 0}
\int_\Interval d\sigma\, \bumpf(\sigma)
\int \chi(\Vz) \,\freg_\varepsilon(\sigma,\Vz)\, d^3\Vz
&=
\lim_{\epsilon\to 0}
\int_\Interval 
\int \chi(\Vz)\bumpf(\sigma) \,\freg_\varepsilon(\sigma,\Vz) \,d^3\Vz\, d\sigma 
=
\int_\Interval 
\int \chi(\Vz)\bumpf(\sigma) \,\fdist(\sigma,\Vz) \,d^3\Vz\, d\sigma 
\\&=
\int_\Interval \bumpf(\sigma) \,d\sigma 
\int \chi(\Vz)\fdist(\sigma,\Vz) \,d^3\Vz
\DEfullstop
\end{align*}
Since this is true for all appropriate $\bumpf(\sigma)$ we have
\begin{align*}
\lim_{\epsilon\to 0}
\int \frac{\freg_\varepsilon(\sigma,\Vz)}{\norm{\Vz}} d^3\Vz
=
\lim_{\epsilon\to 0}
\int \chi(\Vz) \freg_\varepsilon(\sigma,\Vz) d^3\Vz
=
\int \chi(\Vz)\fdist(\sigma,\Vz) d^3\Vz
=
\int \frac{\fdist(\sigma,\Vz)}{\norm{\Vz}} d^3\Vz
\DEfullstop
\end{align*}
Hence (\ref{Intro_lin_Ein_limit}).
\end{Proof}


\subsection{Proofs about the quadrupole}

\begin{Proof}{Proof of \eqref{QP_Zeta_Change_Coords}:
    \jgnchange{change of coordinates for quadrupole}}
\label{pf_ChangCoordZeta}
\jgnchange{This is similar to the proof of
  (\ref{MD_diploe_Change_coords}).} 
Using \eqref{QP_Tab_action} we have 
\begin{align*}
\int_\Interval \zetahat^{\iMahat\iMbhat\iMchat\iMdhat} \,
&\big(\partx_\iMchat\,\partx_\iMdhat \,\TtwoTenhat_{\iMahat\iMbhat}\big)
\big|_{C(\sigma)}\ d\sigma
=
\int_{\Real^4} \THat^{\iMahat\iMbhat}\,\TtwoTenhat_{\iMahat\iMbhat}\,d^4\xhat
=
\int_{\Real^4} T^{\iMa\iMb}\,\TtwoTen_{\iMa\iMb}\,d^4x
=
\int_\Interval \zeta^{\iMa\iMb\iMc\iMd} \,
\big(\partx_\iMc\,\partx_\iMd \,\TtwoTen_{\iMa\iMb}\big) d\sigma
\\&=
\int_\Interval \zeta^{\iMa\iMb\iMc\iMd} \,
\partx_\iMc\,\partx_\iMd \,
\big({\Jaabb}\, \TtwoTenhat_{\iMahat\iMbhat}\big)
\ d\sigma
\\&=
\int_\Interval \zeta^{\iMa\iMb\iMc\iMd}\,
\Big(
\partx_\iMc\,\partx_\iMd \,\big({\Jaabb}\big)\, \TtwoTenhat_{\iMahat\iMbhat}
+
2\, \partx_\iMc \big({\Jaabb}\big)\, \partx_\iMd\,\TtwoTenhat_{\iMahat\iMbhat}
+
{\Jaabb}\ \partx_\iMc\,\partx_\iMd\, \TtwoTenhat_{\iMahat\iMbhat}
\Big)
\ d\sigma
\DEfullstop
\end{align*}
Take each of the terms in turn. For the third term we have
\begin{align*}
\int_\Interval \zeta^{\iMa\iMb\iMc\iMd}\,
&
{\Jaabb}\ \partx_\iMc\,\partx_\iMd\, \TtwoTenhat_{\iMahat\iMbhat}
\ d\sigma
=
\int_\Interval \zeta^{\iMa\iMb\iMc\iMd}\,
{\Jaabb}\ \partx_\iMc\,(J^\iMdhat_\iMd\,\partx_\iMdhat\, \TtwoTenhat_{\iMahat\iMbhat})
\ d\sigma
\\&=
\int_\Interval \zeta^{\iMa\iMb\iMc\iMd}\,
{\Jaabb}\ 
\Big((\partx_\iMc\,J^\iMdhat_\iMd)\,\partx_\iMdhat\, \TtwoTenhat_{\iMahat\iMbhat} + 
J^\iMdhat_\iMd\,\partx_\iMc\,\partx_\iMdhat\, \TtwoTenhat_{\iMahat\iMbhat} \Big)
\ d\sigma
\\&=
\int_\Interval \zeta^{\iMa\iMb\iMc\iMd}\,
{\Jaabb}\ 
(\partx_\iMc\,J^\iMdhat_\iMd)\,\partx_\iMdhat\, \TtwoTenhat_{\iMahat\iMbhat} \,d\sigma
+
\int_\Interval \zeta^{\iMa\iMb\iMc\iMd}\,
{\Jaabb}\ 
J^\iMchat_\iMc\, J^\iMdhat_\iMd\,\partx_\iMchat\,\partx_\iMdhat\,
\TtwoTenhat_{\iMahat\iMbhat}
\ d\sigma
\\&=
\int_\Interval \zeta^{\iMa\iMb\iMc\iMd} \,
{\Jaabb} (\partx_\iMc\,J^\iMdhat_\iMd) \,
\bigg(\int^\sigma \Cdothat^\iMchat\,\partx_\iMchat\,\partx_\iMdhat\,\TtwoTenhat_{\iMahat\iMbhat}\, d\sigma'\bigg)
\ d\sigma
+
\int_\Interval \zeta^{\iMa\iMb\iMc\iMd}\,
{\Jaabb}\ 
J^\iMchat_\iMc\, J^\iMdhat_\iMd\,\partx_\iMchat\,\partx_\iMdhat\,
\TtwoTenhat_{\iMahat\iMbhat}
\ d\sigma
\\&=
-\int_\Interval 
\bigg(\int^\sigma \zeta^{\iMa\iMb\iMc\iMd}\,
{\Jaabb} (\partx_\iMc\,J^\iMdhat_\iMd) \,d\sigma'\bigg)
\Cdothat^\iMchat\,\partx_\iMchat\,\partx_\iMdhat\,\TtwoTenhat_{\iMahat\iMbhat}\, 
\ d\sigma
+
\int_\Interval \zeta^{\iMa\iMb\iMc\iMd}\,
{\Jaabb}\ 
J^\iMchat_\iMc\, J^\iMdhat_\iMd\,\partx_\iMchat\,\partx_\iMdhat\,
\TtwoTenhat_{\iMahat\iMbhat}
\ d\sigma
\DEfullstop
\end{align*}
For the second term we have
\begin{align*}
\int_\Interval \zeta^{\iMa\iMb\iMc\iMd} \,
\partx_\iMc \big({\Jaabb}\big)\, \partx_\iMd\,\TtwoTenhat_{\iMahat\iMbhat}
\ d\sigma
&=
\int_\Interval \zeta^{\iMa\iMb\iMc\iMd} \,
\partx_\iMc \big({\Jaabb}\big)\, J^\iMdhat_\iMd \,\partx_\iMdhat\,\TtwoTenhat_{\iMahat\iMbhat}
\ d\sigma
\\&=
\int_\Interval \zeta^{\iMa\iMb\iMc\iMd} \,
\partx_\iMc \big({\Jaabb}\big)\, J^\iMdhat_\iMd \,
\bigg(\int^\sigma \Cdothat^\iMchat\,\partx_\iMchat\,\partx_\iMdhat\,\TtwoTenhat_{\iMahat\iMbhat}\, d\sigma'\bigg)
\ d\sigma
\\&=
-\int_\Interval 
\bigg(\int^\sigma \zeta^{\iMa\iMb\iMc\iMd}\,
\partx_\iMc \big({\Jaabb}\big)\, J^\iMdhat_\iMd \,d\sigma'\bigg)
\Cdothat^\iMchat\,\partx_\iMchat\,\partx_\iMdhat\,\TtwoTenhat_{\iMahat\iMbhat}\, 
\ d\sigma
\DEfullstop
\end{align*}
For the first term we have
\begin{align*}
\int_\Interval 
\zeta^{\iMa\iMb\iMc\iMd} \,
\partx_\iMc\,\partx_\iMd \,\big({\Jaabb}\big)\,
\TtwoTenhat_{\iMahat\iMbhat}
\ d\sigma
&=
\int_\Interval 
\zeta^{\iMa\iMb\iMc\iMd} \,
\partx_\iMc\,\partx_\iMd \,\big({\Jaabb}\big)\,
\bigg(\int^\sigma \Cdothat^\iMchat\,\partx_\iMchat\,\TtwoTenhat_{\iMahat\iMbhat}\, d\sigma'\bigg)
\ d\sigma
\\&=
-\int_\Interval 
\bigg(\int^\sigma \zeta^{\iMa\iMb\iMc\iMd} \,
\partx_\iMc\,\partx_\iMd \,\big({\Jaabb}\big)\,d\sigma'\bigg)\,
\Cdothat^\iMchat\,\partx_\iMchat\,\TtwoTenhat_{\iMahat\iMbhat}\, 
\ d\sigma
\\&=
-\int_\Interval 
\bigg(\int^\sigma \zeta^{\iMa\iMb\iMc\iMd} \,
\partx_\iMc\,\partx_\iMd \,\big({\Jaabb}\big)\,d\sigma''\bigg)\,
\Cdothat^\iMchat\,
\Big(\int^{\sigma} \partx_\iMchat\,\Cdothat^\iMdhat\,\partx_\iMdhat\,\TtwoTenhat_{\iMahat\iMbhat}\,d\sigma'\Big) 
\ d\sigma
\\&=
\int_\Interval 
\bigg(\int^{\sigma} \Big(\int^{\sigma'} \zeta^{\iMa\iMb\iMc\iMd} \,
\partx_\iMc\,\partx_\iMd \,\big({\Jaabb}\big)\,d\sigma''\Big)\,
\Cdothat^\iMchat\,d\sigma'\bigg) 
\Cdothat^\iMdhat\,\partx_\iMchat\,\partx_\iMdhat\,\TtwoTenhat_{\iMahat\iMbhat}
\ d\sigma
\\&=
\int_\Interval 
\bigg(\Cdothat^\iMdhat\,\int^{\sigma} 
\Big(\Cdothat^\iMchat\,\int^{\sigma'} \zeta^{\iMa\iMb\iMc\iMd} \,
\partx_\iMc\,\partx_\iMd \,\big({\Jaabb}\big)\,d\sigma''\Big)\,
d\sigma'\bigg) 
\partx_\iMchat\,\partx_\iMdhat\,\TtwoTenhat_{\iMahat\iMbhat}
\ d\sigma
\DEfullstop
\end{align*}
Thus adding these terms together we have
\begin{align*}
\int_\Interval \zetahat^{\iMahat\iMbhat\iMchat\iMdhat} \,
&
\big(\partx_\iMchat\,\partx_\iMdhat \,\TtwoTenhat_{\iMahat\iMbhat}\big)
\ d\sigma
=
\int_\Interval \zeta^{\iMa\iMb\iMc\iMd}\,
\Big(
\partx_\iMc\,\partx_\iMd \,\big({\Jaabb}\big)\, \TtwoTenhat_{\iMahat\iMbhat}
+
2\, \partx_\iMc \big({\Jaabb}\big)\, \partx_\iMd\,\TtwoTenhat_{\iMahat\iMbhat}
+
{\Jaabb}\ \partx_\iMc\,\partx_\iMd\, \TtwoTenhat_{\iMahat\iMbhat}
\Big)
\ d\sigma
\\&=
-\int_\Interval 
\bigg(\int^\sigma \zeta^{\iMa\iMb\iMc\iMd}\,
{\Jaabb} (\partx_\iMc\,J^\iMdhat_\iMd) \,d\sigma'\bigg)
\Cdothat^\iMchat\,\partx_\iMchat\,\partx_\iMdhat\,\TtwoTenhat_{\iMahat\iMbhat}\, 
\ d\sigma
+
\int_\Interval \zeta^{\iMa\iMb\iMc\iMd}\,
{\Jaabb}\ 
J^\iMchat_\iMc\, J^\iMdhat_\iMd\,\partx_\iMchat\,\partx_\iMdhat\,
\TtwoTenhat_{\iMahat\iMbhat}
\ d\sigma
\\&\quad
-
2\int_\Interval 
\bigg(\int^\sigma \zeta^{\iMa\iMb\iMc\iMd}\,
\partx_\iMc \big({\Jaabb}\big)\, J^\iMdhat_\iMd \,d\sigma'\bigg)
\Cdothat^\iMchat\,\partx_\iMchat\,\partx_\iMdhat\,\TtwoTenhat_{\iMahat\iMbhat}\, 
\ d\sigma
\\&\quad
+
\int_\Interval 
\bigg(\Cdothat^\iMdhat\,\int^{\sigma} 
\Big(\Cdothat^\iMchat\,\int^{\sigma'} \zeta^{\iMa\iMb\iMc\iMd} \,
\partx_\iMc\,\partx_\iMd \,\big({\Jaabb}\big)\,d\sigma''\Big)\,
d\sigma'\bigg) 
\partx_\iMchat\,\partx_\iMdhat\,\TtwoTenhat_{\iMahat\iMbhat}
\ d\sigma
\\&=
\int_\Interval \bigg(
\zeta^{\iMa\iMb\iMc\iMd}\,
{\Jaabb}\ 
J^\iMchat_\iMc\, J^\iMdhat_\iMd
-
\Cdothat^\iMchat
\int^\sigma \zeta^{\iMa\iMb\iMc\iMd}\,
{\Jaabb} (\partx_\iMc\,J^\iMdhat_\iMd) \,d\sigma'
-
2\Cdothat^\iMchat\int^\sigma \zeta^{\iMa\iMb\iMc\iMd}\,
\partx_\iMc \big({\Jaabb}\big)\, J^\iMdhat_\iMd \,d\sigma'
\\&\qquad\qquad
+
\Cdothat^\iMdhat\,\int^{\sigma} 
\Big(\Cdothat^\iMchat\,\int^{\sigma'} \zeta^{\iMa\iMb\iMc\iMd} \,
\partx_\iMc\,\partx_\iMd \,\big({\Jaabb}\big)\,d\sigma''\Big)\,
d\sigma'\bigg) 
\partx_\iMchat\,\partx_\iMdhat\,\TtwoTenhat_{\iMahat\iMbhat}
\ d\sigma
\DEfullstop
\end{align*}
Hence \eqref{QP_Zeta_Change_Coords} follows by symmetrising $\iMchat$
and $\iMdhat$.
\end{Proof}

\begin{Proof}{Proof that the change of coordinates 
\eqref{QP_Zeta_Change_Coords} is consistent with the gauge-like freedom
\eqref{QP_Zeta_Freedom}}
\label{pf_ChangCoordZeta_Gauge}

First observe that the lower limits in (\ref{QP_Zeta_Change_Coords})
correspond \jgDchange{to the} gauge-like freedom (\ref{QP_Zeta_Freedom}) for
$\zetahat^{\iMahat\iMbhat\iMchat\iMdhat}$.

It is necessary to establish that the gauge-like freedom
(\ref{QP_Zeta_Freedom}) for $\zeta^{\iMa\iMb\iMc\iMd}$ when substituted into
(\ref{QP_Zeta_Change_Coords}) does not \jgDchange{affect} the value of
$\zetahat^{\iMahat\iMbhat\iMchat\iMdhat}$. This is achieved by setting
$\zeta^{\iMa\iMb\iMc\iMd}=\Mone^{\iMb\iMa}\, \Cdot^{\lround\iMc}\,
C^{\iMd\rround} + \Mtwo^{\iMa\iMb\lround\iMc}\,\Cdot^{\iMd\rround}$,
i.e. $\zeta^{\iMa\iMb\iMc\iMd}$ is equivalent to zero, and checking
that $\zetahat^{\iMahat\iMbhat\iMchat\iMdhat}=0$. As they are
independent, we can consider the two terms $\Mone^{\iMb\iMa}\,
\Cdot^{\lround\iMc}\, C^{\iMd\rround}$ and
$\Mtwo^{\iMa\iMb\lround\iMc}\,\Cdot^{\iMd\rround}$ separately.

For the case $\zeta^{\iMa\iMb\iMc\iMd}=\Mone^{\iMb\iMa}\,
\Cdot^{\lround\iMc}\,C^{\iMd\rround}$ we have for the fifth term on
the right hand side of (\ref{QP_Zeta_Change_Coords}) 
\begin{align*}
 \int^{\sigma} 
&
\Cdothat^{\iMdhat}\int^{\sigma'} \Cdot^{\lround\iMc} C^{\iMd\rround} \,
\partx_\iMc\,\partx_\iMd \,\big({\Jaabb}\big)\,d\sigma''\,
d\sigma'
=
 \int^{\sigma} 
\Cdothat^{\iMdhat}\int^{\sigma'} \Cdot^{\iMc} C^\iMd \,
\partx_\iMc\,\partx_\iMd \,\big({\Jaabb}\big)\,d\sigma''\,
d\sigma'
\\&=
 \int^{\sigma} 
\Cdothat^{\iMdhat}\int^{\sigma'} C^\iMd \,
\dfrac{}{\sigma''} \Big(\partx_\iMd
\,{\Jaabb}\Big)
\,d\sigma''\,
d\sigma'
\\&=
 \int^{\sigma} 
\Cdothat^{\iMdhat}\int^{\sigma'} 
\dfrac{}{\sigma''} \Big(
C^\iMd 
\partx_\iMd
\,{\Jaabb}\Big)
\,d\sigma''\,
d\sigma'
-
 \int^{\sigma} 
\Cdothat^{\iMdhat}\int^{\sigma'} 
\Cdot^\iMd 
\partx_\iMd
\,{\Jaabb}
\,d\sigma''\,
d\sigma'
\\&=
 \int^{\sigma} 
\Cdothat^{\iMdhat}
C^\iMd 
\partx_\iMd
\,{\Jaabb}\ 
d\sigma'
-
 \int^{\sigma} 
\Cdothat^{\iMdhat}\int^{\sigma'} 
\dfrac{}{\sigma''}
\,{\Jaabb}
\,d\sigma''\,
d\sigma'
\\&=
 \int^{\sigma} 
\Cdothat^{\iMdhat}
C^\iMd 
\partx_\iMd
\,{\Jaabb}\ 
d\sigma'
-
 \int^{\sigma} 
\Cdothat^{\iMdhat}
\,{\Jaabb}
\,d\sigma'
\\&=
 \int^{\sigma} 
\Cdothat^{\iMdhat}
C^\iMd 
\partx_\iMd
\,{\Jaabb}\ 
d\sigma'
-
 \int^{\sigma} 
\Cdot^\iMd J^{\iMdhat}_\iMd
\,{\Jaabb}
\,d\sigma'
\\&=
 \int^{\sigma} 
\Cdot^\iMd J^{\iMdhat}_\iMd
C^\iMc 
\partx_\iMc
\,{\Jaabb}\ 
d\sigma'
-
C^\iMd J^{\iMdhat}_\iMd
\,{\Jaabb}
+
 \int^{\sigma} 
C^\iMd \dfrac{}{\sigma'} 
(J^{\iMdhat}_\iMd
\,{\Jaabb})
\,d\sigma'
\DEfullstop
\end{align*}
Since
\begin{align*}
\int^\sigma \Cdot^{\iMd} C^\iMc \,
{\Jaabb} \partx_\iMc\,J^{\iMdhat}_\iMd\,d\sigma'
&=
\int^\sigma \Cdot^{\iMd} C^\iMc \,
{\Jaabb} \partx_\iMd\,J^{\iMdhat}_\iMc\,d\sigma'
=
 \int^\sigma \Cdot^{\iMc} C^\iMd \,
{\Jaabb} \partx_\iMc\,J^{\iMdhat}_\iMd\,d\sigma'
\DEcomma
\end{align*}
while for the second term in (\ref{QP_Zeta_Change_Coords})
\begin{align*}
\int_\Interval & \big(\Mtwo^{\iMa\iMb\lround\iMc}\,\Cdot^{\iMd\rround}\big)\,
\Big({\Jaabb} (\partx_\iMc\,J^{\iMdhat}_\iMd)+
2\,\partx_\iMc\,({\Jaabb})\,J^{\iMdhat}_\iMd\Big) \,d\sigma'
\\&=
\tfrac12
\Mtwo^{\iMa\iMb\iMc}\,\int_\Interval
\Cdot^{\iMd}
\Big({\Jaabb} (\partx_\iMc\,J^{\iMdhat}_\iMd)+
2\,\partx_\iMc\,({\Jaabb})\,J^{\iMdhat}_\iMd\Big) \,d\sigma'
+
\tfrac12
\Mtwo^{\iMa\iMb\iMd}\,\int_\Interval\Cdot^{\iMc}
\Big({\Jaabb} (\partx_\iMc\,J^{\iMdhat}_\iMd)+
2\,\partx_\iMc\,({\Jaabb})\,J^{\iMdhat}_\iMd\Big) \,d\sigma'
\\&=
\Mtwo^{\iMa\iMb\iMc}\,\int_\Interval
\Cdot^{\iMd}\,J^{\iMdhat}_\iMd\,(\partx_\iMc\,{\Jaabb}) \,d\sigma'
+
\Mtwo^{\iMa\iMb\iMd}\,\int_\Interval\Cdot^{\iMc}
\Big({\Jaabb} (\partx_\iMc\,J^{\iMdhat}_\iMd)+
\partx_\iMc\,({\Jaabb})\,J^{\iMdhat}_\iMd\Big) \,d\sigma'
\\&=
\Mtwo^{\iMa\iMb\iMc}\,\int_\Interval 
\Cdothat^{\iMdhat}(\partx_\iMc\,{\Jaabb}) \,d\sigma'
+
\Mtwo^{\iMa\iMb\iMd}\,\int_\Interval
\dfrac{}{\sigma'}\Big({\Jaabb} \,J^{\iMdhat}_\iMd\Big)\,d\sigma'
\\&=
\Mtwo^{\iMa\iMb\iMc}\,\int_\Interval 
\Cdothat^{\iMdhat}(\partx_\iMc\,{\Jaabb}) \,d\sigma'
+
\Mtwo^{\iMa\iMb\iMd}\,{\Jaabb} \,J^{\iMdhat}_\iMd
\DEfullstop
\end{align*}
Hence when
$\zeta^{\iMa\iMb\iMc\iMd}=\Mtwo^{\iMa\iMb\lround\iMc}\,\Cdot^{\iMd\rround}$
then $\zetahat^{\iMahat\iMbhat\iMchat\iMdhat}=0$.
\end{Proof}

\begin{Proof}{Proof that \eqref{QP_Zeta_Freedom} incorporates all the
    gauge-like freedom}
\label{pf_Quad_zeta_all_gauge}
Assume $T^{\iMa\iMb}$ is given. 
From (\ref{Ellis_adap_extract_comp}) we know that
the components $\gamma^{\iMa\iMb\iMc\iMd}$ are unique, i.e. have no
gauge-like freedom. Integrating (\ref{QP_gamma_zeta}) we have 
\begin{align*}
\zeta^{\iMa\iMb\iMc\iMd} \to 
\zeta^{\iMa\iMb\iMc\iMd} + 
\sigma \Mone^{\iMb\iMa}\, \delta^\iMc_0\,\delta^\iMd_0
+
\Mtwo^{\iMa\iMb\lround\iMd} \delta^{\iMc\rround}_0
\DEcomma
\end{align*}
which is (\ref{QP_Zeta_Freedom}) in adapted coordinates. Hence
(\ref{QP_Zeta_Freedom}) is incorporates all gauge-like freedom, in adapted
coordinates. Now for a general coordinate system we use
(\ref{QP_Zeta_Change_Coords}). We see in the proof
\ref{pf_ChangCoordZeta_Gauge} in the appendix, that
(\ref{QP_Zeta_Change_Coords}) is consistent with the gauge-like
freedom. Thus there is no additional gauge-like freedom in a general
coordinate system.
\end{Proof}


\newcommand{\non}{\nonumber\\}
\newcommand{\be}{\begin{equation}}
\newcommand{\ee}{\end{equation}}
\def\zetadothat{\hat{\dot\zeta}}
\newcommand{\ga}{\gamma}
\newcommand{\za}{\zeta}

\begin{Proof}{Proof of
    \eqref{QP_gamma_chage_coords_munu}-\eqref{QP_gamma_chage_coords_00}
\jgnchange{The coordinate transformation for adapted coordinates.}
}
\label{pf_gamma_change_coords}
This follows from substituting \eqref{QP_gamma_zeta} into  
\eqref{QP_Zeta_Change_Coords}. 

We set $(x^0,\ldots,x^3)=(\sigma,z^1,z^2,z^3)$ and
$(\xhat^\Ohat\ldots\xhat^{\hat{3}})=
(\hat\sigma,\zhat^{\hat{1}},\zhat^{\hat{2}},\zhat^{\hat{2}})$ into
\eqref{QP_Zeta_Change_Coords} and use the fact that
$\Cdothat^{\iMahat}=\delta^\iMahat_0$. Hence
\eqref{QP_gamma_chage_coords_munu} follows directly.

For \eqref{ChangeCoords_gamma_abmu0} we have from
\eqref{QP_Zeta_Change_Coords}

\begin{align*}
\zetahat^{\iMahat\iMbhat\iSchat\Ohat} 
&=
\zeta^{\iMa\iMb\iMc\iMd}\,
{\Jaabb}\,
J^{\iSchat}_\iMc\, J^{\Ohat}_\iMd
-\tfrac12
\Cdothat^\iSchat\int^\sigma \zeta^{\iMa\iMb\iMc\iMd}\,
\Big({\Jaabb} (\partx_\iMc\,J^{\Ohat}_\iMd)+
2\,\partx_\iMc\,({\Jaabb})\,J^{\Ohat}_\iMd\Big) \,d\sigma'
\\&\qquad
-\tfrac12
\Cdothat^\Ohat\int^\sigma \zeta^{\iMa\iMb\iMc\iMd}\,
\Big({\Jaabb} (\partx_\iMc\,J^{\iSchat}_\iMd)+
2\,\partx_\iMc\,({\Jaabb})\,J^{\iSchat}_\iMd\Big) \,d\sigma'
\\&\quad
+\tfrac12
\Cdothat^{\Ohat}\int^{\sigma} 
\Cdothat^{\iSchat}\int^{\sigma'} \zeta^{\iMa\iMb\iMc\iMd} \,
\partx_\iMc\,\partx_\iMd \,\big({\Jaabb}\big)\,d\sigma''\,
d\sigma'
+\tfrac12
\Cdothat^{\iSchat}\int^{\sigma} 
\Cdothat^{\Ohat}\int^{\sigma'} \zeta^{\iMa\iMb\iMc\iMd} \,
\partx_\iMc\,\partx_\iMd \,\big({\Jaabb}\big)\,d\sigma''\,
d\sigma'
\\&=
\zeta^{\iMa\iMb\iMc\iMd}\,
{\Jaabb}\,
J^{\iSchat}_\iMc\, J^{\Ohat}_\iMd
-\tfrac12
\int^\sigma \zeta^{\iMa\iMb\iMc\iMd}\,
\Big({\Jaabb} (\partx_\iMc\,J^{\iSchat}_\iMd)+
2\,\partx_\iMc\,({\Jaabb})\,J^{\iSchat}_\iMd\Big) \,d\sigma'
\DEfullstop
\end{align*}
Thus from \eqref{QP_gamma_zeta}
\begin{align*}
\gammahat^{\iMahat\iMbhat\iSchat\Ohat} 
&=
\dot\zetahat^{\iMahat\iMbhat\iSchat\Ohat} 
=
(\zeta^{\iMa\iMb\iMc\iMd}\,
{\Jaabb}\,
J^{\iSchat}_\iMc\, J^{\Ohat}_\iMd)\dot{}
-\tfrac12
\zeta^{\iMa\iMb\iMc\iMd}\,
\big({\Jaabb} \,\JJ^{\iSchat}_{\iMc\iMd}+
2\,\partx_\iMc\,({\Jaabb})\,J^{\iSchat}_\iMd\big)
\\&=
(\zeta^{\iMa\iMb00}\,{\Jaabb}\,J^{\iSchat}_0\,
J^{\Ohat}_0)\dot{} 
+
(\zeta^{\iMa\iMb\iSc0}\,{\Jaabb}\,J^{\iSchat}_\iSc\,
J^{\Ohat}_0)\dot{} 
+
(\zeta^{\iMa\iMb0\iSc}\,{\Jaabb}\,J^{\iSchat}_0\,
J^{\Ohat}_\iSc)\dot{} 
+
(\zeta^{\iMa\iMb\iSc\iSd}\,{\Jaabb}\,J^{\iSchat}_\iSc\,
J^{\Ohat}_\iSd)\dot{} 
\\&\quad
-
\tfrac12
\zeta^{\iMa\iMb00}\,
\big({\Jaabb} \,\JJ^{\iSchat}_{00}+
2\,\partx_0\,({\Jaabb})\,J^{\iSchat}_0\big)
-
\tfrac12
\zeta^{\iMa\iMb\iSc0}\,
\big({\Jaabb} \,\JJ^{\iSchat}_{\iSc0}+
2\,\partx_\iSc\,({\Jaabb})\,J^{\iSchat}_0\big)
\\&\quad
-
\tfrac12
\zeta^{\iMa\iMb\iSc0}\,
\big({\Jaabb} \,\JJ^{\iSchat}_{\iSc0}+
2\,\partx_0\,({\Jaabb})\,J^{\iSchat}_\iSc\big)
-
\tfrac12
\zeta^{\iMa\iMb\iSc\iSd}\,
\big({\Jaabb} \,\JJ^{\iSchat}_{\iSc\iSd}+
2\,\partx_\iSc\,({\Jaabb})\,J^{\iSchat}_\iSd\big)
\\&=
(\zeta^{\iMa\iMb\iSc0}\,{\Jaabb}\,J^{\iSchat}_\iSc)\dot{} 
+
(\zeta^{\iMa\iMb\iSc\iSd}\,{\Jaabb}\,J^{\iSchat}_\iSc\,
J^{\Ohat}_\iSd)\dot{} 
\\&\quad
-
\zeta^{\iMa\iMb\iSc0}\,
\big({\Jaabb} \,\JJ^{\iSchat}_{\iSc0}+
{\Jdotaabb}\,J^{\iSchat}_\iSc\big)
-
\tfrac12
\zeta^{\iMa\iMb\iSc\iSd}\,
\big({\Jaabb} \,\JJ^{\iSchat}_{\iSc\iSd}+
2\,\partx_\iSc\,({\Jaabb})\,J^{\iSchat}_\iSd\big)
\\&=
\dot\zeta^{\iMa\iMb\iSc0}\,{\Jaabb}\,J^{\iSchat}_\iSc
+
(\zeta^{\iMa\iMb\iSc\iSd}\,{\Jaabb}\,J^{\iSchat}_\iSc\,
J^{\Ohat}_\iSd)\dot{} 
-
\tfrac12
\zeta^{\iMa\iMb\iSc\iSd}\,
\big({\Jaabb} \,\JJ^{\iSchat}_{\iSc\iSd}+
2\,\partx_\iSc\,({\Jaabb})\,J^{\iSchat}_\iSd\big)
\\&=
\gamma^{\iMa\iMb\iSc0}\,{\Jaabb}\,J^{\iSchat}_\iSc
+
(\gamma^{\iMa\iMb\iSc\iSd}\,{\Jaabb}\,J^{\iSchat}_\iSc\,
J^{\Ohat}_\iSd)\dot{} 
-
\tfrac12
\gamma^{\iMa\iMb\iSc\iSd}\,
\big({\Jaabb} \,\JJ^{\iSchat}_{\iSc\iSd}+
2\,\partx_\iSc\,({\Jaabb})\,J^{\iSchat}_\iSd\big)
\DEcomma
\end{align*}
\jgnchange{where $ \JJ^{\iMahat}_{\iMb\iMc} = \partial_{\iMb}
J^{\iMahat}_{\iMc}$.}


In order to show \eqref{QP_gamma_chage_coords_00} we have from
\eqref{QP_Zeta_Change_Coords}
\begin{align*}
\zetahat^{\iMa \iMb \Ohat\Ohat}
& =
(\Jaabb  J^\Ohat_{\iMc} \, J^\Ohat_{\iMd})\, \zeta^{\iMa \iMb \iMc \iMd}
-
\int^{\sigma} 
\Big(
(\partx_{\iMd}  J^\Ohat_{\iMc})\,J_{\mu \nu}^{\hat{\mu}\hat{\nu}}+
J^\Ohat_{\iMc}\, \partx_{\iMd} J_{\mu \nu}^{\hat{\mu}\hat{\nu}}+
J^{\Ohat}_{\iMd}\, \partx_{\iMc} J_{\mu \nu}^{\hat{\mu}\hat{\nu}}
\Big) \zeta^{\iMa \iMb \iMc \iMd}\,d\sigma'
\\&\qquad
+ \int^\sigma 
\hspace{-.5em} 
d\sigma' \, 
\int^{\sigma'} \partx_{\iMc \iMd} \Jaabb \, \zeta^{\iMa \iMb \iMc \iMd}\, d\sigma''
\DEcomma
\end{align*}
where $\partx_{\iMc \iMd}=\partx_{\iMc}\partx_{\iMd}$. Hence
\begin{equation}
\begin{aligned}
\gammahat^{\hat{\iMa}\hat{\iMb}\Ohat\Ohat}
&=
\tfrac{1}{2}\hat{\ddot{\zeta}}^{\hat{\iMa} \hat{\iMb}\Ohat\Ohat}
\\&=
\tfrac{1}{2}\Big(
\big((\Jaabb\ J^\Ohat_{\iMd} \, J^\Ohat_{\iMc})\, \zeta^{\iMa \iMb \iMc \iMd}\big)\ddot{}
-
\big((
(\partx_{\iMd}  J^\Ohat_{\iMc})\,{\Jaabb}+
J^\Ohat_{\iMc}\, \partx_{\iMd} {\Jaabb}+
J^\Ohat_{\iMd}\, \partx_{\iMc} {\Jaabb}
)
\zeta^{\iMa \iMb \iMc \iMd}\big)\dot{}
+  \partx_{\iMc \iMd}{\Jaabb}\, \zeta^{\iMa \iMb \iMc \iMd}\Big)
\DEfullstop
\end{aligned}
\label{ChangeCoords_zeta_changeNNmm}
\end{equation}
It is important to establish that all the $\zeta^{\iMa \iMb \iMc \iMd}$ on
the right hand side of
\eqref{ChangeCoords_zeta_changeNNmm} can be replaced by the corresponding
$\gamma^{\iMa\iMb\iMc\iMd}$ without using integrals. However since
from \eqref{QP_gamma_zeta}
$\gamma^{\iMa\iMb00}=\tfrac12\ddot\zeta^{\iMa \iMb00}$ and
$\gamma^{\iMa\iMb\iSa0}=\dot\zeta^{\iMa \iMb\iSa0}$ we need to
expand \eqref{ChangeCoords_zeta_changeNNmm} to confirm that no
terms $\zeta^{\iMa \iMb00}$, $\dot\zeta^{\iMa \iMb00}$ or
$\zeta^{\iMa\iMb\iSa0}$ exist on the right hand side.

\begin{align*}
\gammahat^{\hat{\iMa}\hat{\iMb}\Ohat\Ohat}
&=
\tfrac12
\Big((\Jaabb\ J^\Ohat_{\iMd} \, J^\Ohat_{\iMc})\, \zeta^{\iMa \iMb \iMc \iMd}\Big)\ddot{}
-
\Big(\big(
\tfrac12\JJ^\Ohat_{\iMc\iMd}\,{\Jaabb}+
J^\Ohat_{\iMd}\, \partx_{\iMc} {\Jaabb}
\big)
\zeta^{\iMa \iMb \iMc \iMd}\Big)\dot{}
+  (\tfrac12\partx_{\iMc \iMd}{\Jaabb}\, )\zeta^{\iMa \iMb \iMc \iMd}
\\&=
\tfrac12\Big((\Jaabb\ J^\Ohat_{0} \, J^\Ohat_{0})\, 
\zeta^{\iMa \iMb 00}\Big)\ddot{}
+
\Big((\Jaabb\ J^\Ohat_{\iSc} \, J^\Ohat_{0})\, \zeta^{\iMa \iMb \iSc 0}\Big)\ddot{}
+
\tfrac12\Big((\Jaabb\ J^\Ohat_{\iSd} \, J^\Ohat_{\iSc})\, 
\zeta^{\iMa  \iMb \iSc \iSd}\Big)\ddot{}
\\&\quad
-
\Big(\big(
\tfrac12\JJ^\Ohat_{00}\,{\Jaabb}+
J^\Ohat_{0}\, \partx_{0} {\Jaabb}
\big)
\zeta^{\iMa \iMb 00}\Big)\dot{}
-
\Big(\big(
\JJ^\Ohat_{0\iSc}\,{\Jaabb}+
J^\Ohat_{\iSc}\, \partx_{0} {\Jaabb} +
J^\Ohat_{0}\, \partx_{\iSc} {\Jaabb}
\big)
\zeta^{\iMa \iMb \iSc 0}\Big)\dot{}
\\&\quad
-
\Big(\big(
\tfrac12 \JJ^\Ohat_{\iSc\iSd}\,{\Jaabb}+
J^\Ohat_{\iSc}\, \partx_{\iSd} {\Jaabb}
\big)
\zeta^{\iMa \iMb \iSc \iSd}\Big)\dot{}
\\&\quad
+  (\tfrac12\partx_{00}{\Jaabb})\, \zeta^{\iMa \iMb 00}
+  (\partx_{0 \iSc}{\Jaabb})\, \zeta^{\iMa \iMb \iSc 0}
+  (\tfrac12\partx_{\iSc \iSd}{\Jaabb})\, \zeta^{\iMa \iMb \iSc \iSd}
\\&=
\tfrac12(\Jaabb\, 
\zeta^{\iMa \iMb 00})\ddot{}
+
\Big((\Jaabb\ J^\Ohat_{\iSc} \, \zeta^{\iMa \iMb \iSc 0}\Big)\ddot{}
+
\tfrac12\Big((\Jaabb\ J^\Ohat_{\iSd} \, J^\Ohat_{\iSc})\, 
\zeta^{\iMa  \iMb \iSc \iSd}\Big)\ddot{}
-
\Big(
{\Jdotaabb}
\zeta^{\iMa \iMb 00}\Big)\dot{}
\\&\quad
-
\Big(\big(
\JJ^\Ohat_{0\iSc}\,{\Jaabb}+
J^\Ohat_{\iSc}\, {\Jdotaabb}+
\partx_{\iSc} {\Jaabb}
\big)
\zeta^{\iMa \iMb \iSc 0}\Big)\dot{}
-
\Big(\big(
\tfrac12
\JJ^\Ohat_{\iSc\iSd}\,{\Jaabb}+
J^\Ohat_{\iSd}\, \partx_{\iSc} {\Jaabb}
\big)
\zeta^{\iMa \iMb \iSc \iSd}\Big)\dot{}
\\&\quad
+  \tfrac12\,{\Jddotaabb}\, \zeta^{\iMa \iMb 00}
+  (\partx_{\iSc}{\Jdotaabb})\, \zeta^{\iMa \iMb \iSc 0}
+  (\tfrac12\partx_{\iSc \iSd}{\Jaabb})\, \zeta^{\iMa \iMb \iSc \iSd}
\\&=
\tfrac12 \Jddotaabb\,\zeta^{\iMa \iMb 00}
+
\Jdotaabb\,\dot\zeta^{\iMa \iMb 00}
+
\tfrac12
\Jaabb\,\ddot\zeta^{\iMa \iMb 00}
\\&\quad
+
\Jddotaabb\ J^\Ohat_{\iSc} \, \zeta^{\iMa \iMb \iSc 0}
+
\Jaabb\ \JJ^\Ohat_{\iSc00} \, \zeta^{\iMa \iMb \iSc 0}
+
\Jaabb\ J^\Ohat_{\iSc} \, \ddot\zeta^{\iMa \iMb \iSc 0}
+
2\Jdotaabb\ \JJ^\Ohat_{\iSc0} \, \zeta^{\iMa \iMb \iSc 0}
+
2\Jdotaabb\ J^\Ohat_{\iSc} \, \dot\zeta^{\iMa \iMb \iSc 0}
+
2\Jaabb\ \JJ^\Ohat_{\iSc0} \, \dot\zeta^{\iMa \iMb \iSc 0}
\\&\quad
+
\tfrac12\Big((\Jaabb\ J^\Ohat_{\iSd} \, J^\Ohat_{\iSc})\, 
\zeta^{\iMa  \iMb \iSc \iSd}\Big)\ddot{}
-
{\Jddotaabb}
\zeta^{\iMa \iMb 00}
-
{\Jdotaabb}
\dot\zeta^{\iMa \iMb 00}
\\&\quad
-
\big(
\JJ^\Ohat_{00\iSc}\,{\Jaabb}+
2\JJ^\Ohat_{0\iSc}\,{\Jdotaabb}+
J^\Ohat_{\iSc}\, {\Jddotaabb}+
\partx_{0\iSc} {\Jaabb}
\big)
\zeta^{\iMa \iMb \iSc 0}
-
\big(
\JJ^\Ohat_{0\iSc}\,{\Jaabb}+
J^\Ohat_{\iSc}\, {\Jdotaabb}+
\partx_{\iSc} {\Jaabb}
\big)
\dot\zeta^{\iMa \iMb \iSc 0}
\\&\quad
-
\Big(\big(
\tfrac12
\JJ^\Ohat_{\iSc\iSd}\,{\Jaabb}+
J^\Ohat_{\iSd}\, \partx_{\iSc} {\Jaabb}
\big)
\zeta^{\iMa \iMb \iSc \iSd}\Big)\dot{}
\\&\quad
+  \tfrac12\,{\Jddotaabb}\, \zeta^{\iMa \iMb 00}
+  (\partx_{\iSc}{\Jdotaabb})\, \zeta^{\iMa \iMb \iSc 0}
+  (\tfrac12\partx_{\iSc \iSd}{\Jaabb})\, \zeta^{\iMa \iMb \iSc \iSd}
\\&=
\tfrac12
\Jaabb\,\ddot\zeta^{\iMa \iMb 00}
+
\Jaabb\ J^\Ohat_{\iSc} \, \ddot\zeta^{\iMa \iMb \iSc 0}
+
2\Jdotaabb\ J^\Ohat_{\iSc} \, \dot\zeta^{\iMa \iMb \iSc 0}
+
2\Jaabb\ \JJ^\Ohat_{\iSc0} \, \dot\zeta^{\iMa \iMb \iSc 0}
+
\tfrac12\Big((\Jaabb\ J^\Ohat_{\iSd} \, J^\Ohat_{\iSc})\, 
\zeta^{\iMa  \iMb \iSc \iSd}\Big)\ddot{}
\\&\quad
-
\big(
\JJ^\Ohat_{0\iSc}\,{\Jaabb}+
J^\Ohat_{\iSc}\, {\Jdotaabb}+
\partx_{\iSc} {\Jaabb}
\big)
\dot\zeta^{\iMa \iMb \iSc 0}
-
\Big(\big(
\tfrac12
\JJ^\Ohat_{\iSc\iSd}\,{\Jaabb}+
J^\Ohat_{\iSd}\, \partx_{\iSc} {\Jaabb}
\big)
\zeta^{\iMa \iMb \iSc \iSd}\Big)\dot{}
+  (\tfrac12\partx_{\iSc \iSd}{\Jaabb})\, \zeta^{\iMa \iMb \iSc \iSd}
\\&=
\Jaabb\,\gamma^{\iMa \iMb 00}
+
\Jaabb\,J^\Ohat_{\iSc} \,\dot\gamma^{\iMa\iMb\iSa0}
+
\big(
(\Jaabb\ \JJ^\Ohat_{\iSc})\dot{} 
-
\partx_{\iSc} {\Jaabb}
\big)
\gamma^{\iMa \iMb \iSc 0}
\\&\quad
+
\tfrac12\big((\Jaabb\ J^\Ohat_{\iSd} \, J^\Ohat_{\iSc})\, 
\gamma^{\iMa  \iMb \iSc \iSd}\big)\ddot{}
-
\big(\big(
\tfrac12
\JJ^\Ohat_{\iSc\iSd}\,{\Jaabb}+
J^\Ohat_{\iSd}\, \partx_{\iSc} {\Jaabb}
\big)
\gamma^{\iMa \iMb \iSc \iSd}\big)\dot{}
+  (\tfrac12\partx_{\iSc \iSd}{\Jaabb})\, \gamma^{\iMa \iMb \iSc \iSd}
\DEfullstop
\end{align*}
\end{Proof}

\def\DixonII/{{[\textbf{II}]}}
\def\DixonIII/{{[\textbf{III}]}}
\def\DixonIIe#1{{[\textbf{II}{#1}]}}
\def\DixonIIIe#1{{[\textbf{III}{#1}]}}

\section{Dixon's independent components}
\label{ch_Appx_Dixon}
In the following we refer to \cite{DixonII} as \DixonII/ and
\cite{DixonIII} as \DixonIII/.

It may appear that the 20 free components in this article directly
correspond to the 20 independent components of $J^{\iMa\iMb\iMc\iMd}$
given in \DixonIIIe{(1.37)}. This follows because as well as both
having 20 quadrupole components, they both arise from the
divergenceless condition (\ref{Intro_Tab_Div_zero}). 

We can relate the Dixon moments to our moments as follows. In
\DixonIIIe{(10.17)} we see the term $I[\Phi_{\lambda\mu}]$ when we
expand out the right hand side. To unpick
this we use in turn
\DixonIIIe{(10.9)}, \DixonIIIe{(10.6)}, \DixonIIe{(4.5)}, \DixonIIe{(7.4)}
to give
\begin{align*}
I[\Phi_{\lambda\mu}]
&=
(2\pi)^{-4} \int_\Interval d\sigma \int_{T_{C(\sigma)}}\, Dk\,
\tilde{I}^{\lambda\mu}(\sigma,k) \, \tilde\Phi_{\lambda\mu}(C(\sigma),k)
\\&=
(2\pi)^{-4}
\int_\Interval d\sigma \int_{T_{C(\sigma)}}\, Dk\,
 \sum_{n=0}^\infty \frac{(-i)^n}{n!}
k_{\kappa_1}\cdots k_{\kappa_n} I^{{\kappa_1}\cdots{\kappa_n}\lambda\mu}(\sigma) 
\, \tilde\Phi_{\lambda\mu}(C(\sigma),k)
\\&=
(2\pi)^{-4} \sum_{n=0}^N \frac{1}{n!}
\int_\Interval d\sigma 
I^{{\kappa_1}\cdots{\kappa_n}\lambda\mu}(\sigma)
\nabla_{(\kappa_1\ldots\kappa_n)} 
\, \phi_{\alpha\beta}
\\&\qquad\qquad +
(2\pi)^{-4} 
\int_\Interval d\sigma \int_{T_{C(\sigma)}}\, Dk\,
\sum_{n=N+1}^\infty \frac{(-i)^n}{n!}
k_{\kappa_1}\cdots k_{\kappa_n} I^{{\kappa_1}\cdots{\kappa_n}\lambda\mu}(\sigma) 
\, \tilde\Phi_{\lambda\mu}(C(\sigma),k)
\DEcomma
\end{align*}
since from \DixonIIIe{(10.17)} $\Phi_{\lambda\mu}=\textup{Exp}^A
\phi_{\alpha\beta}$. Here the moments
$I^{{\kappa_1}\cdots{\kappa_n}\lambda\mu}$ satisfy the symmetry
conditions \DixonIIIe{(10.3)}, and orthogonality condition
\DixonIIIe{(10.4)}. Thus the first term in the last expression
corresponds to the right hand side of
(\ref{Intro_Tab_Dixon_Multi_action}) if we set
$\zetaMultiDixon^{\lambda\mu{\kappa_1}\cdots{\kappa_n}}
=(-1)^n I^{{\kappa_1}\cdots{\kappa_n}\lambda\mu}$. Although the
orthogonality condition does not completely correspond. 

For the quadrupole Dixon
\DixonIIIe{(1.37)} constructs $J^{\iMc\iMa\iMb\iMd}=\tfrac14\big(
I^{\iMc\iMa\iMb\iMd} 
- I^{\iMa\iMc\iMb\iMd}
- I^{\iMc\iMa\iMd\iMb}
+ I^{\iMa\iMc\iMd\iMb}\big)$. Since this automatically has the
symmetries of the Riemann curvature tensor it has 20 independent
components. Most of these symmetries are imposed because it is
contracted with the Riemann curvature tensor \DixonIIIe{(1.28), (1.29)}.

The key difference is the imposition of the divergenceless
condition. In \DixonIII/ this is achieved by putting the divergence
operator into the argument of $I$ as seen in the term
$I[\tfrac12\Lambda^k\nabla_{*\{\lambda} G_{\kappa\mu\}}]$ in
\DixonIIIe{(10.16)}. As stated as comment (vii) in \DixonIIIe{ page
  109}, this does not lead to any additional algebraic or differential
equations for the $I^{{\kappa_1}\cdots{\kappa_n}\lambda\mu}$. It then
only affects the dynamics of the dipole \DixonIIIe{(1.28),(1.29)}.
By contrast in our treatment we apply the divergence operator directly
to the distribution, and derive the ODEs and free components of the
$\gamma^{\iMa\iMb\iMc\iMd}$. 
In addition our free components do not have these symmetries.

Future work will be to convert the ODEs for $\gamma^{\iMa\iMb\iMc\iMd}$ into ODEs for the Dixon components $\zetaMultiDixon^{\iMa\iMb\iMc\iMd}$. The moments
$t^{{\kappa_1}\cdots{\kappa_n}\lambda\mu}$ in \DixonIIIe{(1.16)} can
then be related to $\zetaMultiDixon^{\iMa\iMb\iMc\iMd}$ via the squeeze
tensors, but with the coordinates adapted to the hypersurfaces
$\Sigma(s)$ \DixonIIIe{ Just after (10.20)}. 
This will enable a direct comparison between \DixonIII/ and the
results in this article.

\end{document}